\documentclass[12pt]{iopart}
\usepackage{iopams}  
\usepackage{graphicx}
\usepackage{setstack}
\usepackage{color}

\def\tw{t_\mathrm{w}}

\newcommand{\vn}[1]{\bi #1}

\begin{document}

\title{Nature of the spin-glass phase at experimental length scales}

\author{R~Alvarez~Ba\~nos$^{1,2}$, A~Cruz$^{1,2}$, L~A~Fernandez$^{1,3}$,
J~M~Gil-Narvion$^{1}$, A~Gordillo-Guerrero$^{1,4}$, M~Guidetti$^{5}$,
A~Maiorano$^{1,6}$, F~Mantovani$^5$,
E~Marinari$^6$,  V~Martin-Mayor$^{1,3}$, J~Monforte-Garcia$^{1,2}$,
A~Mu\~noz~Sudupe$^3$, D~Navarro$^7$, G~Parisi$^6$, S~Perez-Gaviro$^{1,6}$, 
J~J~Ruiz-Lorenzo$^{1,8}$, S~F~Schifano$^5$, B~Seoane$^{1,3}$, A~Tarancon$^{1,2}$,
R~Tripiccione$^5$, D~Yllanes$^{1,3}$}
\address{$^1$ Instituto de Biocomputaci\'on y F{\'{\i}}sica de Sistemas Complejos (BIFI), Zaragoza, Spain.}
\address{$^2$ Departamento de F\'{\i}sica Te\'orica, Universidad de Zaragoza, 50009 Zaragoza, Spain.}
\address{$^3$ Departamento de F\'{\i}sica Te\'orica I, Universidad Complutense, 28040 Madrid, Spain.}
\address{$^4$ Dpto. de Ingenier\'{\i}a El\'ectrica, Electr\'onica y Autom\'atica, Universidad de Extremadura.
               Avda. de la Universidad s/n. 10071. C\'aceres, Spain.}
\address{$^5$ Dipartimento di Fisica Universit\`a di Ferrara and INFN - Sezione di Ferrara, Ferrara, Italy.}
\address{$^6$ Dipartimento di Fisica, SMC of INFM-CNR and INFN, Universit\`a di Roma La Sapienza, 00185 Roma, Italy.}
\address{$^7$ Departamento de Ingenier\'{\i}a, Electr\'onica y Comunicaciones and Instituto de Investigaci\'on
              en Ingenier\'{\i}a de Arag\'on (I3A), Universidad de Zaragoza, 50018 Zaragoza, Spain.}
\address{$^8$ Departamento de F\'{\i}sica, Universidad de Extremadura, 06071 Badajoz, Spain.}
\eads{yllanes@lattice.fis.ucm.es}

\begin{abstract}
We present a massive equilibrium simulation of the three-dimensional
Ising spin glass at low temperatures.  The Janus special-purpose
computer has allowed us to equilibrate, using parallel tempering,
$L\!=\!32$ lattices down to $T\!\approx\! 0.64 T_\mathrm{c}$. We
demonstrate the relevance of equilibrium finite-size simulations to
understand experimental non-equilibrium spin glasses in the
thermodynamical limit by establishing a time-length dictionary.  We
conclude that non-equilibrium experiments performed on a time scale of
one hour can be matched with equilibrium results on $L\approx110$
lattices.  A detailed investigation of the probability distribution
functions of the spin and link overlap, as well as of their
correlation functions, shows that Replica Symmetry Breaking is the
appropriate theoretical framework for the physically relevant length
scales. Besides, we improve over existing methodologies to ensure
equilibration in parallel tempering simulations.
\end{abstract}
\pacs{75.50.Lk, 75.40.Mg, 75.10.Nr} 

\submitto{Journal of Statistical Mechanics}
\maketitle

\section{Introduction}

Spin Glasses (SG) are disordered magnetic alloys that are generally
regarded as particularly convenient model systems for the study of
glassy behaviour~\cite{mydosh:93,fisher:93}. Indeed, ideas originating
in the SG context have been fruitful in the study of structural
glasses, optimisation in computer science, quantum information,
econophysics, etc.

A distinctive feature of SG is that, below their glass temperature,
they remain out of equilibrium even if they are left to relax under
constant experimental conditions for days or weeks. In spite of this,
the {\em equilibrium} properties of their low-temperature phase is
believed to control their non-equilibrium behaviour. Indeed, both
theory~\cite{ballesteros:00,palassini:99} and
experiment~\cite{gunnarsson:91} agree in that the sluggish dynamics is
due to a {\em thermodynamic} phase transition at a critical
temperature $T_\mathrm{c}$, that separates the paramagnetic phase from
a low-temperature one where the spins freeze according to extremely
complex, essentially unpredictable, ordering patterns. Furthermore, it
has been now established that an accurate knowledge of the
thermodynamic equilibrium properties would allow us to predict in
detail many relevant features of their non-equilibrium
relaxation~\cite{franz:98,franz:99}.

There is an already 30-year-old theoretical controversy regarding the
defining properties of the SG phase. On the one hand, the Replica
Symmetry Breaking (RSB) theory that stems from Parisi's solution of
the SG in the mean field approximation~\cite{mezard:87,marinari:00}.
A system well described by the RSB is in a critical state for all
$T<T_\mathrm{c}$, where the surfaces of the magnetic domains are space
filling. On the other hand, the droplet
theory~\cite{mcmillan:84,bray:87,fisher:86,fisher:88} views the SG
phase as a disguised ferromagnet. It provides the solution of SG
models as computed in the Migdal-Kadanoff
approximation~\cite{gardner:84}.  We refer the reader to
\sref{SECT:MODEL} for the detailed predictions of the RSB and
droplet theories for the different physical observables in the SG
phase. The predictions of the somewhat intermediate TNT
theory~\cite{krzakala:00,palassini:00} are discussed also in
\sref{SECT:MODEL}.

Numerical simulations are the main tool that theoretical physicists
have to make progress in the understanding of the SG phase in
$D\!=\!3$ systems. Basically without exceptions, numerical work in
$D\!=\!3$ is best described by RSB theory (see~\cite{marinari:00} for
a review, refs.~\cite{contucci:06,contucci:07b,contucci:09} for recent
work and refs.~\cite{krzakala:00,palassini:00,jorg:08} for some
somewhat dissenting views). Yet, numerical investigations have received
as well severe criticism. It has been claimed that basically all
simulations doable to date are contaminated by critical
effects~\cite{moore:98}. One would need to simulate still larger
systems at still lower temperatures, in order to observe the
asymptotic behaviour corresponding to large enough systems.

Here we present the results of a large-scale simulation performed on
Janus~\cite{janus:06,janus:08}, a special-purpose computer designed
for the simulation of SG.  For this particular task, Janus outperforms
standard computers by several orders of magnitude. We have
devoted (the equivalent of) 200 days of the full Janus computer to an
equilibrium, parallel-tempering simulation of the Ising SG in
$D\!=\!3$.  We have been able to thermalise lattices of size
$L\!=\!32$ down to temperatures $T\!\approx \! 0.64
T_\mathrm{c}$. This is not only a world record, but provides as well
an unprecedented glimpse on the low temperature SG phase.

Our main objectives here have been (see \sref{SECT:MODEL} for
definitions):
\begin{itemize}
\item To perform a precision comparison of equilibrium and
  non-equilibrium spatial correlation functions. It turns out that a
  time-length dictionary exists, which relates with amazing accuracy
  our previous results at finite times~\cite{janus:08b,janus:09b} (on
  non-equilibrium infinite systems) with equilibrium {\em finite}
  lattice {\em sizes}. The unavoidable conclusion is that experimental
  SG are in the dynamical non-equilibrium regimes that correspond to
  equilibrium results on lattices $L\sim 110$. There is no doubt that
  at these length scales, the appropriate effective theory is RSB,
  irrespectively of which of the competing theories is correct for much
  larger $L$.

\item To perform a study of the probability density function (pdf) of
  the spin overlap, and to extrapolate important quantities to the
  thermodynamic limit. So doing, we will gather important information
  about the correlation length in the spin-glass phase.

\item To provide a detailed study of the link overlap. 

\item Last, but not least, to obtain a large set of configurations,
  fireproof thermalised, which will serve as a starting point for more
  sophisticated studies (such as investigation of ultrametricity, or
  temperature chaos). In particular, a detailed study of the spatial
  correlation functions will appear elsewhere~\cite{janus:10b}.

\end{itemize}

The layout of the rest of this paper is as follows. In
\sref{SECT:MODEL} we briefly recall the definition of the
Edwards-Anderson model. In particular, in \sref{SECT:OBSERVABLES} we
describe the observables considered and discuss the scaling behaviour
predicted for them by the different theoretical scenarios.  In
\sref{SECT:PT-THERM}, we describe our simulations and address the
crucial problem of ensuring thermal equilibrium. We have found it most
useful to study the random walk in temperature space performed in our
parallel-tempering simulations (\sref{SECT:THERMALIZATION-CRITERIA}).
In particular, our thermalisation checks significantly expand the
methodology introduced in~\cite{fernandez:09b}. At this point, we are
ready to study in \sref{SECT:P-DE-Q} the pdf of the spin overlap. In
particular, in \sref{sect:picos} we determine through finite size
effects a correlation length in the spin-glass phase. We focus on the
spatial correlation functions in \sref{SECT:COND}, finding
(\sref{SECT:EQUILIBRIUM-DYNAMICS}) crystal-clear indications of the
relevance of our {\em equilibrium} investigations to the {\em
  non-equilibrium} experimental work. 
The properties of  the link overlap are addressed in
\sref{SECT:LINK-OV}.  Our conclusions are presented in
\sref{SECT:CONCLUSIONS}. Technical details are provided in two
appendices.

\section{Model, Observables, Theoretical expectations}\label{SECT:MODEL}

We divide this section in five paragraphs. In
\sref{SECT:MODELDEF} we describe our model. The spin overlap and
related quantities are defined in \sref{SECT:OBSERVABLES}. We
discuss spatial correlation functions in
\sref{SECT:DEF-CORR}. Their non-equilibrium counterparts are
recalled in \sref{SECT:DEF-CORR-DINAMICA}. We address the link
overlap in \sref{SECT:DEF-QLINK}. Even though most of this
section consists of results and definitions well known in the
spin-glass community, we consider it convenient as a quick
reference. We also introduce some specific (and sometimes new or
seldom used) physical quantities for this paper.
  
\subsection{The model}\label{SECT:MODELDEF}
We consider the $D=3$ Edwards-Anderson
model~\cite{edwards:75,edwards:76}. Our dynamical variables are Ising spins
$s_{\vn{x}}\!=\!\pm1$, which are placed on the nodes, $\vn{x}$,
of a cubic lattice of linear size $L$, containing $V=L^3$ sites, and
with periodic boundary conditions. Their interaction is restricted to
lattice nearest neighbours and is given by the Hamiltonian:
\begin{equation}
{\cal H}=-\sum_{\langle \vn{x}\vn{y}\rangle }\ J_{\vn{x},\vn{y}}\, s_{\vn{x}}\, s_{\vn{y}}\,.\label{EA-H}
\end{equation}
Note that the couplings $J_{\vn{x},\vn{y}}$ in the Hamiltonian are themselves stochastic variables:
they take the values $\pm 1$  with $50\%$ probability.
The coupling constants attached to different lattice links
are statistically independent. The physical motivation for working
with a random Hamiltonian is modelling the effects of impurities in a magnetic
alloy.

We shall consider the {\em quenched} approximation: in the time scale
relevant to the spin dynamics, the impurities can be regarded as
static. Hence, we will not allow for any back-reaction of the spins
over the coupling constants. A given realisation of the $\{
J_{\vn{x},\vn{y}}\}$ (a sample, from now on), will be fixed from the
start and considered non-dynamical~\cite{mydosh:93}. 

A random Hamiltonian implies a double averaging procedure. For any
observable $O$ (an arbitrary function of the spins and the coupling
constants), we shall {\em first} compute the thermal average $\langle
O\rangle$ using the Boltzmann weight at temperature $T$ for the
Hamiltonian (\ref{EA-H}). The average over the coupling constants
distribution, $\overline{\langle O\rangle}\,,$ is only taken
afterwards. We will refer sometimes to the second averaging,
$\overline{(\cdot\cdot\cdot)}$, as disorder average.

The reader will notice that the disorder average induces a non-dynamical gauge
symmetry~\cite{toulousse:77}. Let us choose a random sign per site
$\epsilon_{\vn{x}}=\pm 1\,$. Hence, the energy (\ref{EA-H}) is
invariant under the transformation
\begin{equation}
\begin{array}{rcl}
s_{\vn{x}}&\longrightarrow &\epsilon_{\vn{x}}s_{\vn{x}}\,,\\
J_{{\vn{x}},{\vn{y}}}&\longrightarrow &\epsilon_{\vn{x}}\epsilon_{\vn{y}} J_{{\vn{x}},{\vn{y}}}\,.\label{GAUGE-TRANSF}
\end{array}
\end{equation}
Since the gauge-transformed couplings
$\epsilon_{\vn{x}}\epsilon_{\vn{y}} J_{{\vn{x}},{\vn{y}}}$ are just as
probable as the original ones, the quenched mean value of
$\overline{\langle O(\{s_{\vn x}\})\rangle}$ is identical to that of
its gauge average $\sum_{\{\epsilon_{\vn x}=\pm 1\}} \overline{\langle
  O(\{\epsilon _{\vn x} s_{\vn x}\})\rangle}/2^{L^D}\,,$ which
typically is an uninteresting constant value. We show in
\sref{SECT:OBSERVABLES} how to overcome this problem.

We remark as well that the Hamiltonian (\ref{EA-H}) also has a global
$\mathbf{Z}_2$ symmetry (if all spins are simultaneously reversed
$s_{\vn{x}}\to -s_{\vn{x}}$ the energy is unchanged),
corresponding to time-reversal symmetry. This symmetry gets
spontaneously broken in three dimensions upon lowering the temperature
at the SG transition at $T_\mathrm{c}=1.109(10)$~\cite{hasenbusch:08,hasenbusch:08b}.

\subsection{The spin overlap}\label{SECT:OBSERVABLES}

We need observables that remain invariant under the transformation
(\ref{GAUGE-TRANSF}). The Hamiltonian (\ref{EA-H}) provides, of
course, a first example. To make further progress we consider {\em
  real} replicas $\{s_{\vn{x}}^{(1)}\},\{s_{\vn{x}}^{(2)}\}$, copies
of the system that evolve under the same set of couplings $\{
J_{{\vn{x}},{\vn{y}}}\}$ but are otherwise statistically
uncorrelated.\footnote{For the thermal average of any observable
  depending on a {\em single} spin configuration,
  $O(\{s_{\vn{x}}^{(1)}\})$, we have $\bigl\langle
  O(\{s_{\vn{x}}^{(1)}\})\bigr\rangle^2=\bigl\langle
  O(\{s_{\vn{x}}^{(1)}\})\, O(\{s_{\vn{x}}^{(2)}\})\bigr\rangle$.}

Using them we form the {\em overlap field}:
\begin{equation}
q_{\vn{x}}= s_{\vn{x}}^{(1)} s_{\vn{x}}^{(2)}\,,\label{Q-FIELD-DEF}
\end{equation}
which is obviously invariant under (\ref{GAUGE-TRANSF}). 

The Edwards-Anderson order
parameter, the {\em spin overlap}, is the  spatial average of the overlap field:
\begin{equation}
q=\frac{1}{V}\sum_{\vn{x}} q_{\vn{x}}\,.\label{DEF:Q}
\end{equation}
In particular, it yields the (non-connected) spin-glass susceptibility
\begin{equation}
\chi_\mathrm{NC}(T)= V \overline{\langle q^2\rangle}\,,\label{DEF:CHI}
\end{equation}
that diverges at $T_\mathrm{c}$ with the critical exponent
$\gamma$. For all $T<T_\mathrm{c}$, one expects
$\chi_\mathrm{NC}={\cal O}(V)\,$. We shall also consider the Binder
ratio
\begin{equation}\label{eq:Binder}
B(T)=\frac{\overline{\langle q^4\rangle}}{\overline{\langle
    q^2\rangle}^2}\,,\label{DEF:B}
\end{equation}
In particular, for all $T>T_c$, the fluctuations of $q$ are expected
to be Gaussian in the large-$L$ limit, hence $\lim_{L\to\infty} B=3$,
$(T>T_\mathrm{c})$. Its behaviour in the low-temperature phase is
controversial. For a {\em disguised ferromagnet} picture one expects
$B$ to approach $1$ in the limit of large lattices. On the other hand,
for an RSB system one expects $1<B<3$ in the SG phase
($T<T_\mathrm{c}$).  We recall also that one may consider as well the
overlap computed in small boxes, in order to avoid the effect of the
interphases (physical results are equivalent to those obtained with the
standard overlap~\cite{marinari:98c}).

A great deal of attention will be devoted to the probability density function (pdf)
of the overlap
\begin{equation}
\tilde P(q)=\overline{\biggl\langle\delta\Bigl(q - \frac{1}{V}\sum_{\vn{x}} q_{\vn{x}}\Bigr)\biggr\rangle}\,,\label{DEF:PQ-PEINE}
\end{equation}
Note that, in a finite system, the
pdf is not smooth, but composed of $N+1$ Dirac deltas at $q=-1,-\frac{N-2}{N},\ldots,\frac{N-2}{N},1$. Here, we have solved this problem by a convolution of the comb-like
pdf (\ref{DEF:PQ-PEINE}) with a Gaussian of width $1/\sqrt{V}$, ${\cal G}_V(x)=\sqrt{\frac{V}{2\pi}} \mathrm{exp}[-V \frac{x^2}{2}]\,$:
\begin{equation}
P(q=c)
=\int_{-\infty}^{\infty} \mathrm{d}q'\ \tilde P(q')\, {\cal G}_V(c-q')=\overline{\Bigl\langle\, {\cal G}_V\bigl (c-\frac{1}{V}\sum_{\vn{x}} q_{\vn{x}}\bigr)\,\Bigr\rangle}\label{DEF:PQ-SMOOTH}\,.
\end{equation} 
In this way, we basically add the contribution of ${\cal O}(\sqrt{V})$
microscopic values of $q$, belonging to an interval of width $\sim
1/\sqrt{V}$~\cite{fernandez:09}. Note, however, that eqs.~(\ref{DEF:CHI},\ref{DEF:B}) are
computed out of moments of $\tilde P(q)$, rather than of $P(q)$. 

The Edwards-Anderson order parameter $q_\mathrm{EA}$ vanishes for all
$T\geq T_\mathrm{c}$.  Below $T_\mathrm{c}$, in a droplet system,
$P(q)$ collapses in the large-$L$ limit in a pair of Dirac delta
functions of equal weight, centred at $q=\pm q_\mathrm{EA}$.  In an
RSB system, $P(q)$ contains as well a pair of delta functions at
$q_\mathrm{EA}$, but it also has a continuous piece, non-vanishing for
every $q$ such that $-q_\mathrm{EA}<q<q_\mathrm{EA}$.  This is the
origin of the differences in the predictions that both theories make
for $B$ in the low-temperature phase.

We will find it useful to consider as well {\em conditional} expectation
values at fixed $q$. Let $O$ be an arbitrary function of the spins. We define its conditional expectation
\begin{equation}
\mathrm{E}(O|q\!=\!c)=\overline{\Biggl\langle\, O\  {\cal G}_V\biggl(c-\frac{1}{V}\sum_{\vn{x}} q_{\vn{x}}\biggr)\,\Biggr\rangle}\Biggr/ 
\overline{\Biggl\langle\, {\cal G}_V\biggl(c-\frac{1}{V}\sum_{\vn{x}} q_{\vn{x}}\biggr)\,\Biggr\rangle}\,.\label{DEF:q-PROMEDIO}
\end{equation}
Of course, one may easily recover standard expectation values from $\mathrm{E}(O|q)$:
\begin{equation}
\overline{\langle O\rangle}=\int_{-\infty}^{\infty}\mathrm{d}q\ P(q)\,\mathrm{E}(O|q)\,.\label{EC:RECONSTRUCCION}
\end{equation}
Strictly speaking, the integration limits should be
$\pm\infty$. However, truncating the integral to $-1<q<1$, the error is
exponentially small in $L^{D/2}$ (yet, for $L\!=\!8$ and $12$ we
had to extend the limits beyond $\pm 1$).

We can also define the conditional variances as
\begin{equation}\label{eq:var-q}
\mathrm{Var}(O|q=c) = \mathrm{E}(O^2 | q=c) - \mathrm{E}(O|q=c)^2,
\end{equation}
where we have the identity
\begin{equation}\label{eq:var-q-anchura}
\overline{\langle O^2\rangle} - \overline{\langle O\rangle}^2
= \int_{-\infty}^{\infty} \rmd q \ P(q) 
      \left[ \mathrm{Var}(O|q) + \bigl( \mathrm{E}(O|q) - \overline{\langle O\rangle} \bigr)^2\right].
\end{equation}

\subsection{Spatial correlation functions}\label{SECT:DEF-CORR}

The overlap correlation function is
\begin{equation}\label{eq:C4}
C_4(\vn{r})= \frac{1}{V}\sum_{\vn{x}}\ \overline{\langle q_{\vn{x}}\, q_{\vn{x}+\vn{r}}\rangle}\,.
\end{equation}
$C_4(\vn{r})$ decays to zero for large $\vn{r}$ only for
$T>T_\mathrm{c}$. Thus we have considered as well conditional
correlation functions, recall eq.~(\ref{DEF:q-PROMEDIO}):
\begin{equation}\label{eq:C4-q}
C_4(\vn{r}|q)=\mathrm{E}\left(\left.  \frac{1}{V}\sum_{\vn{x}}\,q_{\vn{x}}q_{\vn{x}+\vn{r}}\right|q\right)\,.
\end{equation}
Eq.~(\ref{EC:RECONSTRUCCION}) allows us to recover $C_4(\vn{r})$ from
$C_4(\vn{r}|q)$.

The two main theoretical pictures for the SG phase, the droplet and
RSB pictures, dramatically differ on their predictions for
$C_4(\vn{r}|q)$. Let us discuss them in detail:
\begin{itemize}
\item In the RSB picture, the {\em connected} correlation functions
  tend to zero at large $\vn{r}$. For all
  $q\in[-q_\mathrm{EA},q_\mathrm{EA}]$ we expect the asymptotic
  behaviour
\begin{equation}
C_4(\vn{r}|q)\sim q^2 + \frac{A_q}{r^{\theta(q)}}+\ldots\,,\label{EQ:SCALINGC4Q-LARGE-L}
\end{equation}
where the dots stand for scaling corrections, subleading in the limit
of large $r$.  The exponent $\theta(q)$ in
eq.~(\ref{EQ:SCALINGC4Q-LARGE-L}) has been computed for $D$ larger
than the upper critical dimension
$D_\mathrm{u}=6$:~\cite{dedominicis:98,dedominicis:99}
\begin{eqnarray}
\theta(q=0)&=&D-4\,,\\ 
\theta(0<|q|<q_\mathrm{EA})&=&D-3\,,\\
\theta(|q|=q_\mathrm{EA})&=&D-2\,.
\end{eqnarray}
These mean-field results for $\theta(q)$ become inconsistent for $D<4$
[the correlations should {\em decrease} for large $r$, implying
  $\theta(q)>0$, recall eq.~(\ref{EQ:SCALINGC4Q-LARGE-L})]. An
expansion in $\epsilon=6-D$ suggests that $\theta(q)$ will
renormalise~\cite{dedominicis:06}. Note as well that, at least for
large $D$, $\theta(q)$ is discontinuous at $q=0$. However, we remark
that there are no compelling theoretical arguments supporting the
discontinuity of $\theta(q)$ in $D=3$.  Indeed, recent numerical
studies found no evidence for it~\cite{contucci:09,janus:10b}.  We
finally recall a non-equilibrium computation~\cite{janus:09b} yielding
in $D\!=\!3$:\footnote{We may mention as well three conjectures:
  $\theta(0)=(D-2+\eta)/2$~\cite{dedominicis:06} (that from the
  results in~\cite{hasenbusch:08b}, yields $\theta(0)=0.313(5)$),
  $\theta(0)=1/\hat\nu$ ($\hat\nu$ is the exponent that rules finite
  size effects at $q_\mathrm{EA}$) and
  $\theta(0)+1/\hat\nu=\theta(q_\mathrm{EA})$. There is also
an exact scaling relation $\theta(q_\mathrm{EA}) = 2/\hat\nu$~\cite{janus:10b}.}
\begin{equation}
\theta(q=0)=0.38(2)\,.
\end{equation}
 
\item Quite the opposite to the RSB case, in a system well described
  by a droplet model and for $|q|< q_\mathrm{EA}$, $C_4(\vn{r}|q)$
  does not tend to $q^2$ for large $r$ (we are referring, of course, to the
regime  $1\ll r\ll L$).
  In fact, spin configurations with $|q|< q_\mathrm{EA}$ are spatially
  heterogeneous mixtures of the two pure phases. One should find {\em
    bubbles} or slabs of linear size $\sim L$ of one of the two
  phases, say $q=+q_\mathrm{EA}$, surrounded by a matrix of the
  complementary state (see
  e.g.~\cite{martin-mayor:07,macdowell:06}). It follows that
\begin{equation}\label{eq:C4-droplets}
C_4(\vn{r}|q)= q_\mathrm{EA}^2f_{\vn{r}/r}(r/L)\,,\quad\mathrm{if}\quad
|q|<q_\mathrm{EA}\ \ \mathrm{and}\ \ 1\ll r\ll L\,,
\end{equation}
($f_{\vn{r}/r}(x)$ is a direction-dependent scaling function with
$f_{\vn{r}/r}(0)=1$).  Indeed, the probability that two spins at fixed
distance $r$ belong to domains of opposite orientation is proportional
to $r/L$ in the large-$L$ limit.  On the other hand, precisely at
$|q|=q_{EA}$ but only there, droplet theory predicts that the
connected correlation function vanishes for asymptotically large $r$.
The same behaviour of eq.~(\ref{EQ:SCALINGC4Q-LARGE-L}) was
predicted~\cite{bray:87}. The exponent $\theta(q_\mathrm{EA})$ is
identical to the scaling exponent of the coupling strength, denoted as
$\theta$ or $y$ in the literature, and has a value of 
$\theta(q_\mathrm{EA})\sim0.2$~\cite{bray:87}.
\end{itemize}

\subsection{Non-equilibrium correlation functions}\label{SECT:DEF-CORR-DINAMICA}

Let us recall that non-equilibrium counterparts exist of $q$ and
$C_4(\vn{r}|q)$.  We shall not be computing them here, but we {\em
  will} compare previous computations with our equilibrium results. Hence, we briefly
recall the definitions~\cite{janus:09b}. One considers pairs of times
$\tw$ and $t+\tw$, with $t,\tw>0$, after a sudden quench from a fully
disordered state to the working temperature $T$. The analogous of the
spin overlap is
\begin{equation}
C(t,\tw)=\frac{1}{V}\sum_{\vn{x}}\ \overline{\langle s_{\vn{x}}(\tw) s_{\vn{x}}(t+\tw)\rangle}\,.
\end{equation}
The non-equilibrium spatial correlation function is
\begin{equation}
C_{2+2}(\vn{r};t,\tw)= \frac{1}{V}\sum_{\vn{x}}\ \overline{\langle s_{\vn{x}}(\tw) s_{\vn{x}}(t+\tw) s_{\vn{x}+\vn{r}}(\tw) s_{\vn{x}+\vn{r}}(t+\tw)\rangle}
\end{equation}
At fixed $\tw$, $C(t,\tw)$ monotonically decreases from $C=1$ at $t=0$,
to $C=0$ at $t\to\infty$. Hence, one may consider $C$, rather than
$t$, as an independent variable. We will compare the non-equilibrium
$C_{2+2}(\vn{r};t,\tw)$, computed in very large
lattices~\cite{janus:08b,janus:09b}, with our equilibrium results for
$C_4\left(\vn{r}|q=C(t,\tw)\right)$. To do so, we shall need to relate the finite
{\em time} $\tw$ (on very large lattices) with the finite {\em size} $L$. As we
shall see in \sref{SECT:EQUILIBRIUM-DYNAMICS}, the correspondence between the
non-equilibrium and the equilibrium correlation functions is amazingly
accurate.

\subsection{The link overlap}\label{SECT:DEF-QLINK}

The link overlap is defined as\footnote{Clearly, $\overline{\langle
    Q_\mathrm{link} \rangle}=C_4(1,0,0)$.}
\begin{equation}
Q_\mathrm{link}=\frac{1}{DV}\sum_{\Vert\vn{x}-\vn{y}\Vert=1} q_\vn{x}q_\vn{y}\,.\label{DEF:QLINK}
\end{equation}
It is a more sensitive quantity than the spin overlap to the
differences between a system described by droplet theory or an RSB
system~\cite{marinari:99}. Since it is invariant under time-reversal
symmetry (the global reversal of every spin in either of our two real replicas
$s_{\vn{x}}^{(i)}\longrightarrow -s_{\vn{x}}^{(i)}$) its expectation value
is non-vanishing, even in a finite system at high temperatures. Its pdf
can be defined as we did with the spin overlap, recall
eqs.~(\ref{DEF:PQ-PEINE},\ref{DEF:PQ-SMOOTH}).  In fact, it has been proposed
that the link overlap (rather than the spin overlap) should be
considered as the fundamental quantity to describe the spin-glass
phase below the upper critical
dimension~\cite{contucci:05b,contucci:06}. There are both physical and
mathematical reasons for this:
\begin{itemize}
\item
On the physical side, $Q_\mathrm{link}$ provides an estimate of the
volume of the domains' surfaces. Indeed, consider two configurations
of the overlap field (\ref{Q-FIELD-DEF}) differing only in that a {\em
  domain} of size $\sim L$ has flipped. This will result in a large
change of the spin overlap, $q$. Yet, the only changing contribution
to $Q_\mathrm{link}$ is that of the lattice links crossed by the
domain's surface.  In a droplet theory, where the surface-to-volume
ratio of the domains vanishes in the large-$L$ limit, one does not
expect any $q$ variation of the conditional expectation
$\mathrm{E}(Q_\mathrm{link}|q)$, not even in the $|q|<q_\mathrm{EA}$
region. Hence, the pdf for $Q_\mathrm{link}$ is expected to collapse
to a single-valued delta function in the large-$L$ limit. The
intermediate TNT picture coincides with the droplet theory in this
respect.  For an RSB system, the domains' surfaces are
space filling. Hence, when $q$ suffers a variation of order 1, the
variation of $Q_\mathrm{link}$ will be of order 1, too. Accordingly, a
non-trivial pdf is expected for $Q_\mathrm{link}$, in the limit of
large systems.
\item
On the mathematical side, theorems have been proven for the link
overlap~\cite{contucci:03,contucci:05,contucci:07}, valid for
three-dimensional systems, which are the exact correlate of mean-field
results for the spin overlap.\footnote{The mathematical proof known so
  far is valid only for Gaussian-distributed couplings in
  eq.(\ref{EA-H}). However, physical intuition strongly suggests that
  the theorems are valid in more general cases such as our bimodal
  couplings.} Specifically, the replica equivalence property holds for
the link overlap in three dimensional systems. Replica
equivalence~\cite{parisi:98,parisi:00} is a property of the Parisi
matrix which yields an infinite hierarchy of identities relating
linear combinations of moments of $Q_\mathrm{link}$ in the
large-$L$ limit. A specific example that we shall be using here is
\begin{equation}
\lim_{L\to\infty} \overline{\langle Q_\mathrm{link}\rangle^2}=\lim_{L\to\infty}\left[\,\frac{2}{3} \,\overline{\langle Q_\mathrm{link}\rangle}^2 \ +\ \frac{1}{3}\, \overline{\langle Q_\mathrm{link}^2\rangle} \,\right]\,,\label{EQ:REPLICA-EQUIVALENCE}
\end{equation}
(at finite $L$, the equality is not expected to hold). This is just a particular
case of the family of identities valid for all $k,s=0,1,2,...$
\begin{equation}
\lim_{L\to\infty} \overline{\langle Q_\mathrm{link}^k\rangle \langle Q_\mathrm{link}^s\rangle}=\lim_{L\to\infty}\left[\,\frac{2}{3}\, \overline{\langle Q_\mathrm{link}^k\rangle}\; \overline{\langle Q_\mathrm{link}^s\rangle} \ +\ \frac{1}{3}\, \overline{\langle Q_\mathrm{link}^{k+s}\rangle} \,\right]\,,\label{EQ:REPLICA-EQUIVALENCE-MAS-GENERAL}
\end{equation}
(replica equivalence implies infinitely many relations such as this).
It is amusing that the mathematical proof for the three-dimensional
theorem does {\em not} use Parisi matrices, relying instead on
stochastic stability. Let us stress that ultrametricity implies
replica equivalence, but the converse statement (i.e. replica
equivalence implies ultrametricity) does not hold, in general.\footnote{For the sake of
  completeness, let us recall that replica and overlap equivalence,
  combined, imply ultrametricity~\cite{parisi:00}. In addition,
  replica equivalence and the Ansatz of a generic ultrametricity
  implies ultrametricity just as in the SK
  model~\cite{iniguez:96}. Finally, we point out that replica equivalence
  is tantamount to stochastic stability and a self-averageness
  property.}
\end{itemize}

The distinction between {\em spin} overlap and {\em link} overlap
seems somewhat artificial from the point of view of the mean-field
approximation.  In fact, in the Sherrington-Kirkpatrick model one
easily shows that $Q_\mathrm{link}=q^2$. For finite-connectivity
mean-field models, non-equilibrium numerical computations yield
$Q_\mathrm{link}=a q^2+b$~\cite{fernandez:09f} ($a$ and $b$ are
numerical constants). In $D\!=\!3$ there are also clear indications
that fixing the spin-overlap fixes as well the link overlap: the
conditional variance $\mathrm{Var}(Q_\mathrm{link} | q)$,
eq.~(\ref{eq:var-q}), tends to zero for large lattices,
see~\cite{contucci:06} and \fref{fig:var-qlink},
below. Furthermore, in a TNT or droplet system, the derivative
$\mathrm{d}E(Q_\mathrm{link}|q)/\mathrm{d} q^2$ should vanish in the
large-$L$ limit for all $|q|<q_\mathrm{EA}$ (since there is a single
valid value for $Q_\mathrm{link}$, there can be no $q^2$ dependency
left). Numerical simulations, both in
equilibrium~\cite{contucci:06,contucci:07b} and out of
equilibrium~\cite{janus:08b,jimenez:03}, find so far a non-vanishing
derivative that nevertheless decreases for larger $L$. The
extrapolation to $L=\infty$ is still an open issue, see
\sref{SECT:OVERLAP-EQUIVALENCE}.

We wish to emphasise that $Q_\mathrm{link}$ unveils that the
spin-glass phase is a critical state where minimal perturbations can produce
enormous changes. 
In fact,
let us couple two otherwise independent copies of the system through
$Q_\mathrm{link}$,
\begin{equation}
{\cal H}=-\sum_{\langle \vn{x}\vn{y}\rangle }\ J_{\vn{x},\vn{y}}\, (s^{(1)}_{\vn{x}}\, s^{(1)}_{\vn{y}}+s^{(2)}_{\vn{x}}\, s^{(2)}_{\vn{y}})\ -\ T \epsilon V Q_\mathrm{link}\,.\label{EQ:QLINK-COUPLING}
\end{equation}
In a system described by droplet theory, one expects the link susceptibility
\begin{equation}
\chi_\mathrm{link}\equiv\left.\frac{\partial \overline{\langle
    Q_\mathrm{link} \rangle}}{\partial\epsilon}\right|_{\epsilon=0}=
V\left[\overline{\langle Q_\mathrm{link}^2 \rangle\ -\ \langle Q_\mathrm{link} \rangle^2}\right]\,,\label{DEF:CHI-LINK}
\end{equation}
to remain finite in the large-$L$ limit, for all $T<T_\mathrm{c}$
(precisely at $T_\mathrm{c}$, a critical divergence might
arise). Hence  $\overline{\langle
  Q_\mathrm{link} \rangle}_\epsilon= \overline{\langle Q_\mathrm{link}
  \rangle}_{\epsilon=0}+ \epsilon \chi_\mathrm{link}+\ldots\,$ in a droplet or
TNT system.

On the other hand, in the mean-field approximation, one finds for RSB
systems a discontinuity with
$\epsilon$~\cite{franz:92}:
\begin{eqnarray}
\overline{\langle Q_\mathrm{link}\rangle}_{\epsilon>0}&=&
  \mathrm{E}(Q_\mathrm{link}|q=q_\mathrm{EA})\ +\ a_+ \sqrt{\epsilon}+\ldots\,,\\
\overline{\langle Q_\mathrm{link}\rangle}_{\epsilon<0}&=&\ 
\mathrm{E}(Q_\mathrm{link}|q=0)- \ a_- \sqrt{-\epsilon}+\ldots\,.
\end{eqnarray}
Actually, the mean-field computation was carried out for the {\em
  spin} overlap, yet, in mean-field models,
$Q_\mathrm{link}$ is essentially $q^2$, hence we can borrow their
result.  We should emphasise that the situation is even more critical
than for standard first-order phase transitions:
$\chi_\mathrm{link}(\epsilon)$ diverges when $\epsilon\to 0\,$ (just
as if the specific heat of liquid water approaching its boiling
temperature showed a divergence!).

Below the upper critical dimension, there has been very little
investigation of $\chi_\mathrm{link}$ (see, however,
ref.~\cite{marinari:99}). In fact, eq.(\ref{EQ:REPLICA-EQUIVALENCE})
has interesting implications in this respect. Let us rewrite it in the
equivalent form
\begin{equation}
\lim_{L\to\infty} \left[\,\overline{\langle Q_\mathrm{link}^2\rangle}\ -\ \overline{\langle Q_\mathrm{link}\rangle^2}\,\right]=\frac{2}{3} \, \lim_{L\to\infty}\left[\, \overline{\langle Q_\mathrm{link}^2\rangle} \ -\ \overline{\langle Q_\mathrm{link}\rangle}^2\, \right]\,,\label{REP-EQUIVALENCE-SENCILLA}
\end{equation}
In an RSB system, the right-hand side of
eq.~(\ref{REP-EQUIVALENCE-SENCILLA}) is positive (since
$Q_\mathrm{link}$ may take values on a finite interval). Yet,
eq.~(\ref{DEF:CHI-LINK}), the lhs of (\ref{REP-EQUIVALENCE-SENCILLA})
is nothing but the large-$L$ limit of $\chi_\mathrm{link}/L^D$. Hence,
RSB implies $\chi_\mathrm{link}\sim L^D$, as expected for first-order
phase transitions (see e.g.~\cite{amit:05}).

We note that for droplet, or TNT systems,
eq.~(\ref{REP-EQUIVALENCE-SENCILLA}) is merely an empty
$0\!=\!\frac{2}{3}\times 0$ statement, just as for RSB systems in
their paramagnetic phase.  Hence we have found of interest to study
the dimensionless ratio
\begin{equation}\label{eq:R-link}
R_{\mathrm{link}}=\frac{\overline{\langle Q_\mathrm{link}^2 \rangle\ -\ \langle Q_\mathrm{link} \rangle^2}}{\overline{\langle Q_\mathrm{link}^2 \rangle}\ -\ \overline{\langle Q_\mathrm{link} \rangle}^2}\,.
\end{equation}
eq.~(\ref{REP-EQUIVALENCE-SENCILLA}) implies that, for an RSB system
on its large-$L$ limit, $R_{\mathrm{link}}=\frac{2}{3}$ for all
$T<T_\mathrm{c}$. For a droplet or TNT system any value $0\leq
R_{\mathrm{link}}\leq 1$ is acceptable. In fact, the high-temperature
expansion for the $D\!=\!3$ EA model tells us that, in the large-$L$
limit, $R_{\mathrm{link}}=1-{\cal O}(T^{-2})$. 

We finally recall that the Chayes et
al. bound~\cite{chayes:86,maiorano:07} may seem to imply that
$\chi_\mathrm{link}$ can diverge at most as $L^{D/2}$, rather than as
$L^D$ as required by RSB. The way out of the paradox is a little
technical.\footnote{Imagine generalising model (\ref{EA-H}) in the
  following sense: the coupling is $J_{\vn{x}\vn{y}}=+1$ with
  probability $p$ (and $J_{\vn{x}\vn{y}}=-1$ with probability $1-p$),
  so that our model is just the particular instance $p=0.5$. One may
  follow ref.~\cite{chayes:86} to show that $\partial\overline{\langle
    Q_\mathrm{link}\rangle}/\partial p$ diverges at most as
  $L^{D/2}$. However, the critical value of $\epsilon$ would still be
  $\epsilon=0$ for $p$ in a finite range around $p=0.5$ (this is the
  crucial point: in the standard argument~\cite{chayes:86,maiorano:07}
  one would require that the critical value of $\epsilon$ vary when
  $p$ moves away from $p=0.5$). Hence, the rate of divergence of
  $p$-derivatives does not convey information on the rate of
  divergence of $\epsilon$-derivatives.}

\section{Numerical methods}\label{SECT:PT-THERM}

We describe here our numerical simulations. We describe the simulation
organisation on Janus in \sref{SECT:JANUS}. We explain our choice
of parameters for the parallel tempering simulation in
\sref{SECT:PT-PARAMETERS}. An absolutely crucial issue is that of
thermalisation criteria, \sref{SECT:THERMALIZATION-CRITERIA}. We
largely extend here the methods first introduced in
ref.~\cite{fernandez:09b}, which allows us to distribute on a rational basis
the computational resources according to the difficulty in
thermalising each particular sample. At variance with
ref.~\cite{fernandez:09b}, which was restricted to the critical region,
we are here probing the deep spin-glass phase, hence more demanding
criteria need to be met. The statistical data analysis is described in
\sref{SECT:MONTECARLO-EVALUATION}. Finally, in
\sref{SECT:THERMALIZATION-TESTS} we describe some more
traditional thermalisation tests.

\subsection{The Janus computer}\label{SECT:JANUS}

Our Monte Carlo simulations have been carried out on the Janus special-purpose
machine. Information about Janus' hardware as well as some
details of low-level programming can be found in
\cite{janus:06,janus:08,janus:09}. Janus is built out of 256 computing
cores (Virtex-4 LX200 FPGAs) arranged on 16 boards. With the code used
for this paper, each core updates $3\times 10^{10}$ spins per second
with a heat bath algorithm. The 16 FPGAs on a board communicate with a
host PC via a 17th on-board control FPGA.

The controlling PC generates the couplings $\{J_{\vn{x}\vn{y}}\}$,
initialises the Janus random number generators, and provides as well
the starting spin configurations.  All the required data is
transmitted to the FPGAs (one FPGA per {\em real} replica) that carry
out both the Heat Bath (HB) updating and the Parallel Tempering (PT)
temperature exchange. Due to the special architecture of Janus, the PT
step is not costless, as we previously need to compute the total
energy for each temperature. We thus equilibrate the computational
cost of both updates by performing several HB sweeps before a PT
temperature swap is attempted.  Fortunately, selecting a modest number of
HB sweeps per PT update hardly affects the efficiency. After a
suitable number of PT cycles, spin configurations of all replicas are
copied to PC memory to take measurements.  The measurement process
on the PC is easily parallelised with the next simulation block in
Janus so that the PC is always ready for the next reading.

During the simulation, we store on disk information about the PT
dynamics (temperature random walk and acceptance rates),
configuration energies, and measurements related to the overlap and
link overlap fields. We also store full spin configurations every
several measurement steps (usually a hundred)  to be
later used for {\it offline\/} measurements (see \sref{SECT:MONTECARLO-EVALUATION})
 or as a checkpoint for
continuing the simulation if needed.

In a few specific cases (namely one $L=24$ sample and four $L=32$
samples) the time required to fulfil our thermalisation criteria was
exceedingly long, more than six months. For these samples we have
accelerated the simulation by increasing the level of parallelism. We
have used a special low-level code that transfers the PT procedure to
the control FPGA. This has allowed us to distribute the set of
temperatures along several FPGAs on a board, speeding up the
simulation accordingly.

For the smaller lattices ($L\le 12$) we substitute the communication
with Janus by a call to a simulation routine in the PC. Although these
simulation are much less demanding, we go down to very small
temperatures. As a consequence, the total cost is not negligible and
we have used a PC cluster to complete the simulations.

\subsection{Choosing parameters for Parallel Tempering}\label{SECT:PT-PARAMETERS}

The key point in a parallel-tempering \cite{hukushima:96,Marinari:98b}
simulation consists in ensuring that each configuration spends enough
time at high temperatures  so that its memory can be erased. Since we
intend to study the physics of the Edwards-Anderson spin glass at very
low temperatures, our simulations are necessarily very long. Because
of this, we do not need to reach temperatures as high as those used in
critical point studies. We can perform a quantitative analysis using
the known behaviour of the heat-bath dynamics above the critical
point.

Following \cite{ogielski:85}, the equilibrium autocorrelation time in the
thermodynamic limit is taken from a power law to a critical divergence
\begin{equation}
\tau_\mathrm{HB}(T)\sim(T-T_\mathrm{c})^{-z\nu}\,.
\label{TAU_DE_T}
\end{equation} 
For instance, for the maximum temperature used in our largest lattice
($L=32$) Ogielski found $\tau_\mathrm{HB}(T)\sim 10^5$
\cite{ogielski:85}. This is several orders of magnitude shorter than
our shortest simulations (see \tref{tab:parameters}).

The choice of the minimum temperature was taken so that the whole
simulation campaign took about 200 days of the whole Janus machine
and so that $T_\mathrm{c} - T_\mathrm{min}\sim L^{-1/\nu}$.
With 4000 samples for $L=16, 24$ and 1000
for $L=32$, this resulted in $T_\mathrm{min}=0.479, 0.625$ and 0.703,
respectively. Smaller lattices, $L=8,12$, were simulated on
conventional computers. In all cases, we simulated four independent
real replicas per sample.

As to the other parallel-tempering parameters, namely the number and
distribution  of intermediate temperatures and the frequency of the parallel
tempering updates, the choice is more arbitrary. We dedicated several weeks of
the machine to test several combinations trying, if not to optimise our
decision, at least to avoid clearly bad choices.

Specifically, we varied the number $N_T$ of temperatures keeping the
acceptance of the parallel-tempering update between 7\% and 36\%. This
corresponds to an increase of roughly a factor of two in $N_T$. Noticing that
the computational effort is proportional to $N_T$, we found that the efficency
hardly changed, even for such a wide acceptance range. Eventually, we chose a
compromise value of about 20\% in the acceptance, resulting in the parameters
quoted on \tref{tab:parameters}. This both avoided unconventionally low
acceptances and saved disk space.

In contrast to conventional computers, Janus needs about as much time
to do a parallel-tempering update than a heat-bath one. Therefore,
while it is customary to perform both updates with the same frequency,
after testing frequencies in the range 1--100 we have chosen to do a
parallel-tempering update each 10 heath-bath ones. In fact, even if
the time to do a parallel-tempering step were negligible, we have
checked that doing a single heat-bath between parallel temperings
would produce a practically immeasurable gain. We note, finally, that
this issue was investigated as well in ref.~\cite{bittner:08} (in
that work clear conclusions were not reached, as far as the $D\!=\!3$
Edwards-Anderson model at low temperatures and large $L$ is
concerned).

\begin{table}
\centering
\caption{Parameters of our parallel-tempering simulations. In all
  cases we have simulated four independent real replicas per
  sample. The $N_T$ temperatures are uniformly distributed between
  $T_\mathrm{min}$ and $T_\mathrm{max}$ (except for the runs of the
  first row, which have all the temperatures of the second one plus
  $T=0.150$ and $T=0.340$).  In this table $N_\mathrm{mes}$ is the
  number of Monte Carlo Steps between measurements (one MCS consists
  of 10 heat-bath updates and 1 parallel-tempering update). The
  simulation length was adapted to the thermalisation time of each
  sample (see section~\ref{SECT:THERMALIZATION-CRITERIA}). The table
  shows the minimum, maximum and medium simulation times
  ($N_\mathrm{HB}$) for each lattice, in heat-bath steps.  Lattice
  sizes $L=8,12$ were simulated on conventional PCs, while sizes
  $L=16,24,32$ were simulated on Janus. Whenever we have two runs with
  different $T_\mathrm{min}$ for the same $L$ the sets of simulated samples
  are the same for both. The total spin updates for all lattice sizes sum
  $1.1\times 10^{20}$.}
\label{tab:parameters}
\begin{tabular*}{\columnwidth}{@{\extracolsep{\fill}}ccccrlllcc}
\br
 $L$ &  $T_{\mathrm{min}}$ &  $T_{\mathrm{max}}$ &
  $N_T$ &  $N_\mathrm{mes}$ &  $N_\mathrm{HB}^\mathrm{min}$ &
\multicolumn{1}{c}{ $N_\mathrm{HB}^\mathrm{max}$}& \multicolumn{1}{c}{ $N_\mathrm{HB}^\mathrm{med}$} &
 $N_\mathrm{s}$ & System\\
\mr
 8 & 0.150 & 1.575 & 10  & $10^3$ & $5.0\!\times\! 10^6$ & $8.30\!\times\!10^8$    & $7.82\!\times\!10^6$ & 4000 & PC    \\ 
 8 & 0.245 & 1.575 &  8  & $10^3$ & $1.0\!\times\! 10^6$ & $6.48\!\times\!10^8$    & $2.30\!\times\!10^6$ & 4000 & PC    \\
12 & 0.414 & 1.575 & 12  & $5\!\times\! 10^3$ & $1.0\!\times\! 10^7$ & $1.53\!\times\!10^{10}$ & $3.13\!\times\!10^7$ & 4000 & PC \\
16 & 0.479 & 1.575 & 16  & $10^5$ & $4.0\!\times\! 10^8$ & $2.79\!\times\!10^{11}$ & $9.71\!\times\!10^8$ & 4000 & Janus \\
24 & 0.625 & 1.600 & 28  & $10^5$ & $1.0\!\times\! 10^9$ & $1.81\!\times\!10^{12}$ & $4.02\!\times\!10^9$ & 4000 & Janus \\
32 & 0.703 & 1.549 & 34  & $2\!\times\!10^5$ & $4.0\!\times\! 10^9$ & $7.68\!\times\!10^{11}$ & $1.90\!\times\!10^{10}$ & 1000 & Janus \\
32 & 0.985 & 1.574 & 24  & $2\!\times\!10^5$ & $1.0\!\times\! 10^8$ & $4.40\!\times\!10^9$ & $1.16\!\times\!10^8$ & 1000 & Janus \\
\br
\end{tabular*}
\end{table}

\subsection{Thermalisation criteria}\label{SECT:THERMALIZATION-CRITERIA}

In order to optimise the amount of information 
one can obtain given a computational budget, 
the length of the simulations must be carefully 
selected. It is well known that sample-to-sample 
fluctuation is the main source of statistical error.
Thus, we want to simulate each sample for 
the shortest time that ensures thermalisation.

The most common robust thermalisation check consists
in the determination of the autocorrelation times
for physical observables~\cite{sokal:97}. However, in order
for this determination to be precise one needs 
a much longer simulation than needed to thermalise 
the system (e.g., while ten exponential autocorrelation
times can be enough to thermalise the system, 
we need an at least ten times longer simulation to determine
this autocorrelation time).
Notice that this is not an issue in ordered systems,
where one employs very long simulations in order to
reduce statistical errors.

The typical practical recipe to assess thermalisation 
for disordered systems consists in studying the time evolution
of the disorder-averaged physical observables. In particular, 
the so-called $\log_2$-binning procedure uses the evolution
of the time averages along the intervals 
$I_n = (2^{-(n+1)}N_\mathrm{HB}, 2^{-n} N_\mathrm{HB}]$.
The system is considered to be thermalised if
the first few intervals are compatible.

This procedure is not optimal, because the thermalisation time is
wildly dependent on the sample. Thus, a simulation time long enough to
thermalise the slowest samples will be excessive for most of the rest.
Perhaps even more frightening, the average over samples may well hide
that a few samples, the very worst ones, are still quite far from
equilibrium.

Fortunately the use of parallel tempering 
presents us with the possibility to use
the dynamics of the temperature random walk
to learn about the thermalisation scale
for each sample. In fact, in order to ensure
thermalisation each of the participating 
configurations must cover the whole temperature
range. Here, expanding on a method first used
in~\cite{fernandez:09b}, we have promoted this idea to a fully
quantitative and physically meaningful level.

\begin{figure}[t]
\centering
\includegraphics[height=\linewidth,angle=270]{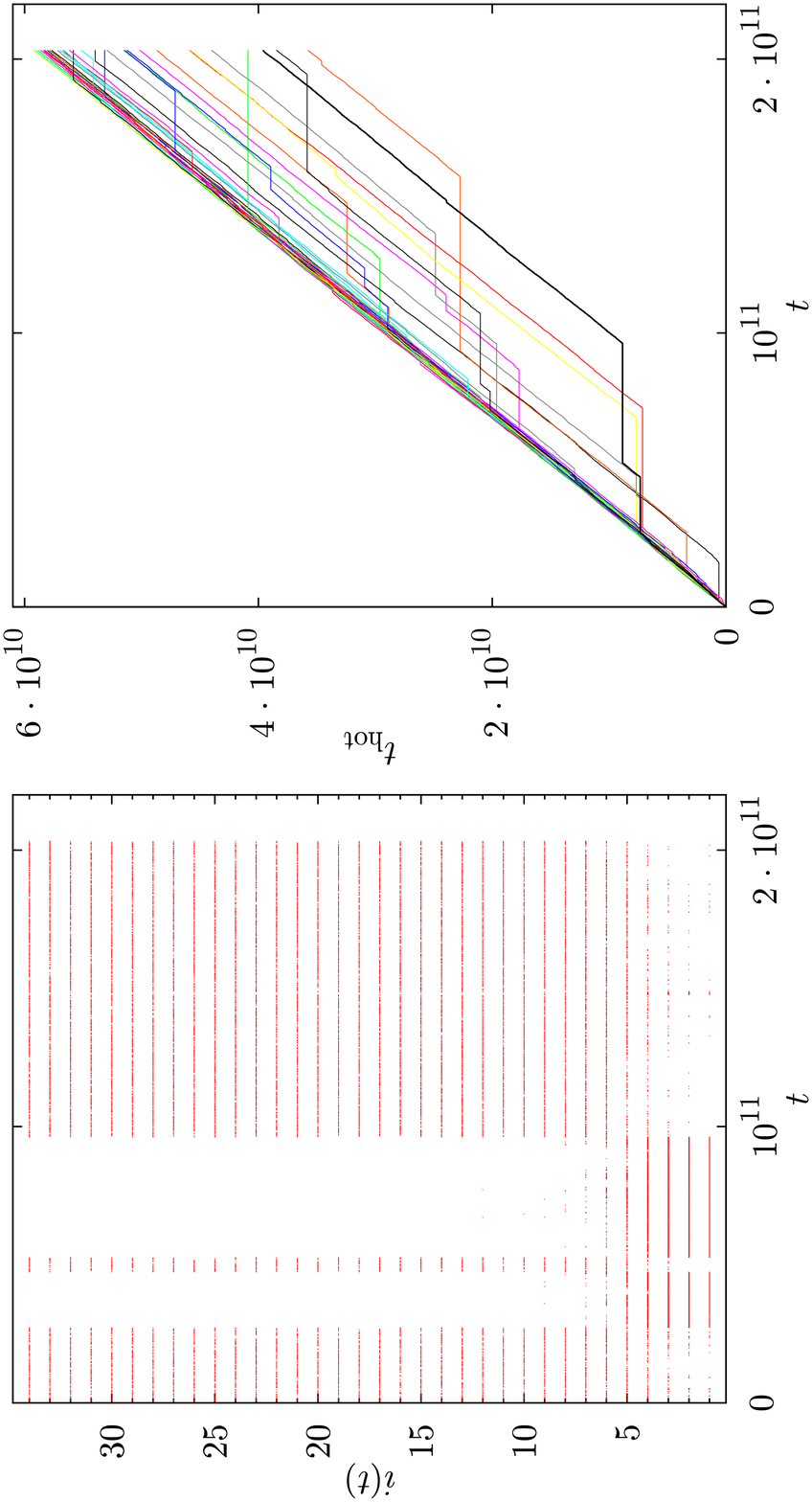}
\caption{We plot (left panel) the temperature index of a fixed
  configuration of an $L=32$ sample as a function of the number of HB
  sweeps.  We plot one point every $5$ million HB sweeps.
  The critical temperature corresponds to
  $i_\mathrm{c}=17$. This specific sample has
  $\tau_\mathrm{exp}=1.75\times 10^{10}$ HB sweeps. In the right
  panel, we show the time that all configuration of a same replica
  spends in the paramagnetic phase.}
\label{fig:historiabetas}
\end{figure}

Let us consider the ordered set of $N_{T}$ temperatures
$\{T_1,\ldots,T_{N_T}\}$ and let us suppose that
$T_{i_\mathrm{c}-1}<T_\mathrm{c}\leq T_{i_\mathrm{c}}$.  In
\fref{fig:historiabetas}---left we show an instance of the
random walk of the temperature index, $i(t)\in\{ 1,2,\ldots, N_T\}$,
performed by one of the $N_T$ copies of the system considered in the
parallel tempering.  The random walk is clearly
not Markovian, as the system remembers for a long time that it belongs
to the high (low) temperature phase. This effect is also demonstrated
in \fref{fig:historiabetas}---right, where we plot the time spent
  over $T_\mathrm{c}$ as a function of the simulation time (mind the
  long plateaux).

To make these arguments quantitative, we shall use the standard tools
of correlated time series~\cite{amit:05,sokal:97}. We need a mapping defined
on the $1,\ldots, N_T$ range of temperature indices so that
\begin{eqnarray}
f(i) \geq 0,  \qquad \forall i  \geq i_\mathrm{c}, \label{eq:f1}\\
f(i)  <  0,    \qquad \forall i < i_\mathrm{c},    \label{eq:f2}\\
\sum_{i=1}^{N_T} f(i)  = 0. \label{eq:suma-f}
\end{eqnarray}
It is also convenient that $f$ be monotonic. Because we have
chosen the same number of temperatures above and below $T_\mathrm{c}$, 
a simple linear $f$ is suitable, but the method works with any function
fulfilling the above conditions.

\begin{figure}
\centering
\includegraphics[height=\linewidth,angle=270]{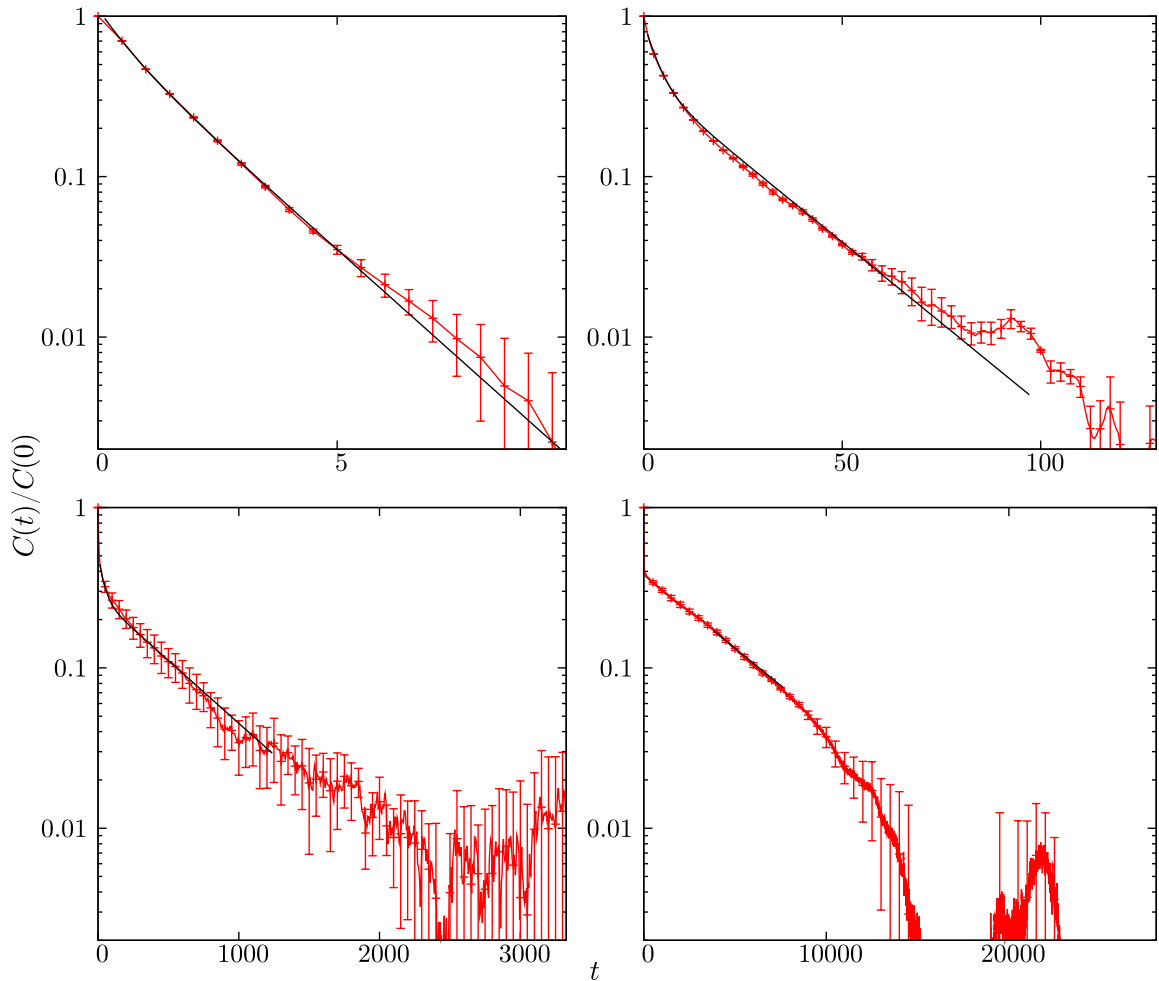}
\caption{Autocorrelation functions for samples with
  $\tau_\mathrm{exp}$ of different orders of magnitude. We have plotted
  the range $[0,6\tau_\mathrm{exp}]$. We include the automatic double
  exponential fit, see~\ref{sec:protocol}. In the last panel the fit
  fails due to the strong downwards fluctuation and our programme has
  chosen a restricted interval for a fit to a single exponential. In
  order to avoid cluttering the graphs, we have only plotted a few
  times (the actual correlation functions have many more points). The
  horizontal axis is in units of $10^6$ heat-bath updates.}
\label{fig:corr}
\end{figure}

For each of the participating configurations we
consider the time evolution $i_t$ of the temperature 
index. We define the equilibrium autocorrelation function as
\begin{equation}
C(t) = \frac{1}{N_\mathrm{HB}-t_0-t}\sum_{t'=t_0}^{N_\mathrm{HB}-t} f(i_{t'}) f(i_{t'+t}) ,
\end{equation}
where $t_0$ is long enough to ensure that the temperature random walk has
reached a steady regime. Due to condition~(\ref{eq:suma-f}), we avoid
subtracting the squared mean value of $f$ in this definition.
From the normalised $\hat C (t) = C(t) /C(0)$, see e.g. \fref{fig:corr}, 
we can define the integrated correlation times:
\begin{equation}\label{eq:tau-int}
\tau_\mathrm{int} = \frac12 + \sum_{t=0}^{W} \hat C(t),
\end{equation}
where $W$ is a self-consistent window that avoids the divergence
in the variance of $\tau_\mathrm{int}$.

The great advantage of these functions over the physical observables
is that we can average over the $N_T$ configurations
in the parallel tempering.\footnote{Even if these are not completely 
statistically independent, the averaged autocorrelation has a much
smaller variance. In addition, the need to simulate several replicas
provides independent determinations of $C(t)$, which permits a further
error reduction and an estimate of the statistical errors.} 

This procedure works surprisingly well, not only giving reliable
estimates of the integrated time but even providing the, more
physical but notoriously difficult to measure, exponential 
autocorrelation time. Indeed, the correlation function
admits an expansion on exponentially decaying modes
\begin{equation}\label{eq:corr}
\hat C(t) = \sum_i A_i\ \rme^{-t/\tau_{\mathrm{exp},i}},\quad \sum_i A_i=1.
\end{equation}

\begin{figure}
\centering
\includegraphics[height=0.7\linewidth,angle=270]{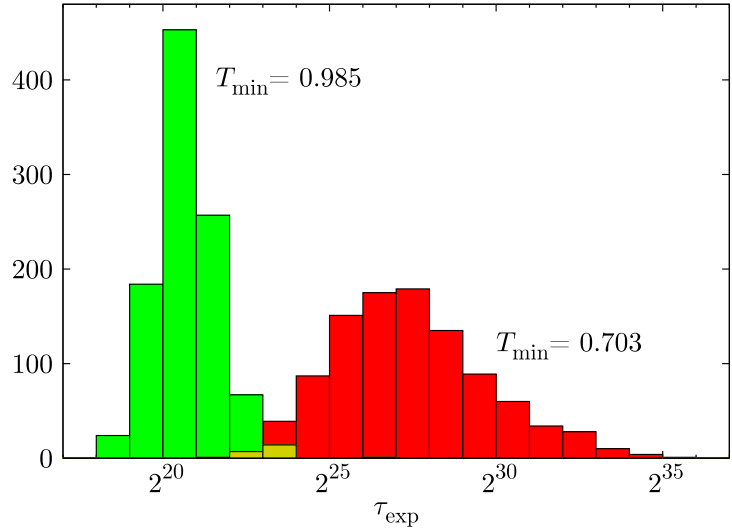}
\caption{Histogram of exponential autocorrelation times for our simulations of
the $L=32$ lattice (1000 samples).}
\label{fig:histograma-taus}
\end{figure}

\begin{figure}
\centering
\includegraphics[height=0.7\linewidth,angle=270]{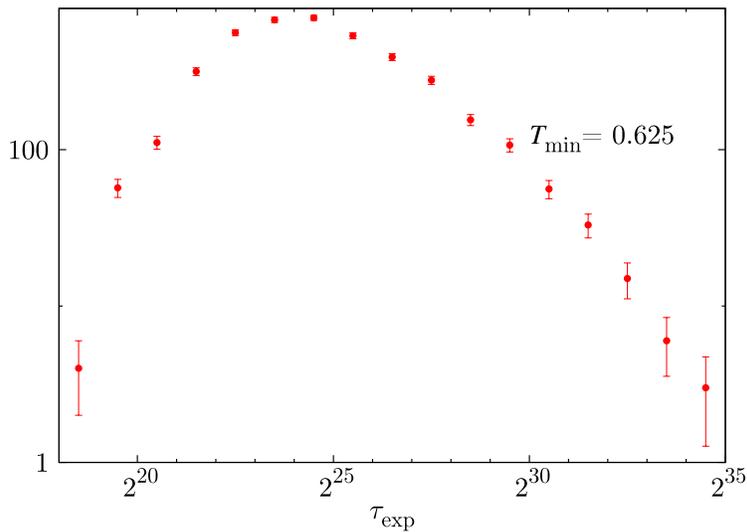}
\caption{Logarithm of the histogram of exponential autocorrelation
  times for our simulations of the $L=24$ lattice (4000 samples). Mind
  the  behaviour of the long-times tail.}
\label{fig:histograma-taus_L24_log}
\end{figure}
In this representation, the exponential time $\tau_\mathrm{exp}$ is
the largest of the $\tau_{\mathrm{exp},i}$.\footnote{The number of
  modes equals the dimension of the dynamical matrix of the Monte
  Carlo Markov process, which in our case is $(N_T!)\times 2^{N_T
    V}$.}  Barring symmetry considerations, this exponential time
should be the same for all random variables in the simulation,
including the physical observables.

The relative sizes of the $A_i$, and hence $\tau_\mathrm{int}$, depend to a
certain extent on the particular choice of $f$. Notice, however, that
criteria~(\ref{eq:f1}--\ref{eq:f2}) select a family of functions that
hopefully reduce the amplitude of the irrelevant fast modes. In any case,
$\tau_\mathrm{exp}$ has a physical meaning independently of these somewhat
arbitrary considerations.

In practice, the simulations are too long (up to $N_\mathrm{HB}\sim 10^{12}$)
to consider all the $f(i_t)$ individually and we
have to introduce some data binning, averaging over a large 
number of consecutive measurements. As it turns out, this is 
not a very limiting issue for two reasons. On the one hand, 
as long as these bins are much shorter than $\tau$, there
is no real information loss. On the other hand, one can reconstruct
any polynomial $f$ up to degree $k$ ---in particular our linear $f$--- by saving the sums of
the first $k$ powers of the $i_t$. 

Even after this binning, we have worked with time series 
with a length of up to several million, so in order to 
compute the autocorrelation we have used a Fast Fourier Transform
algorithm~\cite{frigo:05}.

The details of the chosen thermalisation protocol can be found in
\ref{sec:protocol}. We summarise by saying that our main thermalisation
criterion is ensuring that $N_\mathrm{HB}> 12 \tau_\mathrm{exp}$
($2\tau_\mathrm{exp}$ are discarded and the remaining $10\tau_\mathrm{exp}$
are used to measure and study $\hat C(t)$).

In \fref{fig:corr} we plot several autocorrelation functions showing how
the data quality allows for an exponential fit. We have chosen randomly
4 samples with very different exponential autocorrelation 
times: $6.5\times 10^6$, $8.8\times10^7$, $1.5\times 10^9$ and $1.8\times 10^{10}$.

To summarise the distribution of the exponential autocorrelation times we have
computed a histogram. Due to the large dispersion of these quantities we have
chosen $\log_2 \tau_\mathrm{exp}$ as a variable. In
\fref{fig:histograma-taus} we show the results for the two runs
performed in $L=32$ (see table~\ref{tab:parameters}). Notice the dramatic
increase of the $\tau_\mathrm{exp}$ when decreasing the minimum temperature
of the simulation. The smooth shape of the curves defined by the histogram is
a further test of our procedure for determining autocorrelation times.

In \fref{fig:histograma-taus_L24_log} we plot the logarithm of
the histogram in the $L=24$ case to show the exponential behaviour of
the long-times tail. This result gives confidence that rare events,
with very large (logarithms of) autocorrelation times, are at least
exponentially suppressed. We have not made efforts to measure with
precision the small autocorrelation times as they are immaterial
regarding thermalisation, which is ensured by the minimum number of
iterations performed for all samples.

\subsection{Monte Carlo evaluation of observables}\label{SECT:MONTECARLO-EVALUATION}

We present now some technical details about our evaluation of
mean values, functions of mean values and error estimation.

Some of the observables considered in this work were obtained by means
of an online analysis: the internal energy, the link overlap, powers
of the spin overlap ($q,q^2,q^4$), and Fourier transforms of the
correlation function $C_4(\vn{r})$ for selected momenta. These
quantities were computed as Monte Carlo time averages along the
simulation. Note that the length of the simulation is sample
dependent, something that would be a nuissance in a multispin coding
simulation, but not in Janus were each sample is simulated
independently.  The disorder averaging followed the Monte Carlo
one. Statistical errors were computed using a jackknife method over
the samples, see for instance~\cite{amit:05}.

However, when designing the simulation, one cannot anticipate all
quantities that would be interesting, or these can be too expensive to be
computed in runtime. In particular, we did not compute the
conditional correlation functions $C_4(\vn{r}|q)$. Fortunately, an
offline analysis of the stored configurations has allowed us to
estimate them. We had to overcome a difficulty, though, namely the
scarcity of stored configurations.  In fact, for the samples that were
simulated only for the minimum simulation time, we had only
$N_\mathrm{conf} \sim 100$ configurations stored on disk (ranging from
$N_\mathrm{conf}=10$ for $L=12$ to $N_\mathrm{conf} = 200$ in the case
$L=32$).  We regard the second half (in a Monte Carlo time sense) of
these configurations as fireproof thermalised. Yet, when forming the
overlap field, eq.~(\ref{Q-FIELD-DEF}), one needs only that the two
spin configurations, $\{s_{\vn{x}}^{(1)}\}$ and
$\{s_{\vn{x}}^{(2)}\}$, be thermalised and independent. Clearly
enough, as long as the two configurations belong to different real
replicas and belong to the second half of the Monte Carlo history they
will be suitable. There is no need that the two configurations were
obtained at the same Monte Carlo time (as it is done for the online
analyses). Furthermore, the four real replicas offer us 6 pair
combinations.  Hence, we had at least $6\times (N_\mathrm{conf}/2)^2
\sim 10000$ (60000 for $L=32$) measurements to estimate the overlaps
and the correlation functions. We used the Fast Fourier Transform to
speed up the computation of the spatial correlations.  For those
samples that had more configurations (because their total simulation
time exceeded $N_\mathrm{min}^\mathrm{HB}$), we considered
nevertheless $N_\mathrm{conf}/2$ configurations evenly spaced along
the full second half of the simulation. When some quantity, for
instance the $P(q)$, could be computed in either way, online or
offline, we have compared them. The two ways turn out to be not only
compatible, but also equivalent from the point of view of the
statistical errors. As an example of this let us compute the following
quantity:
\begin{equation}
\sigma_\mathrm{link}^2 = \overline{\langle Q_\mathrm{link}^2\rangle} - \overline{\langle Q_\mathrm{link}\rangle}^2.
\end{equation}
For $L=32$, $T = 0.703$, the value of $\sigma_\mathrm{link}^2$ computed from 
online measurements of $Q_\mathrm{link}$ and $Q_\mathrm{link}^2$ is
\begin{equation}
V \sigma_\mathrm{link, online}^2 = 50.88(90).
\end{equation}
We could now recompute this value from offline measurements of $Q_\mathrm{link}$
and $Q_\mathrm{link}^2$. Instead, we are going to use eq.~(\ref{eq:var-q-anchura}),
which involves the intermediate step of computing
conditional expectation values and variances at fixed $q$ 
and then integrating with $P(q)$. This will serve as a test both 
of the offline measurements' precision  and of our Gaussian convolution method
for the definition of clustering quantities. The result is
\begin{equation}\label{eq:sigma-link}
V \sigma_\mathrm{link, conf}^2 = 50.81(90),
\end{equation}
The precision of $\sigma_\mathrm{link, online}^2$ and $\sigma_\mathrm{link, conf}^2$
is the same and the difference less than $10\%$ 
of the error bar, even though we only analysed $100$ configurations per 
sample for the second one. Of course, both determinations are very highly 
correlated, so the uncertainty in their difference is actually much smaller than their individual
errors. Computing the difference for each jackknife block we see that
\begin{equation}
V [\sigma_\mathrm{link,conf}^2 - \sigma_\mathrm{link, online}^2] = -0.065(79),
\end{equation}
which is indeed compatible with zero.

A subtle point regards non-linear functions of thermal mean values
that are later on averaged over the disorder. In this work, the only
instance is $\chi_\mathrm{link}$, see eq.~(\ref{DEF:CHI-LINK}). Care
is needed to estimate such non-linear functions because a naive
evaluation would be biased, and the bias might be sizeable compared to
the statistical errors~\cite{ballesteros:97}. This problem does not
arise in non-linear functions such as eq.~(\ref{DEF:q-PROMEDIO}),
which are computed on observables only {\em after} the double
averaging process over the thermal noise and over the samples. The
problem and several solutions are discussed in~\ref{AP:NON-BIAS} (see
also~\cite{hasenbusch:08b}).

A final issue is the comparison of data computed in different system
sizes at the {\em same} temperatures.  Unfortunately the grids of
temperatures that we used for the different $L$ differ. Hence we have
interpolated our data by means of a cubic spline.

\subsection{Thermalisation tests}\label{SECT:THERMALIZATION-TESTS}

We will consider in this subsection thermalisation tests directly based
on physically interesting quantities.

\begin{figure}[b]
\centering
\includegraphics[height=0.7\linewidth,angle=270]{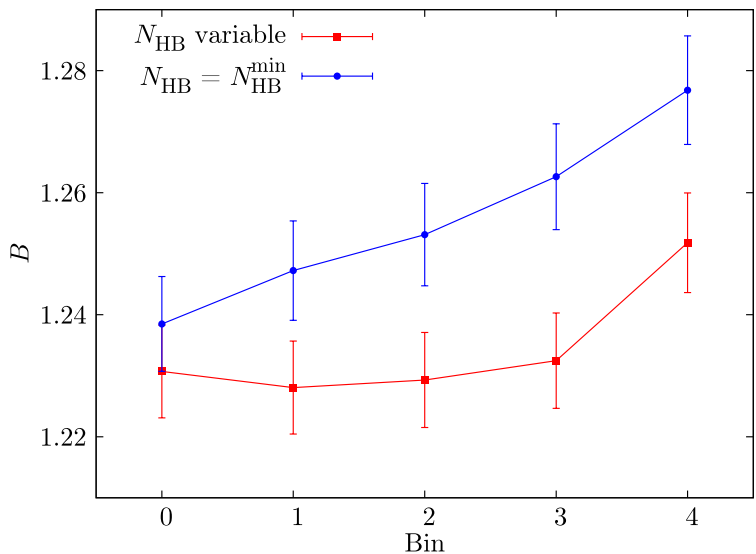}
\caption{Evolution of the Binder parameter for $L=32$, $T=0.703$ using
  $\log_2$ binning (0 = second half, 1 = second quarter, ...). The blue curve
  (circles) is the result of stopping at step 1 of our thermalisation
  protocol (i.e., all samples simulated for a fixed time of $4\!\times\! 10^9$
  heat-bath updates). The red curve (squares) is the result of completing all
  the steps, which implies an increase of roughly 150\% in simulation time.}
\label{fig:log2}
\end{figure}

We start with the traditional $\log_2$-binning procedure. We choose
the Binder parameter for the overlap, see eq.~(\ref{DEF:B}), which is
specially sensitive to rare events.  In \fref{fig:log2} we show
the results for $B(T_\mathrm{min})$ for $L=32$, considering only the first
$4\times 10^9$ Heat Bath steps of each of our 1000 samples, as if 
all the simulations were $N_\mathrm{min}^{\mathrm{HB}}$ heat-bath steps long (blue
line). We could not affirm that even the last two bins were stable within
errors. Things change dramatically if we consider Monte Carlo
histories of a length proportional to the exponential autocorrelation
time. Note that, thanks to our choice of $N^\mathrm{HB}_\mathrm{min}$ in
\tref{tab:parameters}, the simulation time for most samples
has not increased.  If we first rescale data according to the
total simulation length (itself proportional to the autocorrelation
time) and average for equal {\it rescaled} time, the $\log_2$-binning
procedure gives 4 steps of stability within errors. That is to say: we
obtain the Binder parameter without thermalisation bias just
discarding 1/16 of the history (and taking up to 1/8).  Regarding the
Binder parameter our requirement of $12\tau_\mathrm{exp}$ is excessive.

In retrospect (see \fref{fig:log2}), shorter simulations would
have produced indistinguishable physical results for most observables.
We do not regret our choices, however, as we plan to use these
thermalised configurations in the future~\cite{janus:xx} for very
delicate analyses (such as temperature chaos), which are much more
sensitive to thermalisation effects.

\begin{figure}
\centering
\includegraphics[height=0.7\linewidth,angle=270]{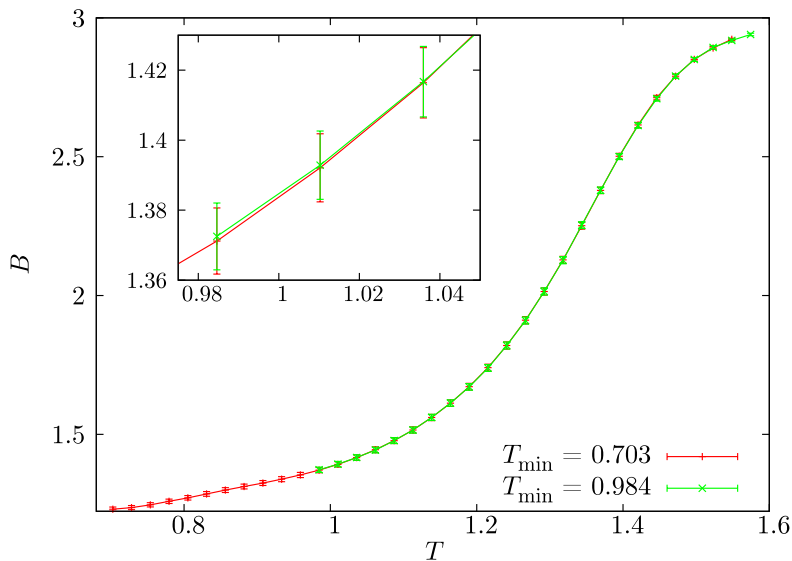}
\caption{Binder ratio as a function of the temperature for $L=32$. The good
  overlap between two different simulations (one of them in the much easier
  critical region) is a further thermalisation check. We use the same set of
  1000 samples.}
\label{fig:Binder-T}
\end{figure}

A different test can be performed by comparing the difficult
low-temperature simulations of our largest lattice with simulations in
the critical region of the {\em same samples}. A faulty thermalisation
(for instance, a configuration remains trapped at low temperatures)
could be observable as inconsistencies in the values of quantities in
common temperatures. In \fref{fig:Binder-T} we show the Binder
parameter as a function of temperature for the two simulations with
$L=32$ (see \tref{tab:parameters}). The agreement between both
simulations is excellent.

\begin{figure}
\centering
\begin{minipage}{.49\linewidth}
\includegraphics[height=\linewidth,angle=270]{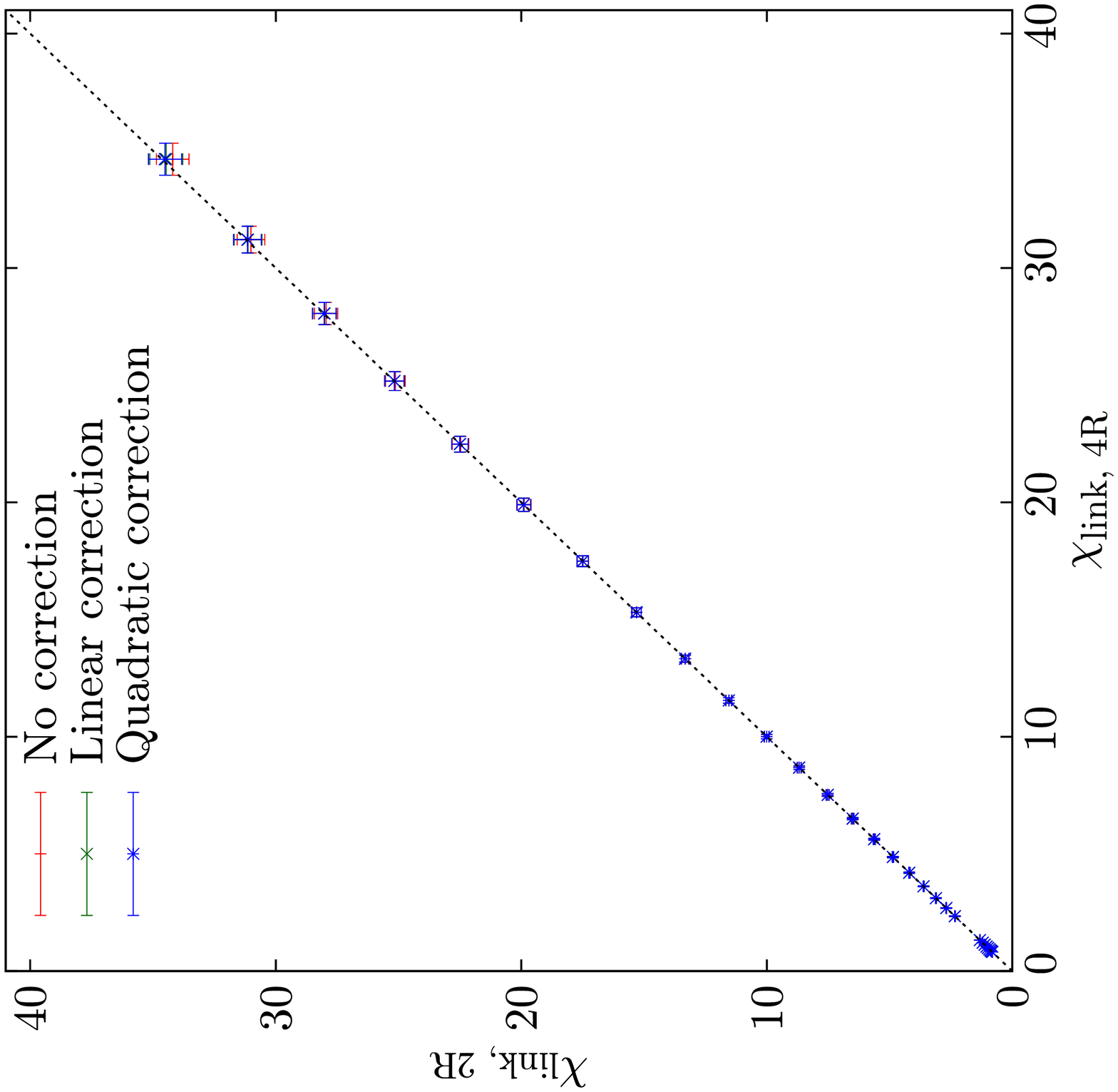}
\end{minipage}
\begin{minipage}{.49\linewidth}
\includegraphics[height=\linewidth,angle=270]{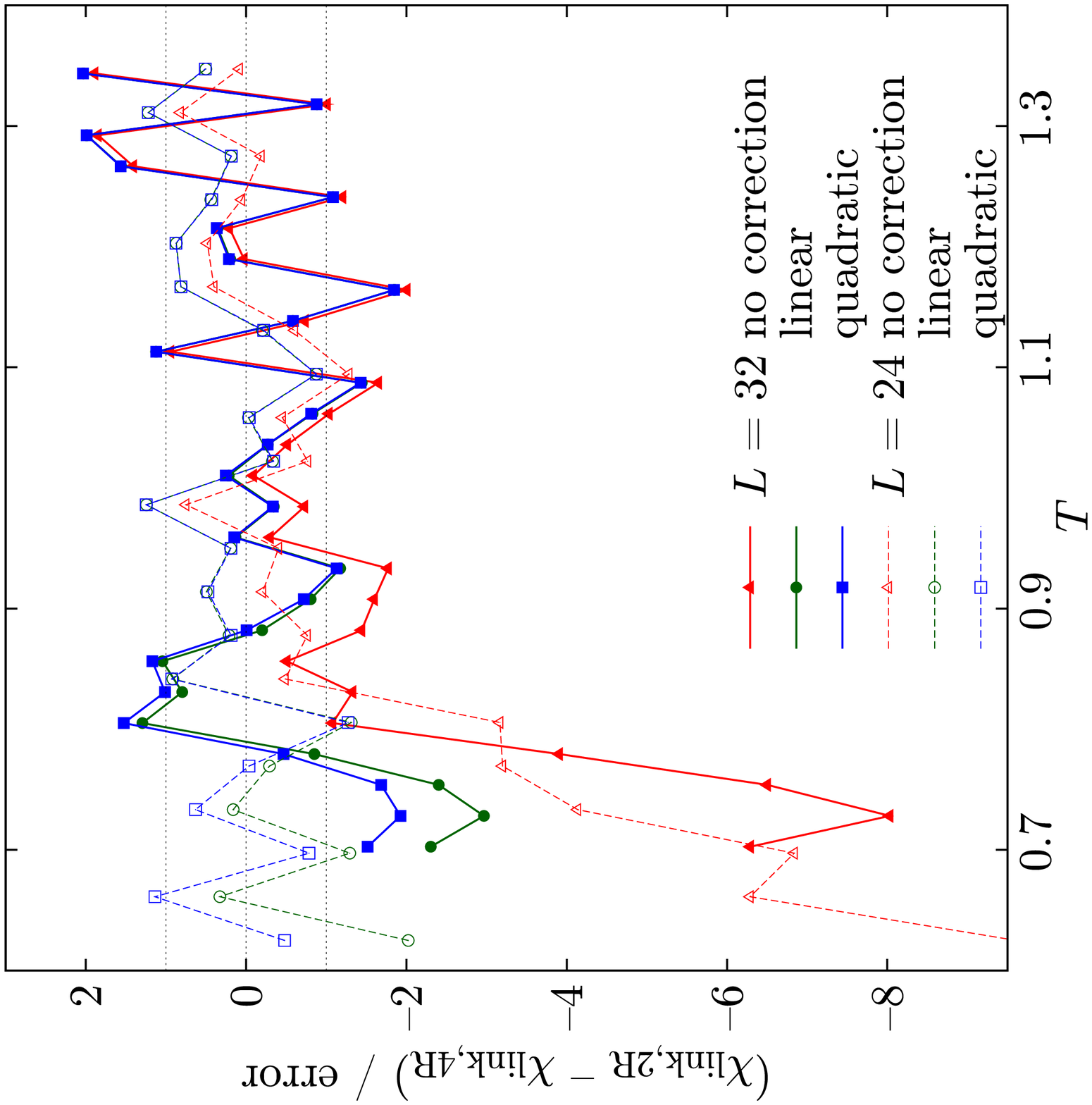}
\end{minipage}
\caption{Bias correction in the computation of $\chi_\mathrm{link}$,
  eq.~(\ref{DEF:CHI-LINK}).  On the left panel we plot the two-replica
  estimators $\chi_\mathrm{link, 2R}$ as a function of the unbiased
  four-replica estimator $\chi_{\mathrm{link, 4R}}$,
  eq.~(\ref{CHI-4R}), for all our temperatures in the $L = 32$
  lattice.  The two-replica estimators $\chi_\mathrm{link, 2R}$ are
  computed with no bias correction, eq.~(\ref{CHI-2R-NO}), with linear
  corrections, eq.~(\ref{CHI-2R-LINEAR}), and with quadratic
  corrections, eq.~(\ref{CHI-2R-QUADRATIC}).  The right panel
  displays, for the three two-replica estimators, their difference
  with the four-replica estimator {\em in units of the statistical
    error for that difference}, as a function of temperature. We show
  our data for $L=32$ and $L=24$. Note that the statistical error in
  the {\em difference} between two estimators is largely reduced (as
  compared to individual errors) due to dramatic data correlation.}
\label{fig:diff-chi-link}
\end{figure}
A very different test on the statistical quality of our data is the
comparison of the values of $\chi_\mathrm{link}$ obtained using the
different possible estimators for $\langle
Q_\mathrm{link}\rangle^2$. We have an unbiased estimator if we use
$Q_{\mathrm{link,4R}}^{(2)}$, see eq.~(\ref{BIAS-CORRECTION2}), the linearly
bias-corrected estimator $Q_\mathrm{link,linear}^{(2)}$ in
eq.~(\ref{BIAS-CORRECTION-LINEAR}), and the quadratically bias-corrected
estimator $Q_\mathrm{link,quadratic}^{(2)}$ in
eq.~(\ref{BIAS-CORRECTION-QUADRATIC}). The different determinations
are equal only if the total simulation time (in each sample) is much
longer than the integrated autocorrelation time for $Q_\mathrm{link}$.
As we see in \fref{fig:diff-chi-link}--left, only computing
$\chi_\mathrm{link}$ from the biased estimator
$[Q_\mathrm{link}]_{2/2}^2$ results in a measurable bias. Once bias
correction is taken into account, differences are only a fraction of
the statistical error for each estimator. Nevertheless, the different
statistical estimators are dramatically correlated. Hence, their
difference might be significant. In
\fref{fig:diff-chi-link}--right we plot these differences for
$L=24$ and $L=32$ as a function of temperature, in units of the
statistical error for that difference. As we see, at the lowest
temperatures for $L=32$, the bias for the estimate of
$\chi_\mathrm{link}$ obtained from $Q_\mathrm{link,linear}^{(2)}$ is
still measurable. Only the estimate from
$Q_\mathrm{link,quadratic}^{(2)}$ is statistically compatible with the
unbiased estimator. Since our data fully complies with our
expectations, we consider the above analysis as a confirmation of our
expectation
$N\gg\tau_\mathrm{int,Q_\mathrm{link}},\tau_\mathrm{exp}\,.$

\begin{figure}
\centering
\includegraphics[height=\linewidth,angle=270]{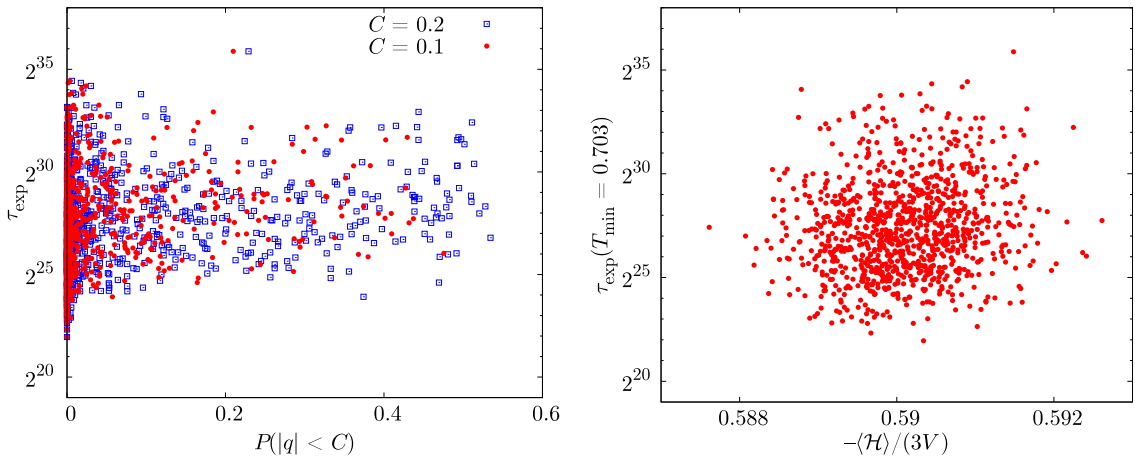}
\caption{Scatter plot of the exponential autocorrelation time ($L=32$)
  versus the probability of the overlap being less than a small quantity (left)
  and the energy (right). We do not observe correlation between the
  thermalisation times and these physically relevant quantities.}
\label{fig:tau-vs-prob}
\end{figure}

We carefully avoided to make decisions during thermalisation based on
the values of physical quantities. However, one could worry about the
possibility of important statistical correlations between the
temperature random walk and interesting quantities. Such correlation
could originate some small biases that would be difficult to
eliminate. Fortunately, we have not found any correlation of this
type. In \fref{fig:tau-vs-prob} we show the correlation between
$\tau_\mathrm{exp}$ and two important quantities: probability of the
overlap being small and the energy.

\section{The overlap probability density}\label{SECT:P-DE-Q}
In this section we study the pdf of the spin overlap. This is a
particularly interesting quantity because, as we saw in
\sref{SECT:MODEL}, it has a qualitatively different behaviour in
the droplet and RSB pictures of the spin-glass phase.

\begin{figure}[b]
\centering
\begin{minipage}{.48\linewidth}
\includegraphics[height=\linewidth,angle=270]{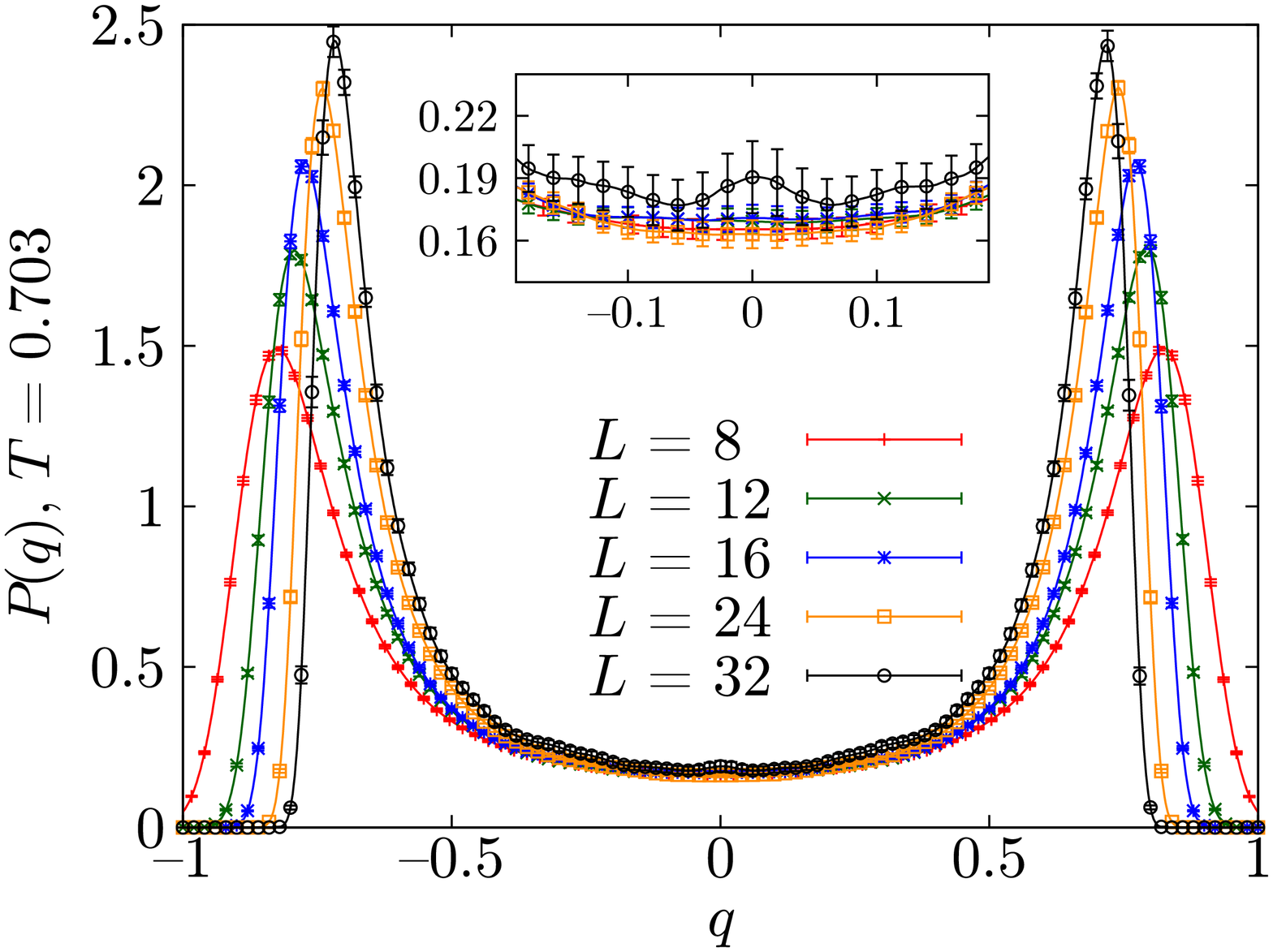}
\end{minipage}
\begin{minipage}{.48\linewidth}
\includegraphics[height=\linewidth,angle=270]{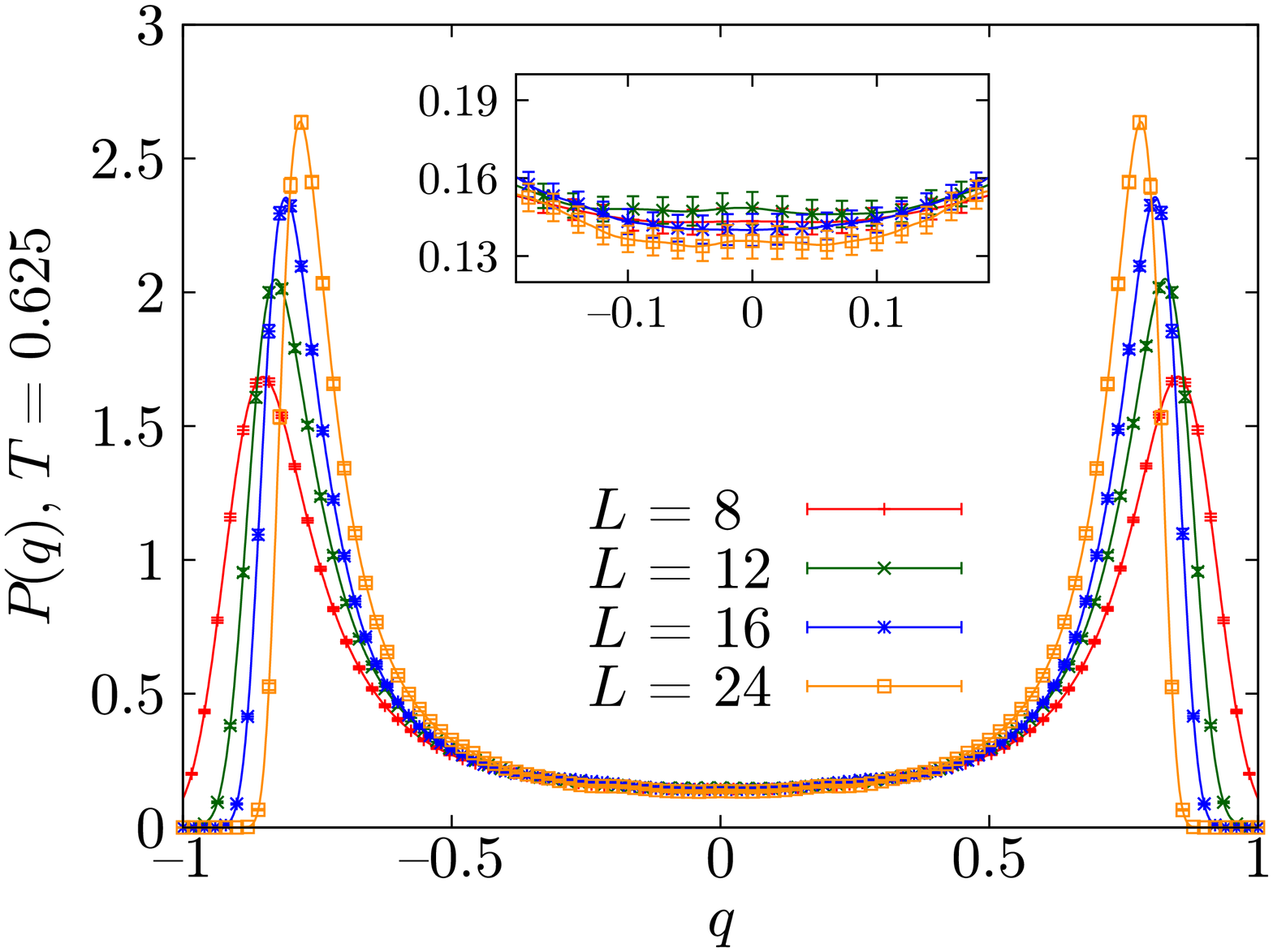}
\end{minipage}
\caption{Overlap  probability density function $P(q)$, eq.~(\ref{DEF:PQ-SMOOTH}), at $T=0.625$ and $T=0.703$.
Notice that for the central sector of $q\sim0$ the curves for the different system sizes quickly
reach a plateau with $P(q) > 0$.}
\label{fig:Pq}
\end{figure}

We have plotted $P(q)$ for $T=0.703$ (the lowest for $L=32$) and
$T=0.625$ (the lowest for $L=24$) in \fref{fig:Pq}. Notice that
the convolution of the comb-like $\tilde P(q)$,
eq.~(\ref{DEF:PQ-PEINE}), with the Gaussian function,
eq.~(\ref{DEF:PQ-SMOOTH}), has yielded a very smooth
$P(q)$. Initially, one would expect the peaks of this pdf to grow
narrower and closer together as $L$ increases, eventually becoming two
Dirac deltas at $\pm q_\mathrm{EA}$. The shift in position is clearly
visible in the figures, but a more careful analysis is needed to
confirm that the peaks are indeed getting sharper
(\sref{sect:picos}).  In addition, the probability in the $q=0$
sector should either go to zero (droplet) or reach a stable non-zero
value (RSB). Even if a visual inspection of \fref{fig:Pq} seems to
favour the second scenario, we shall need a more quantitative analysis
to draw conclusions.

In the remainder of this section we undertake such a quantitative
characterisation of $P(q)$ and, in particular, its thermodynamical
limit. To this end, we will study the evolution of $P(q=0)$ with $T$
and $L$ (\sref{sect:P0}); the extrapolation to infinite volume of
the Binder cumulant (\sref{sect:Binder}) and finally the
evolution of the shape and position of the peaks with the system's
size (\sref{sect:picos}).

\subsection{The $q=0$ sector}\label{sect:P0}
\begin{figure}
\centering
\begin{minipage}{.48\linewidth}
\includegraphics[height=\linewidth,angle=270]{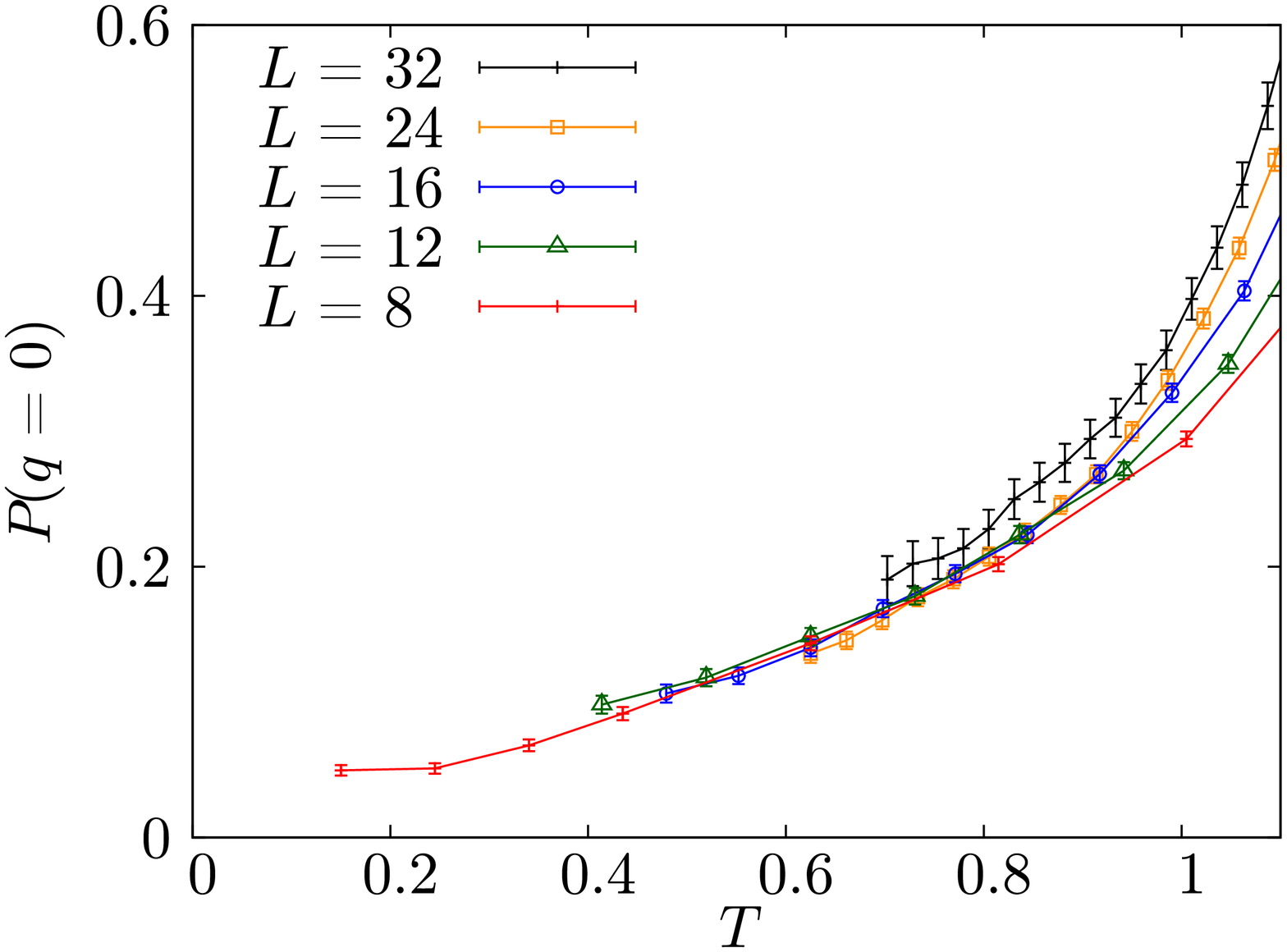}
\end{minipage}
\begin{minipage}{.48\linewidth}
\includegraphics[height=\linewidth,angle=270]{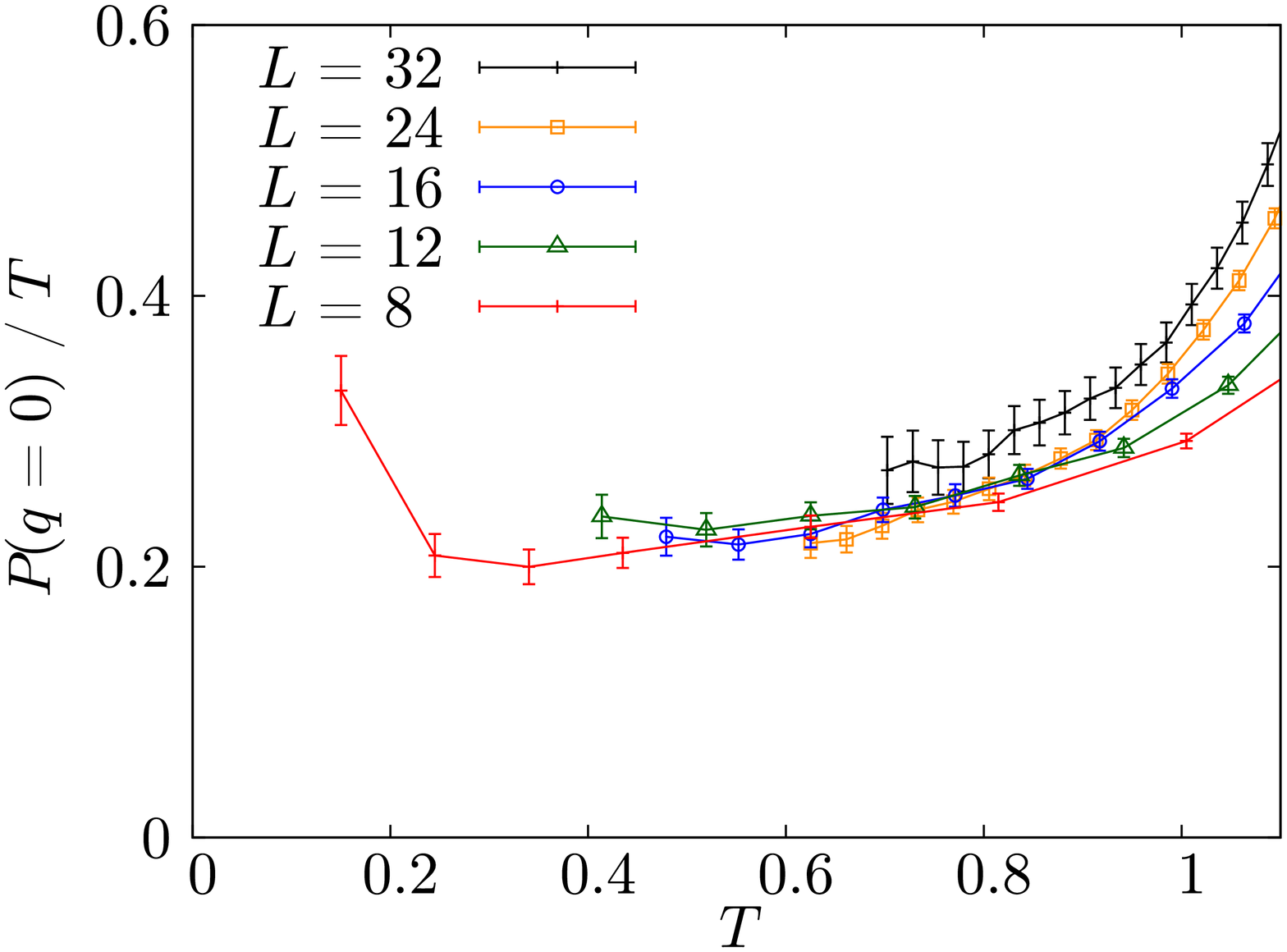}
\end{minipage}
\caption{Overlap density distribution function at zero overlap as a function
  of temperature. We observe an enveloping curve with a linear behaviour, 
as expected in an RSB setting.}
\label{fig:zero-overlap}
\end{figure}
We have plotted in \fref{fig:zero-overlap}---left the probability
density at $q=0$ as a function of $T$ for all our lattices. There
clearly is an enveloping curve in the region $T<T_\mathrm{c}$ with a
decreasing, but positive, value of $P(0)$. In a mean-field
setting~\cite{mezard:87} we expect this probability density to go to
zero linearly in $T$. In order to check this, we have plotted $P(0)/T$
against $T$ in \fref{fig:zero-overlap}---right. As we can see, this
expectation is fulfilled. For a similar study
see~\cite{katzgraber:01}. We remark that the seemingly out of control
value of $P(0)$ for our lowest temperature in $L=8$ is an artifact of
the binary nature of the couplings (a finite system always has a
finite energy gap). Indeed, in~\cite{palassini:01}, the finite size
behaviour of $P(0)$ for the Edwards-Anderson model with binary
couplings was studied as a function of temperature. Finite-size
effects on $P(0)$ turned out to be stronger close to $T=0$ than at
finite temperature.

From a droplet model point of view, Moore et al.~\cite{moore:98}
have argued that the apparent lack of a vanishing limit
for $P(0)$ in numerical work in the 1990s was
an artifact of critical fluctuations. In fact, at $T_\mathrm{c}$,
$P(0)$ diverges as $L^{\beta/\nu}$ while droplet theory
predicts that, for very large lattices, it vanishes as $L^{-\zeta}$, with $\zeta\sim0.2$,
for all $T<T_\mathrm{c}$. These authors rationalise the numerical
findings as a crossover between these two limiting behaviours. 
However, a numerical study at very low temperatures
(so the critical regime is avoided) found for moderate system sizes a non-vanishing 
$P(0)$~\cite{katzgraber:01}. Furthermore, we compute in \sref{sect:picos}
a characteristic length for finite-size effects in the spin-glass phase, which turns 
out to be small at $T=0.703$.

\subsection{The Binder cumulant}\label{sect:Binder}
\begin{figure}[t]
\centering
\begin{minipage}{.48\linewidth}
\includegraphics[height=\linewidth,angle=270]{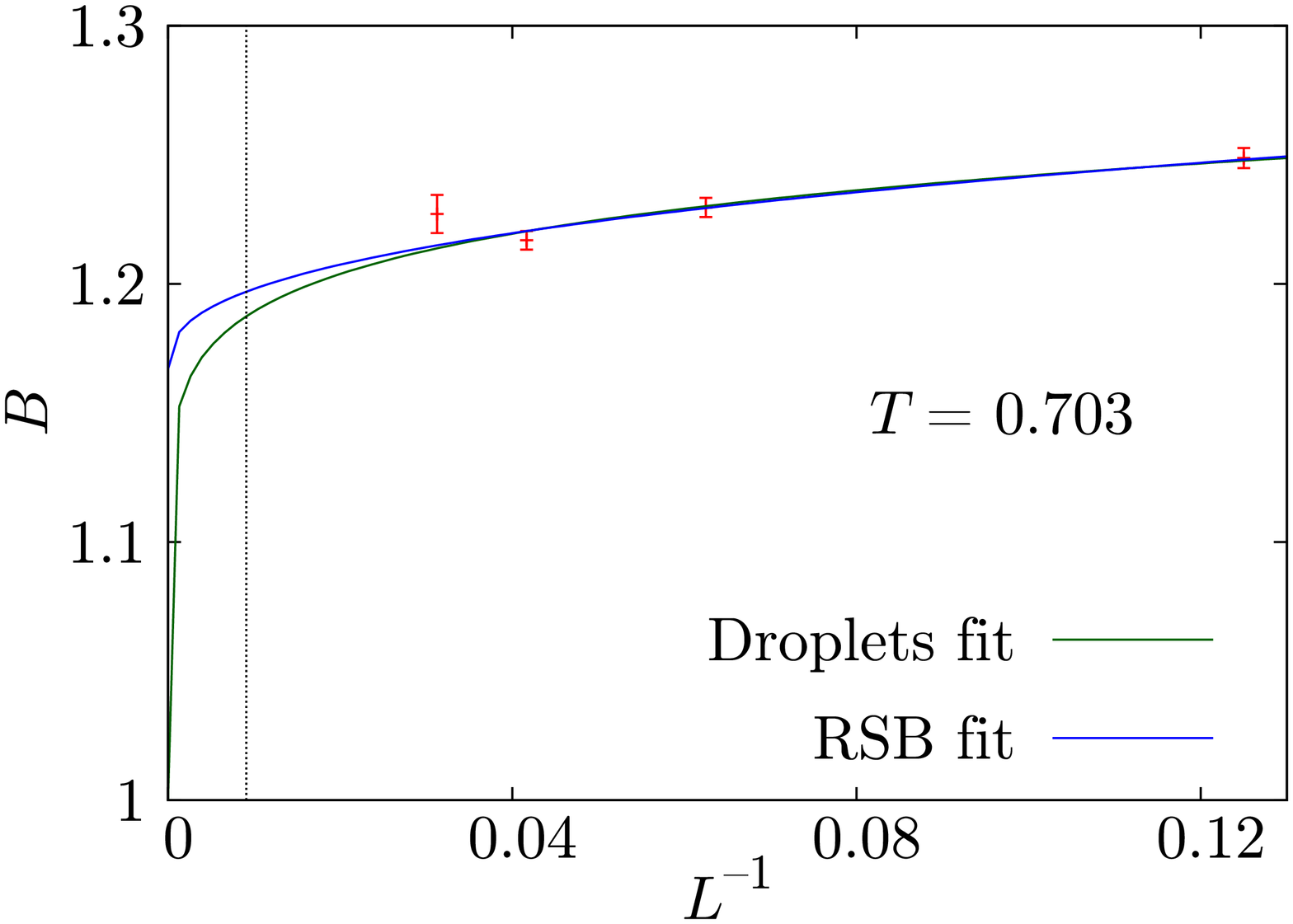}
\end{minipage}
\begin{minipage}{.48\linewidth}
\includegraphics[height=\linewidth,angle=270]{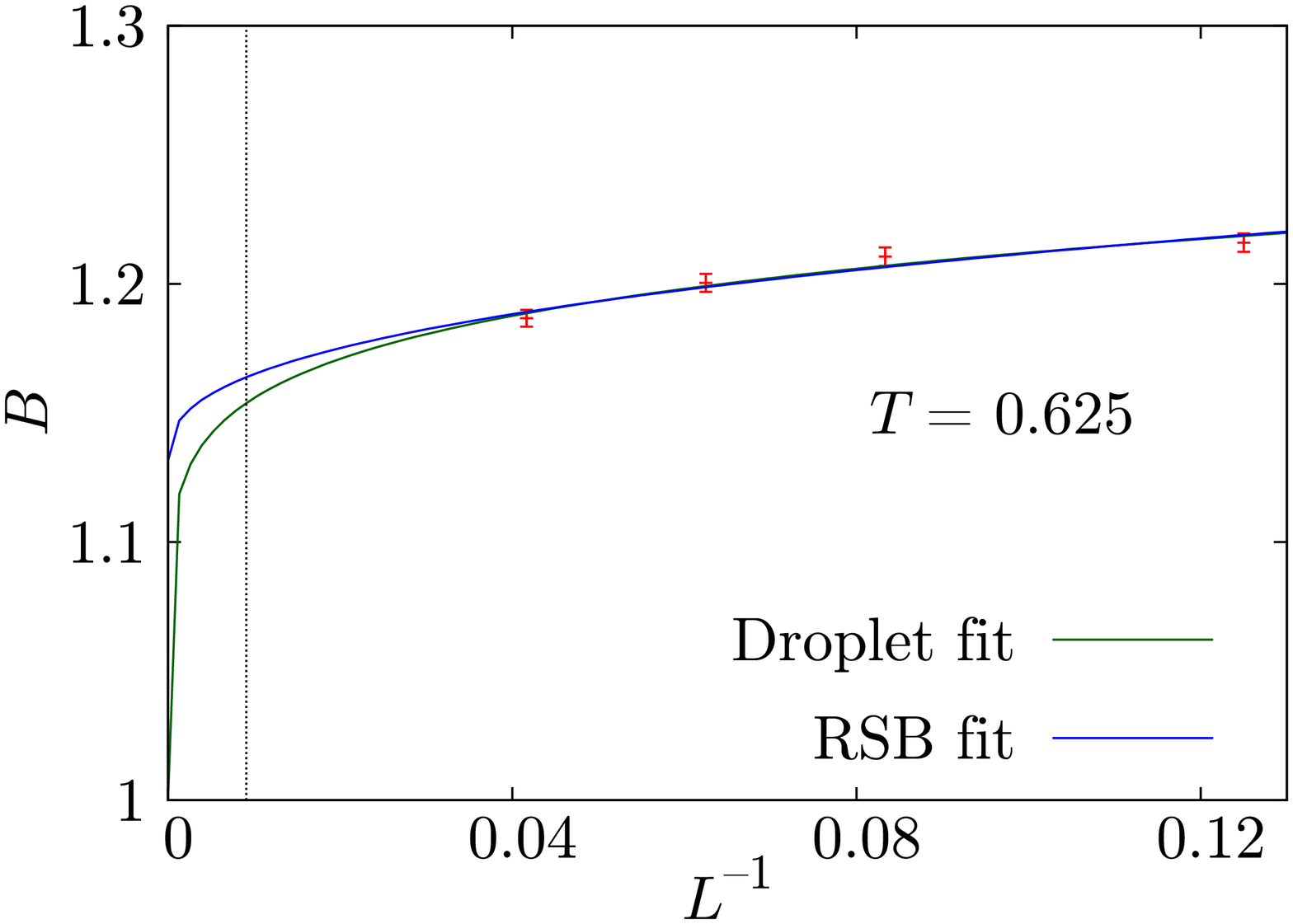}
\end{minipage}
\caption{Infinite volume extrapolation of the Binder parameter at $T=0.703$ and $T=0.625$ and 
fits to the behaviour expected in the RSB, eq.~(\ref{eq:Binder-RSB}), and droplet, eq.~(\ref{eq:Binder-droplet}), 
pictures. See \tref{tab:Binder}. For the experimentally relevant scale of $L=110$ (dotted vertical line,
see \sref{SECT:EQUILIBRIUM-DYNAMICS}) both fits are well above the $B=1$ value of a coarsening system.}
\label{fig:Binder-L}
\end{figure}

We have plotted the Binder cumulant~(\ref{eq:Binder}) for $T=0.625,0.703$ as a function of the system size 
in \fref{fig:Binder-L}. As discussed in \sref{SECT:OBSERVABLES}, the evolution (and thermodynamical limit) 
of this observable is different in the droplet and RSB pictures:
\begin{eqnarray}
\mathrm{Droplet:}\qquad B(T;L) &=& 1 + a L^{-\zeta},\label{eq:Binder-droplet} \\
\ \ \, \quad\mathrm{RSB:}\qquad B(T;L) &=& c + d L^{-1/\hat\nu},\label{eq:Binder-RSB}
\end{eqnarray}
where $1/\hat\nu=0.39(5)$~\cite{janus:10b}. Since it is compatible
with our best estimate for the replicon exponent, $\theta(0)=0.38(2)$,
we prefer to use the second, more accurate value (there is some
analytical ground for this identification~\cite{janus:10b}).  We will
attempt to distinguish between these two behaviours by fitting our
data to (\ref{eq:Binder-droplet}) and (\ref{eq:Binder-RSB}).

These two-parameter fits are plotted in \fref{fig:Binder-L} and the
resulting parameters are gathered in \tref{tab:Binder}.  In the case
of the RSB fit, Eq~(\ref{eq:Binder-RSB}), we have included two error
bars: the number enclosed in parentheses $(\,\cdot\,)$ comes from the
statistical error in a fit fixing $1/\hat\nu$ to $\theta(0)$ and the
one inside square brackets $[\,\cdot\,]$ is the systematic error due
to the uncertainty in $\theta(0)$.

\begin{table}[b]
\caption{Scaling of the Binder parameter and fit to the behaviour expected in the droplet, eq.~(\ref{eq:Binder-droplet}),  and RSB pictures, eq.~(\ref{eq:Binder-RSB}).}\label{tab:Binder}
\begin{tabular*}{\columnwidth}{@{\extracolsep{\fill}}cccccccc}
\cline{2-8}
& \multicolumn{3}{c}{\bfseries Droplet fit} & &  \multicolumn{3}{c}{\bfseries RSB fit} \\
\hline
\multicolumn{1}{c}{ $T$} & \multicolumn{1}{c}{ $\chi^2/\mathrm{d.o.f.}$} & \multicolumn{1}{c}{ $a$} & \multicolumn{1}{c}{ $\zeta$} & &
 \multicolumn{1}{c}{ $\chi^2/\mathrm{d.o.f.}$} & \multicolumn{1}{c}{ $c$} & \multicolumn{1}{c}{ $d$}  \\
\hline
0.703 & 3.78/3 & 0.312(17) & 0.110(17) & & 3.44/3 & 1.165(12)[34] & 0.186(34)[03]\\
0.625 & 2.00/2 & 0.289(16) & 0.134(21) & & 2.73/2 & 1.128(11)[33] & 0.193(28)[03]\\
\hline
\end{tabular*}
\end{table}

As it turns out, both fits have acceptable values of $\chi^2$ per
degree of freedom (d.o.f.). However, the evolution of $B$ with $L$ is
very slow, so in order to accommodate the limit value of
$B(L\to\infty)=1$ consistent with the droplet picture, we have needed
a very small exponent ($\zeta \sim 0.12$, smaller 
than the droplet prediction of $\zeta\approx0.2$~\cite{bray:87}). On the other hand,
according to droplet theory~\cite{bray:87}, the connected spatial
correlation function at $q\!=\!q_\mathrm{EA}$ decays as
$1/r^{\zeta}$. A direct study~\cite{janus:10b},
however, indicates that these correlations decay as $1/r^{0.6}$. 

The reader may find it disputable, from an RSB point of view, that a
single power law should govern finite size effects. It would be rather
more natural that corrections were of order
\begin{equation}
\frac{1}{L^{\theta_\mathrm{eff}(L)}}= \int_0^1 \mathrm{d}q\, \frac{P(q)}{L^{\theta(q)}}\,.
\end{equation}
It turns out, however, that $\theta(q)$ hardly depends on $q$ (except
on the neighbourhood of $q_\mathrm{EA}$), see~\cite{janus:10b} and
Sect.~\ref{SECT:COND}. The neighbourhood of $q_\mathrm{EA}$ would
produce a subleading correction of order $1/L^{0.6}$.

In any case, see \sref{SECT:EQUILIBRIUM-DYNAMICS}, we remark that the
relevant regime for comparison with experimental work is
$L\approx110$, where both the RSB and the droplet fits predict that
$B(T,L)$ is well above $1$ (see \fref{fig:Binder-L}).

\subsection{The peaks of $P(q)$, $q_\mathrm{EA}$, and finite size effects}\label{sect:picos}

One of the features of the $P(q)$ about which droplet and RSB agree
is the fate of its two symmetric peaks as we approach the
thermodynamical limit.  These should grow increasingly narrow and
shift their position until they eventually become two Dirac deltas at
$q = \pm q_\mathrm{EA}$. The actual value of $q_\mathrm{EA}$ is
notoriously difficult to compute~\cite{janus:09b,perez-gaviro:06,iniguez:97},
see, however, \cite{janus:10b}.

Characterising the evolution of these peaks as we increase the system
size is the goal of this section. We start by defining
$q_\mathrm{EA}(L)$ as the position of the maximum of $P(q;L)$ (since
the pdf is symmetric, we shall consider all overlaps to be positive in
the remainder of this section).  Thanks to the Gaussian smoothing
procedure described in eq.~(\ref{DEF:PQ-SMOOTH}), this maximum is very
well defined. We compute its position by fitting the peak to a
third-order polynomial (notice that the peaks are very asymmetric).

In order to further describe the peaks, we will also employ the half-widths $\sigma^{(\pm)}$ at 
half height $\bigl[P(q^{(\pm)}) = P(q_\mathrm{EA}(L))/2\bigr]$:
\begin{equation}\label{eq:sigma}
\sigma^{(\pm)} = \bigl| q^{(\pm)} - q_\mathrm{EA}(L)\bigr|\,
\end{equation}
where $q^{(-)} < q_\mathrm{EA}(L)< q^{(+)}$.

\begin{figure}[t]
\centering
\begin{minipage}{.48\linewidth}
\includegraphics[height=\linewidth,angle=270]{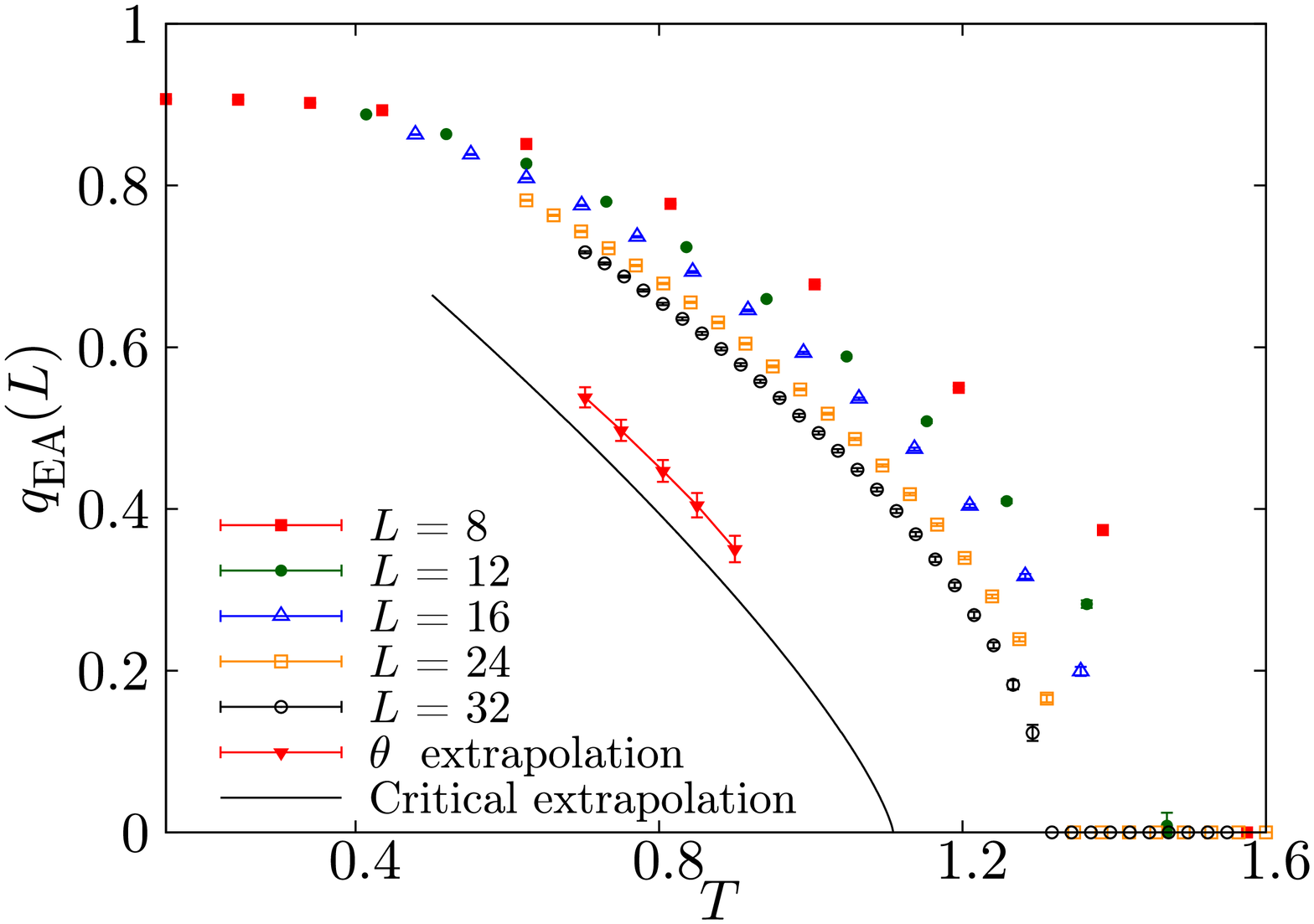}
\end{minipage}
\begin{minipage}{.48\linewidth}
\includegraphics[height=\linewidth,angle=270]{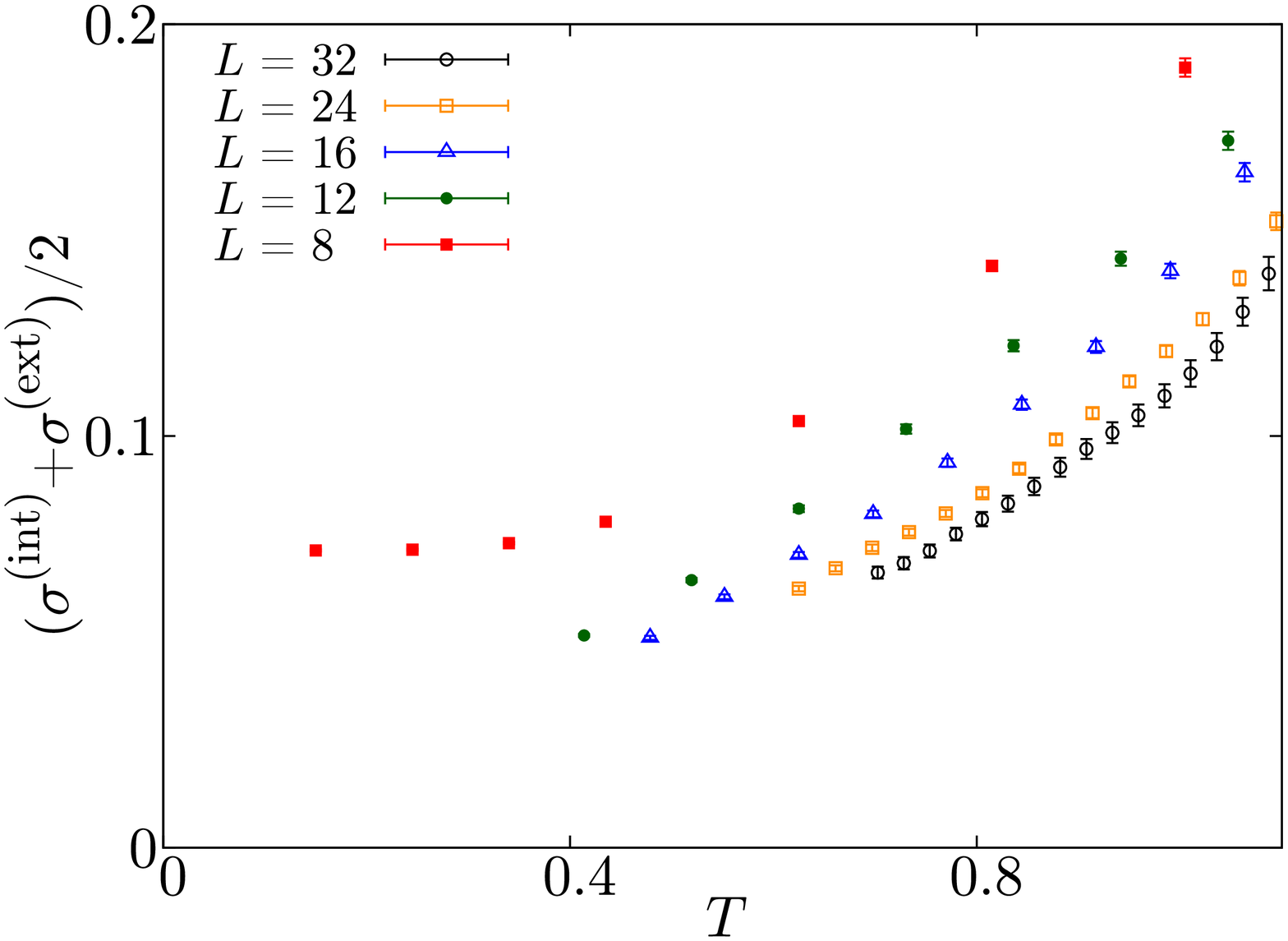}
\end{minipage}
\caption{\emph{Left:} $q_\mathrm{EA}(L)$ as a function of the
  temperature. We include two different infinite-volume
  extrapolations: using the replicon exponent, eq.~(\ref{eq:qEA-inf})
  and \tref{tab:qEA}, and the one obtained from finite-size
  scaling arguments in the critical region,~eqs.~(\ref{eq:qEA-FSS})
  and~(\ref{eq:qEA-FSS2}).  \emph{Right:} Width of the peaks of $P(q)$,
  eq.~(\ref{eq:sigma}), as a function of $T$ for all our lattice
  sizes.}
\label{fig:qEA-T}
\end{figure}

We have plotted these parameters as a function of temperature
in \fref{fig:qEA-T}. On \tref{tab:sigma}  we
can see that the width of the peaks does decrease with a power law in $L$,
although very slowly. The product $\sigma P(q_\mathrm{EA}(L))$  has a 
small dependence on $L$.

We can now extrapolate $q_\mathrm{EA}(L)$ to find the order parameter
in the thermodynamical limit. A finite-size scaling
study~\cite{janus:10b} shows that
\begin{equation}\label{eq:qEA-inf}
q_\mathrm{EA} (L,T) = q_\mathrm{EA}^\infty(T) \biggl[1+\frac{A(T)}{L^{1/\hat\nu}}\biggr],\quad A(T)=[L_\mathrm{c}(T)]^{1/\hat\nu}\,,
\end{equation}
where $1/\hat\nu=0.39(5)$.  Yet, as discussed after
eq.~\eref{eq:Binder-RSB}, we prefer to identify $1/\hat\nu$ with the
replicon exponent, $\theta(0)=0.38(2)$.  A disagreeing reader merely
needs to double the error estimate in the extrapolation of
$q_\mathrm{EA}$. Note that one should not attempt a three-parameter
fit to eq.~(\ref{eq:qEA-inf}), as there are too few degrees of
freedom. An independent estimate of $1/\hat\nu$ is required. Similar
extrapolations were attempted in~\cite{iniguez:96}, with smaller
system sizes ($L\leq16$) and a lesser control over $1/\hat\nu$.

\begin{table}[t]
\caption{Width $\sigma=\bigl(\sigma^{(+)} +\sigma^{(-)}\bigr)/2$ of the peaks in
$P(q)$ and fit to a power law $\sigma(L)=AL^B$ in the range $[L_\mathrm{min}, 32]$.
We also include the product $\sigma P(q_\mathrm{EA}(L))$.}
\label{tab:sigma}
\begin{tabular*}{\columnwidth}{@{\extracolsep{\fill}}cccccc}
\cline{2-6}
& \multicolumn{2}{c}{$T = 0.703$}&
& \multicolumn{2}{c}{$T = 0.805$}\\
\hline
\multicolumn{1}{c}{$L$ }  & $\sigma$ & $\sigma P(q_\mathrm{EA}(L))$& &
$\sigma$ & $\sigma P(q_\mathrm{EA}(L))$\\
\hline
8  & 0.1177(20) &0.1784(10) & & 0.1391(25)& 0.1833(10) \\   
12 & 0.0963(21) &0.1740(12) & & 0.1165(25)& 0.1809(12) \\
16 & 0.0817(16) &0.1696(11) & & 0.1001(22)& 0.1756(11) \\
24 & 0.0735(16) &0.1690(12) & & 0.0860(19)& 0.1728(12) \\
32 & 0.0668(29) &0.1631(23) & & 0.0798(34)& 0.1669(22) \\
\hline
$L_\mathrm{min}$         & 16         & & & 16     \\
$\chi^2/\mathrm{d.o.f.}$ & 0.43/1     & & &1.13/1  \\
$B$                      & $-0.278(28)$ & & & $-0.346(30)$ \\
\hline
\end{tabular*}
\end{table}
\begin{table}[h]
\caption{Extrapolation to infinite volume of $q_\mathrm{EA}(L,T)$
using the replicon exponent, eq.~(\ref{eq:qEA-inf}). We also include
the confidence interval previously obtained in a non-equilibrium study~\cite{janus:09b}.\label{tab:qEA}}
\lineup
\begin{tabular*}{\columnwidth}{@{\extracolsep{\fill}}ccc}
\hline
\multicolumn{1}{c}{\bfseries $L$ } & \multicolumn{1}{c}{\bfseries $T = 0.703$}
& \multicolumn{1}{c}{\bfseries $T = 0.805$}\\
\hline
8  & 0.82461(83)  & 0.7818(11)\0 \\
12 & 0.79333(85)  & 0.7412(11)\0 \\
16 & 0.77300(75)  & 0.71681(95)  \\
24 & 0.74027(71)  & 0.67905(83)  \\
32 & 0.7174(14)\0 & 0.6535(16)\0 \\
\hline
$L_\mathrm{min}$         & 16           & 16     \\
$\chi^2/\mathrm{d.o.f.}$ & 1.83/1       & 0.98/1 \\
$q_\mathrm{EA}$          & 0.538[11](6) & 0.447[12](6) \\
Bounds from \cite{janus:09b} & $0.474 \leq q_\mathrm{EA}\leq 0.637$ & $0.368\leq q_\mathrm{EA} \leq 0.556$ \\
\hline
\end{tabular*}
\end{table}
\begin{table}[h]
\caption{Determination of $L_\mathrm{c}$ in eq.~\eref{eq:qEA-inf} for
  several temperatures below $T_\mathrm{c}$. Errors are given as in
  \tref{tab:qEA}.  The characteristic length $L_\mathrm{c}(T)$ scales
  as a correlation length when $T$ approaches $T_\mathrm{c}$
  ($\nu\approx2.45$ from~\cite{hasenbusch:08b}).  We warn the reader
  that the $\chi^2/\mathrm{d.o.f.}$ for the fits at $T=0.85$ and
  $0.90$ are, respectively, $2.6/1$ and $2.7/1$.}\label{tab:Lc}
\lineup
\begin{tabular*}{\columnwidth}{@{\extracolsep{\fill}}lclc}
\hline
\multicolumn{1}{c}{$T$ } & \multicolumn{1}{c}{$L_\mathrm{c}^{1/\hat\nu}$} & \multicolumn{1}{c}{$L_\mathrm{c}$}
& \multicolumn{1}{c}{$L_\mathrm{c}(T_\mathrm{c}-T)^\nu$}\\
\hline
0.703 & 1.253[10](32) & \01.78[4](11)  & 0.197[4](13)\\
0.75  & 1.448[12](34) & \02.58[6](16)  & 0.210[4](13)\\
0.805 & 1.731[14](44) & \04.08[9](27)  & 0.221[5](15)\\
0.85  & 2.023[16](54) & \06.09[13](42) & 0.222[5](15)\\
0.90  & 2.514[21](66) & 10.63[22](71) & 0.230[5](15)\\
\hline
\end{tabular*}
\end{table}
We present the values of $q_\mathrm{EA}(L)$ and the result of a fit to
eq.~(\ref{eq:qEA-inf}) on \tref{tab:qEA}.  As we can see,
the errors due to the uncertainty in the
exponent, denoted by $[\,\cdot\,]$, are greater than those caused by
the statistical error in the individual points, $(\,\cdot\,)$. In
fact, our data admit good fits for a very wide range of values in
$1/\hat\nu$. For instance, if we try to input the value of the
exponent obtained in the droplet-like extrapolation of the Binder
parameter, $\zeta\sim0.12$ (see eq.~(\ref{eq:Binder-droplet}) and
\tref{tab:Binder}), we still obtain a good fit, even though the
extrapolated value for $q_\mathrm{EA}$ is almost zero at $T=0.703$ and
negative at $T=0.805$. Therefore, using the droplet exponent $\zeta$
the spin-glass phase  would be non-existent. 

Also included in \tref{tab:qEA} is the confidence interval for
this observable computed from non-equilibrium considerations
in~\cite{janus:09b}. Notice that the equilibrium values are much more
precise, but consistent. The extrapolations included in this table
(and analogous ones for other values of $T$) are plotted on
\fref{fig:qEA-T}.

We remark that the estimate of $q_\mathrm{EA}$ from
eq.~(\ref{eq:qEA-inf}) is fully compatible with the results of a
Finite-Size Scaling analysis of the conditional correlation
functions~\cite{janus:10b}.

Interestingly enough the estimate of $q_\mathrm{EA}$ provides a
determination of the correlation-length in the spin glass phase. The
reader might be surprised that a correlation length can be defined in
a phase where correlations decay algebraically. Actually, finite size
effects are ruled by a crossover length
$L_\mathrm{c}(T)$~\cite{josephson:66}, that scales as a correlation
length (i.e. $L_\mathrm{c}(T)\propto (T_\mathrm{c}-T)^{-\nu}$). In
fact, one would expect $q_\mathrm{EA}(T,L)/q_\mathrm{EA}(T)=1+
h[L/L_\mathrm{c}(T)]$. The only thing we know about the crossover
function is that it behaves for large $x$ as $h(x)\sim
x^{-1/\hat\nu}$. Making the simplest ansatz $h(x)= x^{-1/\hat\nu}$,
the amplitude for the finite-size corrections in eq.~\eref{eq:qEA-inf}
can be interpreted as a power of the crossover length
$L_\mathrm{c}(T)$, \tref{tab:Lc}.  We note that our determination of
$L_\mathrm{c}(T)$ really scales as a bulk correlation length, with
$T_\mathrm{c}$ and $\nu$ taken from~\cite{hasenbusch:08b}. It turns
out to be remarkably small at $T=0.703$.

The above argument tells us that good determinations of
$q_\mathrm{EA}(T)$ are possible, provided that $L\gg
L_\mathrm{c}(T)$. Yet, finite size scaling can be used as well to
extrapolate $q_\mathrm{EA}(T,L)$ to the large-volume limit, even
closer to $T_\mathrm{c}$ where $L$ becomes {\em smaller} than
$L_\mathrm{c}$. This somehow unconventional use of finite size scaling
was started in
Refs.~\cite{luescher:91,kim:93,caracciolo:95,caracciolo:95b}, and has
also been used in the spin-glass context~\cite{palassini:99,jorg:06}.
Most of the times, these ideas are used in the paramagnetic phase, but
we show below how to implement them in the low-temperature phase.

Close to $T_\mathrm{c}$, we know that
\begin{equation}\label{eq:qEA-inf-FSS}
q_\mathrm{EA}^\infty(T) = \lambda (T_\mathrm{c}-T)^\beta [1+ \mu (T_\mathrm{c}-T)^{\omega\nu}+\ldots]\,.
\end{equation}
We have excellent determinations of $T_\mathrm{c}$ and $\beta$ from the work in~\cite{hasenbusch:08b}, 
so we need only to estimate the amplitude $\lambda$. In fact, Wegner's confluent
corrections $(T_\mathrm{c}-T)^{\omega\nu}$ are small close to $T_\mathrm{c}$.
To proceed, we note that finite-size scaling tells us that
\begin{equation}\label{eq:qEA-FSS}
q_\mathrm{EA}(L,T) = L^{-\beta/\nu} F(x)[1+ L^{-\omega} G(x)+\ldots],\qquad x= L^{1/\nu} (T_\mathrm{c}-T),
\end{equation}
where the critical exponents are (from~\cite{hasenbusch:08b}),
\begin{equation}
\nu = 2.45(15),\qquad \beta = 0.77(5),\qquad \omega = 1.0(1).
\end{equation}
In order to connect eq.~\eref{eq:qEA-FSS} with the infinite-volume limit in
eq.~\eref{eq:qEA-inf-FSS} the asymptotic 
behaviour of the scaling functions $F(x)$ and $G(x)$ must be for large $x$
\begin{equation}
F(x) \sim x^\beta,\qquad G(x)\sim x^{\omega\nu}.
\end{equation}

\begin{figure}[t]
\centering
\includegraphics[height=0.7\linewidth,angle=270]{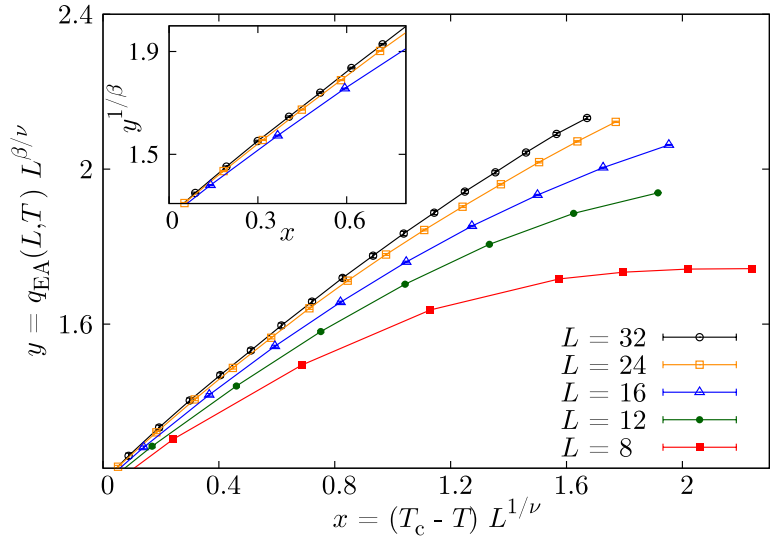}
\caption{Scaling plot of $y=q_\mathrm{EA}(L,T)L^{\beta/\nu}$ in the critical region below $T_\mathrm{c}$, 
following eq.~(\ref{eq:qEA-FSS}) and using the values given in~\cite{hasenbusch:08b}
for the critical exponents and $T_\mathrm{c}$. \emph{Inset:} Close-up of the region near $T_\mathrm{c}$ 
in the representation of eq.~(\ref{eq:qEA-FSS2}), showing a linear behaviour for large $L$.}
\label{fig:qEA-FSS}
\end{figure}

The resulting scaling plot is represented on \fref{fig:qEA-FSS}. 
 Varying the values of $T_\mathrm{c}$
and the critical exponents inside their error margins does not make significant  
changes in the plot. Notice 
how the curves collapse for small values of the scaling variable $x$ and large $L$, 
but how for our lowest temperatures scaling corrections become important. In fact, 
eq.~\eref{eq:qEA-FSS} implies that when the temperature is lowered away from $T_\mathrm{c}$
the amplitude for scaling corrections grows 
as $x^{\omega\nu} \approx x^{2.45}$.

In order to estimate the amplitude $\lambda$ we 
shall concentrate on the small-$x$ region where finite-size scaling
corrections are smallest. Disregarding scaling corrections in eq.~\eref{eq:qEA-FSS},
\begin{equation}\label{eq:qEA-FSS2}
\bigl(q_\mathrm{EA}(L,T)  L^{\beta/\nu}\bigr)^{1/\beta}=F(x)^{1/\beta} \ \underset{x\to\infty}\longrightarrow\ x.
\end{equation}
The inset of \fref{fig:qEA-FSS} shows that we reach this asymptotic behaviour
for $L\geq24$. Then, using the simplest parameterisation, $F(x) = (\lambda^{1/\beta} x+B)^\beta$,
\begin{equation}\label{eq:qEA-FSS3}
q_\mathrm{EA}(L,T) = \lambda (T_\mathrm{c}-T)^{\beta} \left[ 1+ \frac{\beta B}{\lambda^{1/\beta} (T_\mathrm{c}-T) L^{1/\nu}}+\ldots\right]\ .
\end{equation}
We can fit our $L=32$ data for $x<0.4$ (where the curves for $L=24$ and $L=32$
are compatible) and use the resulting value of $\lambda$ to extrapolate in
eq.~(\ref{eq:qEA-FSS3}) to infinite volume. This extrapolation is represented
as a function of $T$ on \fref{fig:qEA-T}. It is clear that this critical
extrapolation differs with the extrapolation from \eref{eq:qEA-inf} at most by
two standard deviations. The difference, if any, could be explained as
Wegner's confluent corrections. However, to make any strong claim on confluent
corrections, one would need to estimate the error in the critical
extrapolation. Unfortunately, we have found that this error estimate is quite
sensitive to the statistical correlation between $T_\mathrm{c}$, $\nu$, and
$\beta$ (as far as we know, the corresponding covariance matrix has not been
published).

One could be tempted to compare eq.~\eref{eq:qEA-FSS3} with eq.~\eref{eq:qEA-inf}
and conclude $\hat\nu=\nu$. We observe that, at the numerical level, 
$\nu=2.45(15)$~\cite{hasenbusch:08b} and $\hat\nu = 2.6(3)$~\cite{janus:10b}. However, 
we do not regard this as fireproof. Indeed, it is a consequence of our somewhat arbitrary 
parameterisation $F(x) = (\lambda^{1/\beta} x+B)^\beta$. To investigate this issue 
further, the small-$x$ region is not enough. One is interested in the asymptotic behaviour
of $F(x)$ for large $x$ where unfortunately corrections to scaling are crucial. A careful
study of the crossover region can be done only by considering corrections to scaling
both at the critical temperature (at $q=0$) and below the critical temperature (at $q=q_\mathrm{EA}$).

Finally, the reader could worry about the applicability of
\eref{eq:qEA-inf-FSS} well below $T_\mathrm{c}$. The issue has been
considered recently within the framework of droplet
theory~\cite{moore:10}. It was found that \eref{eq:qEA-inf-FSS} is
adequate for all $T<T_\mathrm{c}$ (actually, no Wegner's scaling
corrections were discussed in~\cite{moore:10}). Thus, the fact that
our data are describable as scaling behavior with leading Wegner's
correction does not imply that they are not representative of the low
temperature phase.

\section{Conditional correlation functions}\label{SECT:COND}
\begin{figure}[t]
\centering
\includegraphics[height=\linewidth,angle=270]{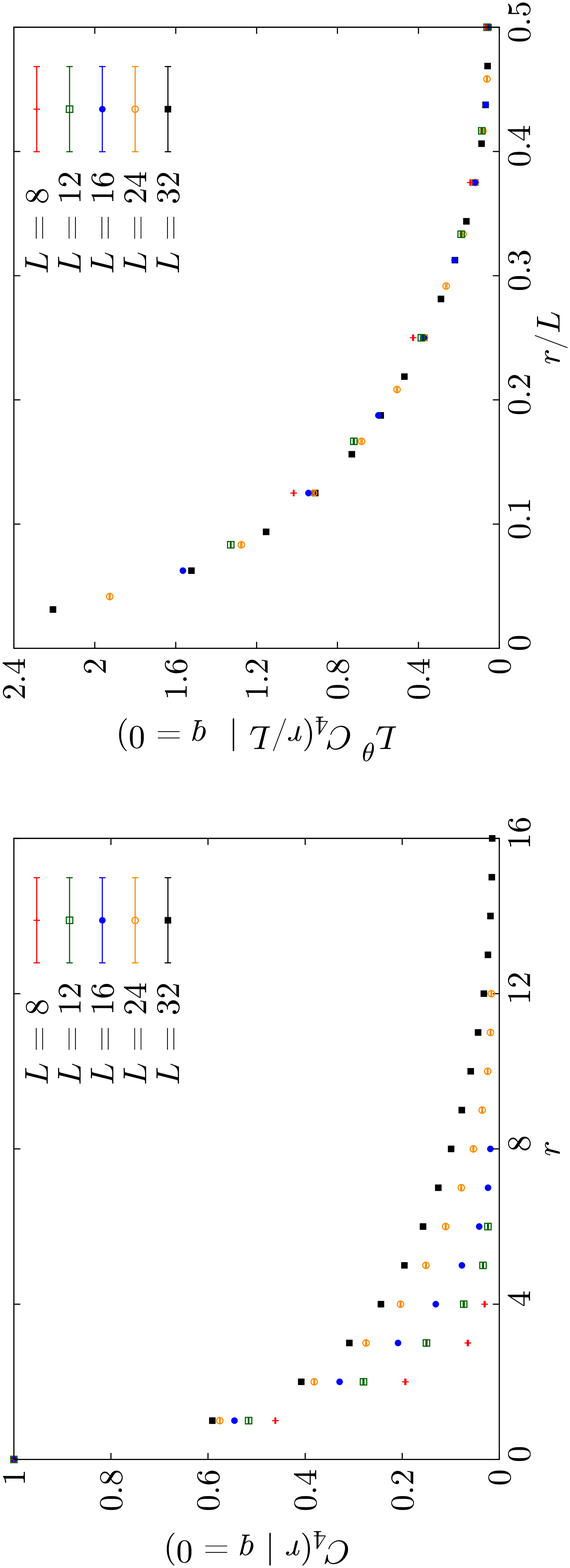}
\caption{Spatial correlation function $C_4\bigl(\, (r,0,0)|q=0\bigr)$ at $T=0.703$. 
We show on the right panel a rescaled version using the replicon
exponent $\theta=0.38$ and the scaling variable $r/L$.}
\label{fig:C4}
\includegraphics[height=0.7\linewidth,angle=270]{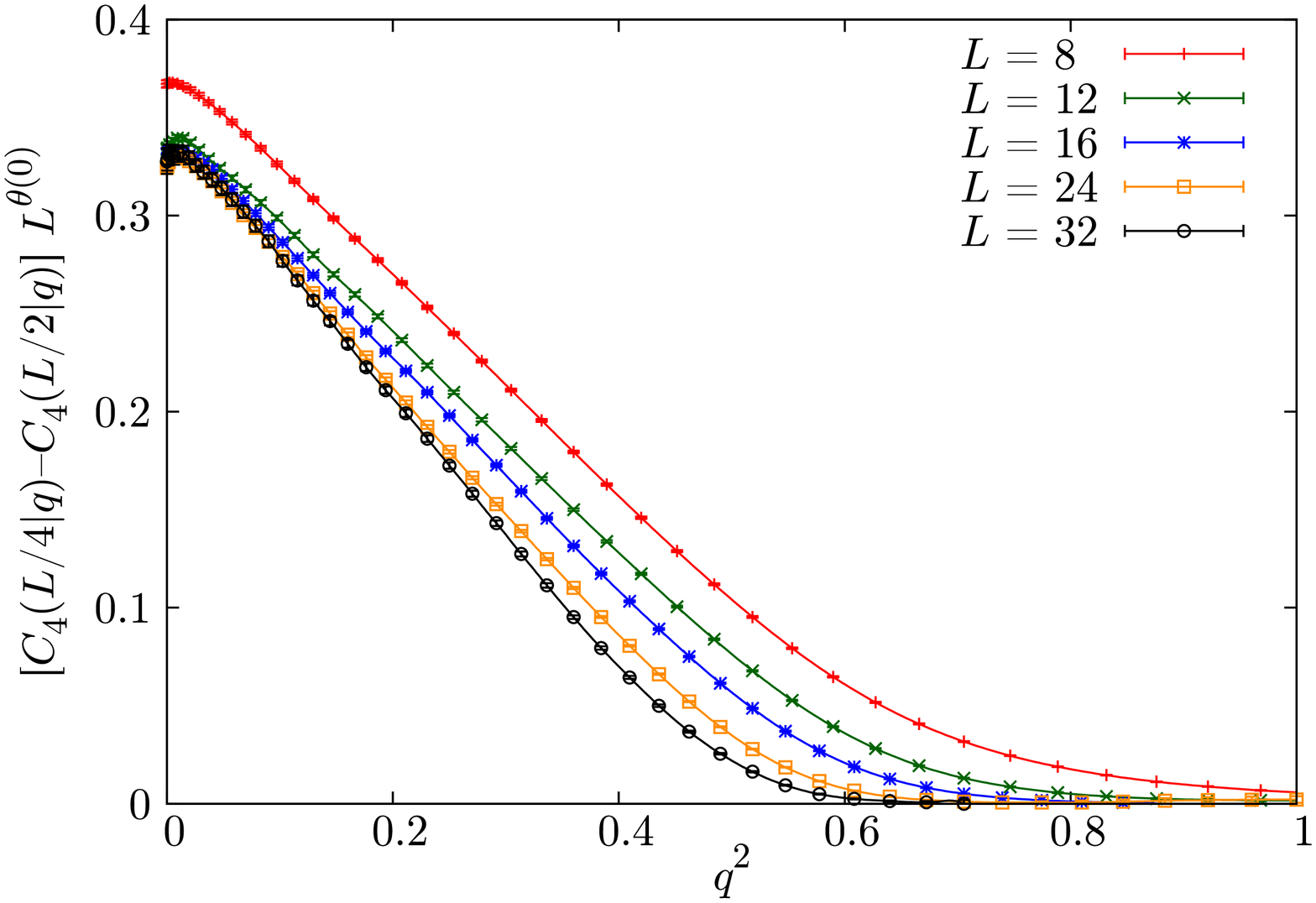}
\caption{Subtracted correlation function,
  eq.~\eref{eq:C4-substraida}, in units of $1/L^{\theta(0)}$ as
  function of $q^2$. We took the non-equilibrium determination of the
  replicon exponent,
  $\theta(0)=0.38(2)$~\cite{janus:09b}.}\label{fig:C4-substraida}
\end{figure}
Let us consider the conditional spatial correlation function
$C_4(r|q)$, eq.~(\ref{eq:C4-q}). A thorough study in the Fourier
space is performed in~\cite{janus:10b}. Here, we provide some
complementary information, concentrating on real space and
considering as well the statistical fluctuations on the correlators.

We first concentrate on $q=0$, the region where the droplet 
and RSB theory most differ. In \fref{fig:C4}---left
we show $C_4(r|q=0)$ for $T=0.703$, which is seen to
tend to zero for large $r$. Furthermore, if we use the
droplet  scaling of eq.~(\ref{eq:C4-droplets}),
we see that we need to rescale the correlation function
by a factor $L^{\theta(0)}$, with $\theta(0)=0.38(2)$ the replicon
exponent, in order to collapse the curves. 

As for other values of $q$, we may consider the differences
\begin{equation}\label{eq:C4-substraida}
C_4(r=L/4|q)-C_4(r=L/2|q)\sim\frac{1}{L^{\theta(q)}}\,,
\end{equation}
where the subtraction takes care of the large-$r$ background in
$C_4(r|q)$. As we show in \fref{fig:C4-substraida}, the subtracted
correlation function scales in the range $q^2<0.2$ as
$L^{-\theta(0)}$.  This implies that the connected correlation
functions $C_4(r|q)-q^2$ decay algebraically for large $r$ (a similar
conclusion was reached in~\cite{contucci:09}).  On the other hand, for
$q^2= q_\mathrm{EA}^2\approx 0.3$, the exponent $\theta(q)$ is
definitively larger than $\theta(0)$ (a detailed analysis indicates
$\theta(q_\mathrm{EA})\sim 0.6$~\cite{janus:10b}). The crossover
from the scaling $C_4(r=L/4|q)-C_4(r=L/2|q)\sim 1/L^{\theta(0)}$ to
$C_4(r=L/4|q)-C_4(r=L/2|q)\sim 1/L^{\theta(q_\mathrm{EA})}$ can be
described by means of Finite Size Scaling~\cite{janus:10b}.

\begin{figure}
\begin{minipage}{.5\linewidth}
\includegraphics[height=\linewidth,angle=270]{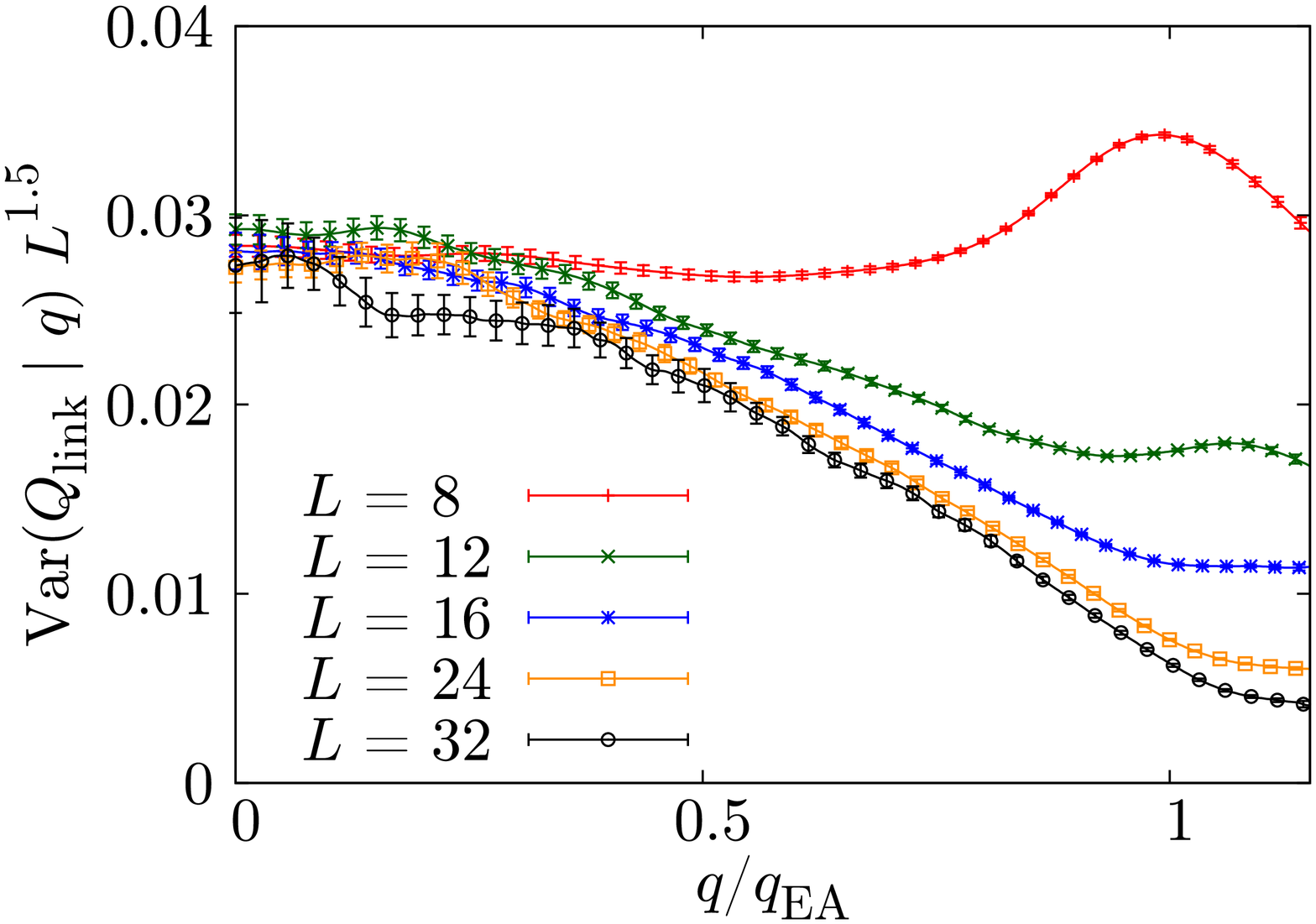}
\end{minipage}
\begin{minipage}{.5\linewidth}
\includegraphics[height=\linewidth,angle=270]{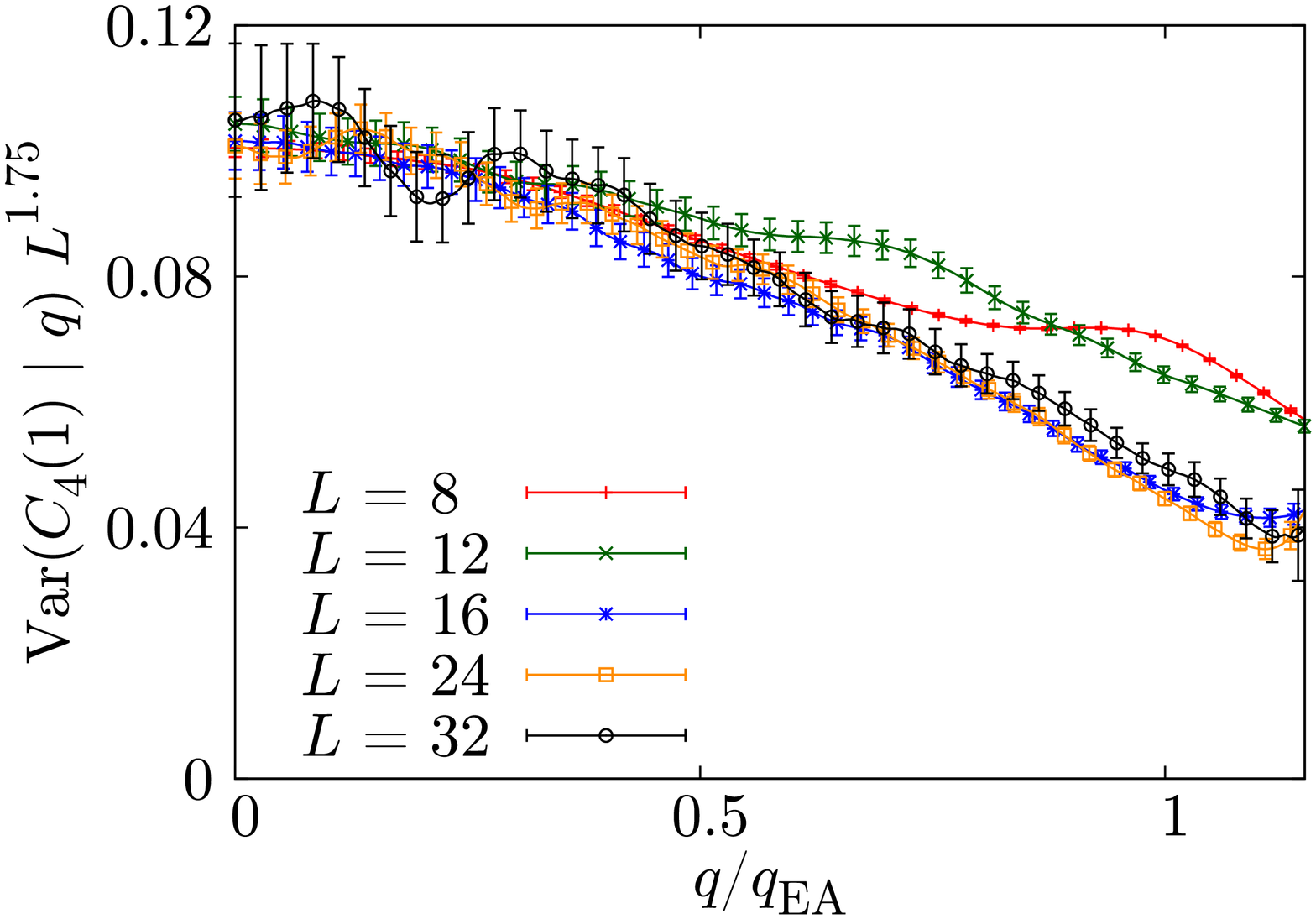}
\end{minipage}

\begin{minipage}{.5\linewidth}
\includegraphics[height=\linewidth,angle=270]{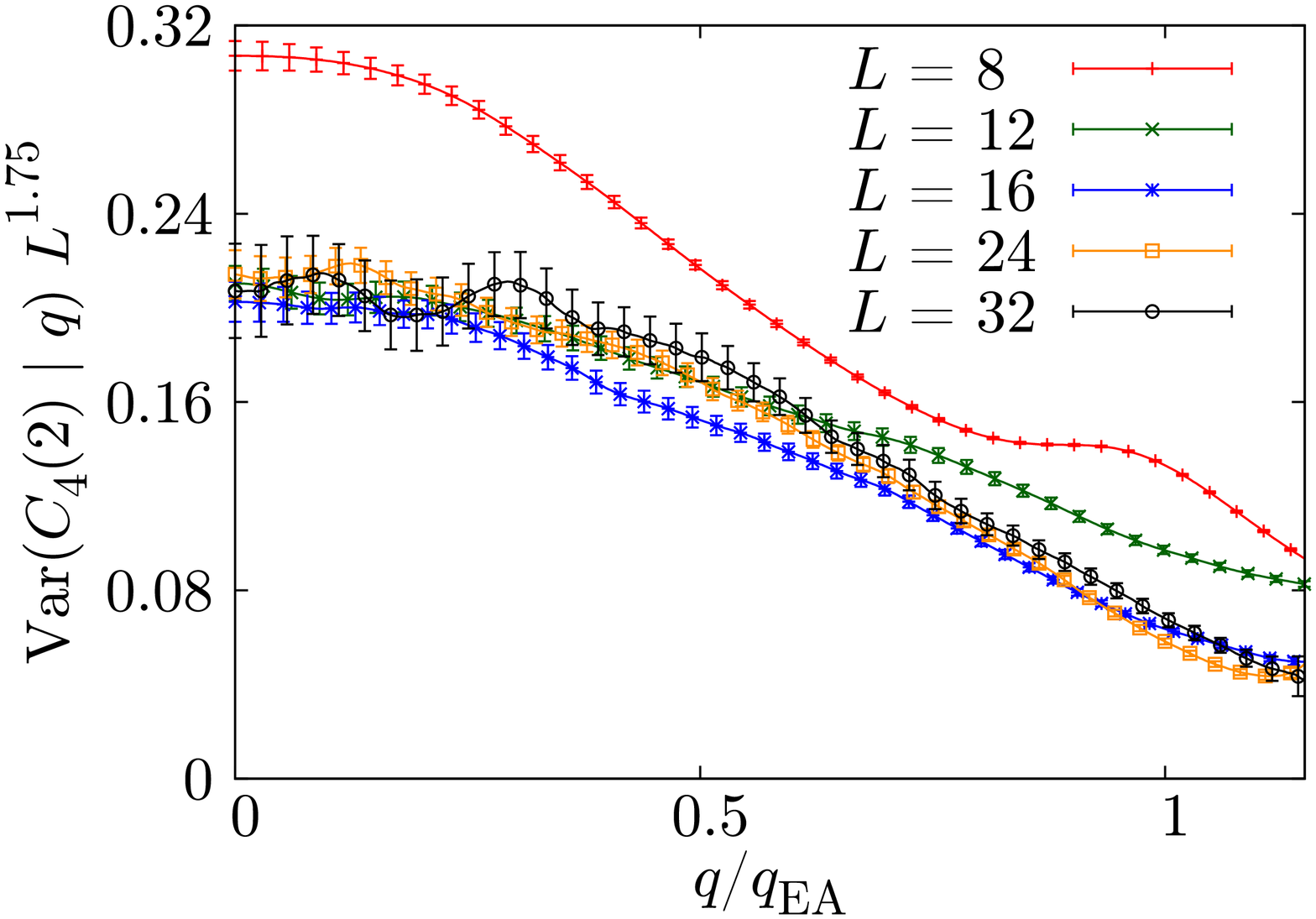}
\end{minipage}
\begin{minipage}{.5\linewidth}
\includegraphics[height=\linewidth,angle=270]{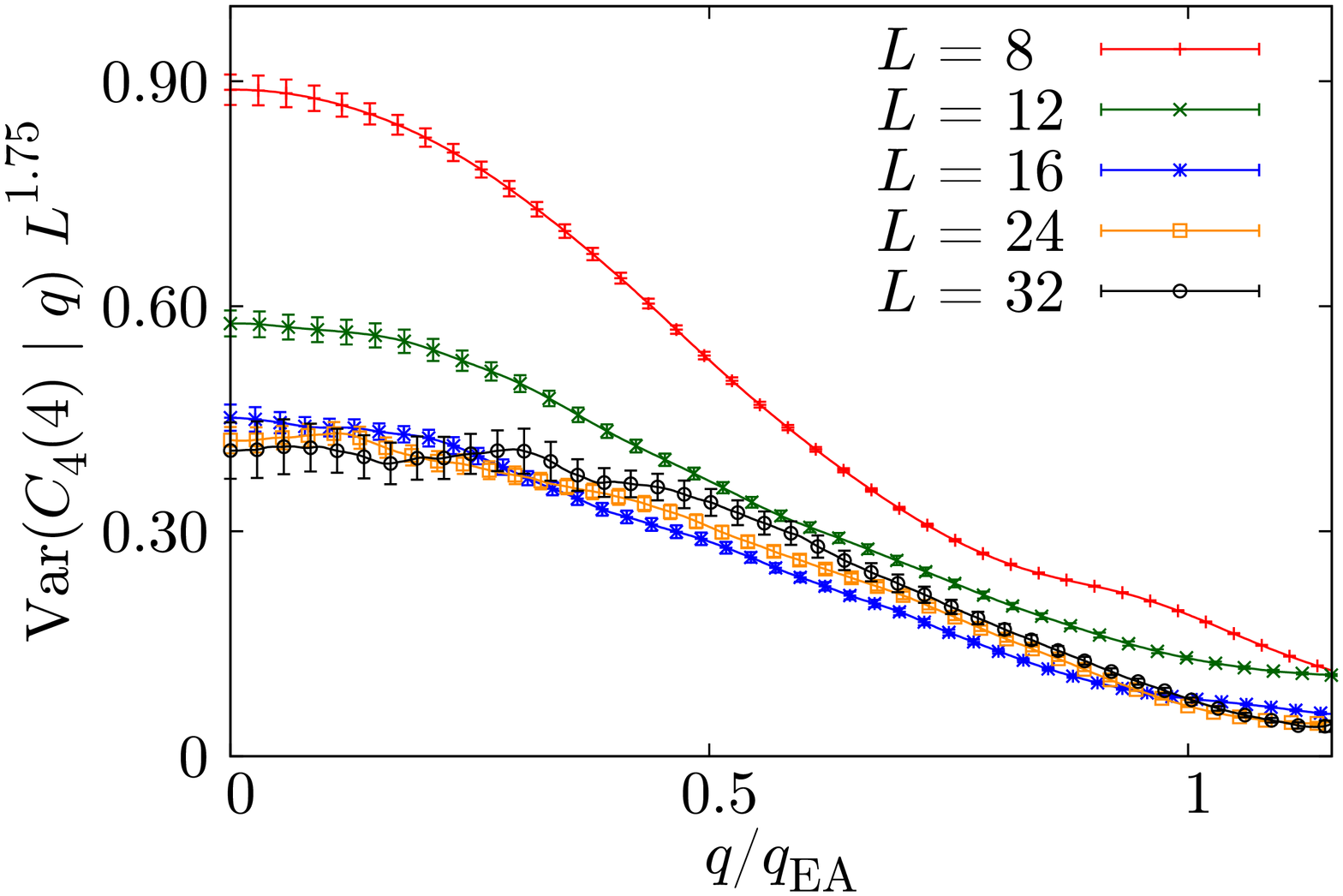}
\end{minipage}
\caption{Plots of the conditional variance at fixed $q$ of $Q_\mathrm{link}$ 
and $C_4(r)$ at $T=0.703$, rescaled by appropriate powers of $L$ (we chose
exponents that provided a good scaling at $q=0$). The abcissas correspond to
$q$ in units of $q_\mathrm{EA}(L,T=0.703)$.}\label{fig:var-qlink}
\end{figure}

Recalling that $\overline{\langle Q_\mathrm{link}\rangle} =
C_4(r\!=\!1)$, we can consider the spatial correlation as a sort of
generalisation of the link overlap. In this sense it is worth
recalling that in a mean-field setting fixing $q^2$ also fixes
$Q_\mathrm{link}$. In a three-dimensional RSB system one would,
therefore, expect the conditional variance
$\mathrm{Var}(Q_\mathrm{link} | q)$, eq.~(\ref{eq:var-q}), to tend to
zero for large lattices~\cite{contucci:06}. The first panel of
\fref{fig:var-qlink} demonstrates that this is the case in our
simulations, where we find that $\mathrm{Var}(Q_\mathrm{link}|q) \sim
L^{-D/2}$. We can extend this result to $r>1$ by considering the
conditional variances of $C_4$. Notice that, unlike $Q_\mathrm{link}$,
$C_4$ is already defined as an averaged quantity in eq.~(\ref{eq:C4})
and not as a stochastic variable, so speaking of its variance is
either trivial or an abuse of language. However, to avoid clutter, we
have maintained the notation $\mathrm{Var}\bigl(C_4(r) |q\bigr)$, as
its intended meaning is clear. These are are plotted in
\fref{fig:var-qlink}, where we see that they decrease even
faster than $\mathrm{Var}(Q_\mathrm{link}|q)$, with a power of $L$
that does not seem to depend on $r$.

\section{Non-equilibrium vs. equilibrium}\label{SECT:EQUILIBRIUM-DYNAMICS}
In reference~\cite{janus:08b}, we suggested the existence of a
time-length dictionary, relating results in the thermodynamical limit
at finite time $\tw$ with equilibrium results for finite size $L$. The
matching for $T=0.7$ was $L \approx 3.7 \xi(\tw)$, where $\xi(\tw)$ is
the coherence length at time $\tw$.  The comparison there was
restricted to $L\leq 20$. The expectation value
$\mathrm{E}(Q_\mathrm{link} | q)$ was confronted with the correlation
function $C_{2+2}(r=1)$, recall the definitions in
section~\ref{SECT:DEF-CORR-DINAMICA}.  We also predicted that the
equilibrium data for $L=33$ would match our non-equilibrium results
for $\tw=2^{32}$. Using the same time-length dictionary our $L=32$
simulations would correspond to $\tw \approx 2^{31}$ and those for
$L=24$ would correspond to $\tw \approx 2^{26}$.

\begin{figure}
\centering
\begin{minipage}{.48\linewidth}
\includegraphics[height=\linewidth,angle=270]{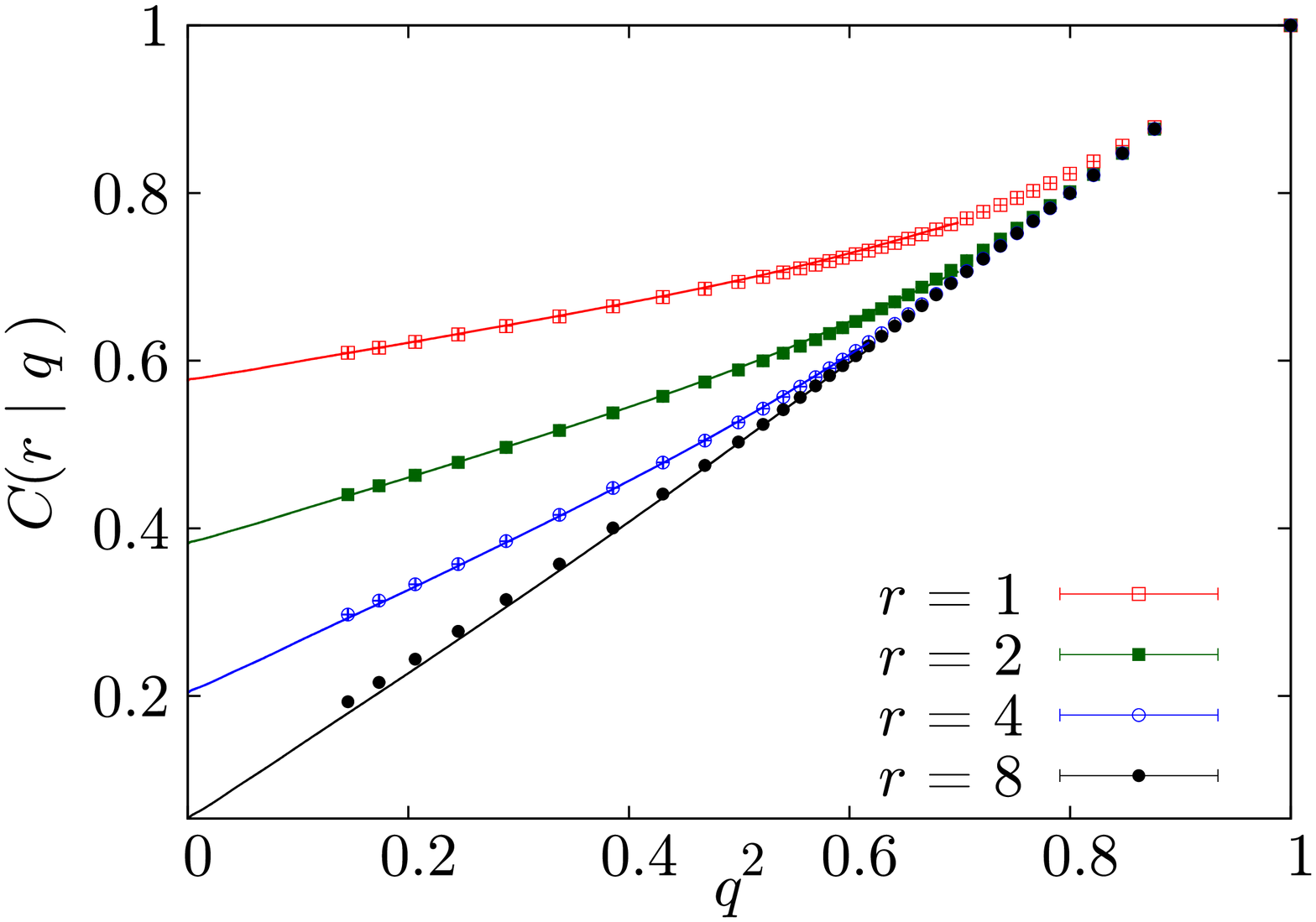}
\end{minipage}
\begin{minipage}{.48\linewidth}
\includegraphics[height=\linewidth,angle=270]{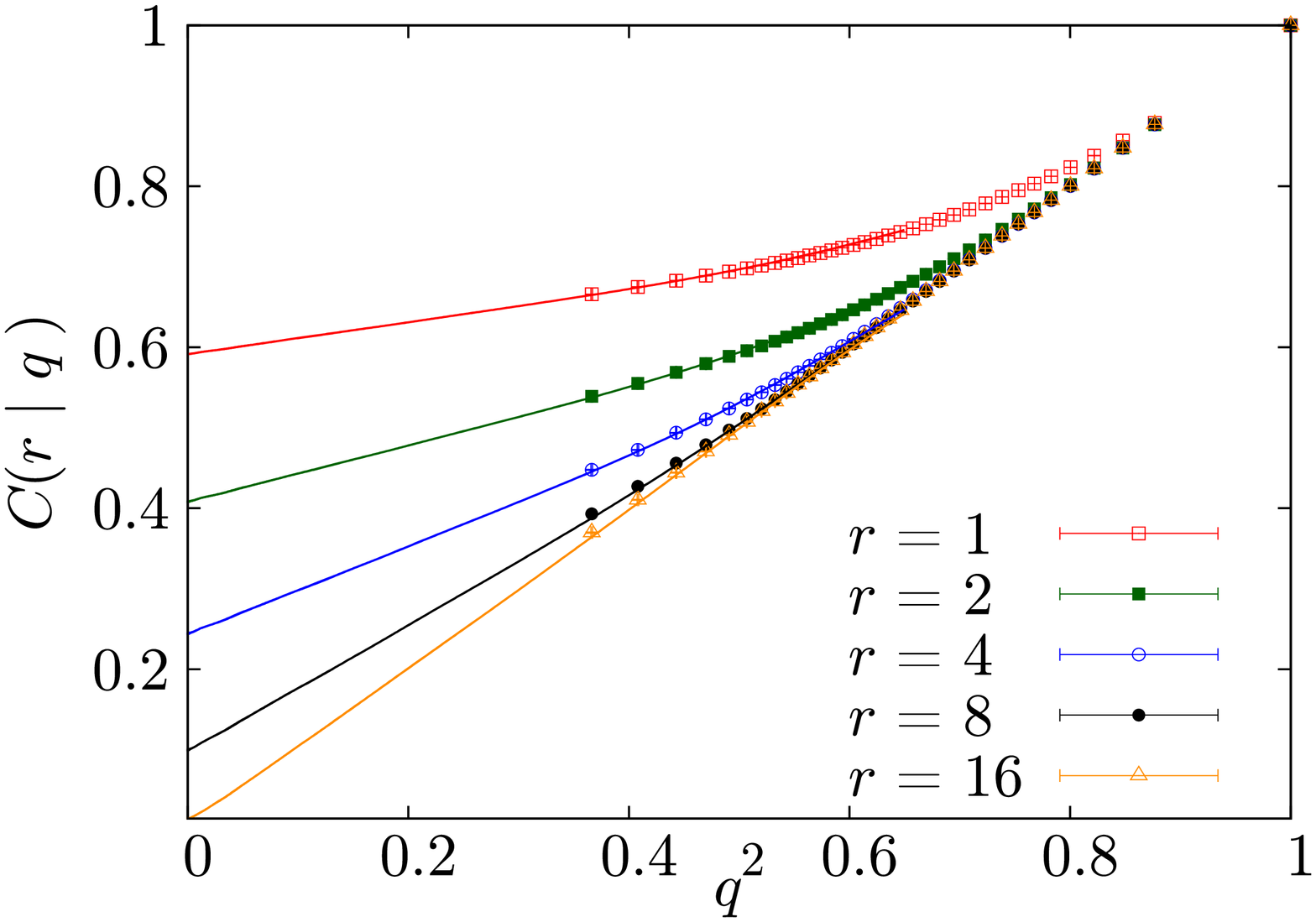}
\end{minipage}
\caption{Equilibrium $C_4(r|q)$ as a function of $q^2$ (lines)
for $L=24$ (left) and $L=32$ (right) lattices at $T=0.703$. 
We compare with non-equilibrium data from~\cite{janus:09b} (points) of $C_{2+2}(r,t,\tw)$ 
as a function of $C^2(t,\tw)$ for $\tw=2^{26}$ (left) and $\tw=2^{31}$ (right),
(see \sref{SECT:DEF-CORR-DINAMICA} for definitions). The errors in 
both sets of data are comparable, and smaller than the point size.}
\label{fig:Estatica-vs-Dinamica}
\end{figure}

Now, recalling that $\overline{\langle Q_\mathrm{link}\rangle}$ is
merely $C_4(r\!=\!1)$, it is natural to extend this correspondence between
$C_4(r)$ and $C_{2+2}(r)$ to $r>1$. Of course, care must be exercised
because $C_4$ in a finite lattice cannot be computed beyond $r=L/2$,
while $C_{2+2}$ is defined for arbitrary $r$.  However, the matching
is very accurate, even for $r$ dangerously close to $L/2$, see
\fref{fig:Estatica-vs-Dinamica}. It is interesting to point out that
the off-equilibrium results of~\cite{janus:08b} and our equilibrium
simulations have similar precision, even though the latter required
about twenty times more computation time on Janus, not to mention a
much more complicated simulation protocol. In this sense we arrive at
the conclusion that simulating the dynamics may be the best way to
obtain certain equilibrium quantities. On the other hand, only the
equilibrium simulations give access to the crucial $C(t,\tw)=0$ physics.

We may now wonder about the experimentally relevant scale of one hour
($\tw \sim 3.6 \times 10^{15}$, taking one MC step as one
picosecond~\cite{mydosh:93}).  Assuming a power-law behaviour,
$\xi(\tw) = A \tw^{1/z(T)}$, with
$z(0.64T_\mathrm{c})=11.64(15)$~\cite{janus:09b}, we conclude that the
correspondence is 1 hour $\longleftrightarrow L \approx 110$. Note,
see for instance \fref{fig:Binder-L}, that $L=110$ is close enough to
$L=32$ to allow a safe extrapolation.

Let us finally stress that the modified droplet scaling for
$\xi(\tw)$~\cite{bouchaud:01} would predict that one hour of physical
time would correspond to equilibrium data on $L$ even smaller than
110. Indeed, according to these authors the time needed to reach some
coherence length $\xi(t_w)$ grows as
\begin{equation}
\tw \sim \tau_0 \xi^{z_\mathrm{c}} \exp\left(\frac{ Y(T) \xi^\psi}{T}\right)\,,
\label{MS}
\end{equation}
where $\tau_0$ is the microscopical time associated to the dynamics;
$z_\mathrm{c}$ is the dynamical critical exponent computed at the
critical point; $\psi$ is the exponent that takes the free energy
barriers into account (from the dynamical point of view) and $Y(T)=Y_0
(1-T/T_\mathrm{c})^{\psi \nu}$, with the $\nu$ exponent being the
static critical exponent linked to the coherence length. Near the
critical point $Y(T) \to 0$ and the power law critical dynamics is
recovered. On the other hand, if we stay below $T_\mathrm{c}$,
Eq. (\ref{MS}) predicts an algebraic grow of $t_\mathrm{w}$ with
$\xi(\tw)$ only for very small coherence lengths. However, as the
coherence length grows, the time needed to reach it diverges exponentially on $\xi(\tw)$.

\section{The link overlap}\label{SECT:LINK-OV}

We shall address here three separated problems: overlap equivalence
(\sref{SECT:OVERLAP-EQUIVALENCE}), replica equivalence
(\sref{SECT:REPLICA-EQUIVALENCE}), and the scaling of the link
susceptibility (\sref{SECT:LINK-SUSCEPTIBILITY}).

\subsection{Overlap equivalence}\label{SECT:OVERLAP-EQUIVALENCE}

As we have discussed previously, it has been
proposed~\cite{contucci:05b,contucci:06} that attention should be
shifted from the spin-overlap (the primary object for mean-field
systems) to the link overlap (the would-be primary object below the
upper critical dimension). Two requirements should be met for this
change of variable to be feasible:
\begin{enumerate}
\item The conditional variance $\mathrm{Var}(Q_\mathrm{link}|q)$ must vanish
in the large $L$ limit.
\item The conditional expectation $\mathrm{E}(Q_\mathrm{link}|q)$ should be
a strictly increasing function of $q^2$.
\end{enumerate}

The scaling with $L$ of $\mathrm{Var}(Q_\mathrm{link}|q)$,
\sref{SECT:COND}, does suggest that the first requirement
holds. We shall investigate here the second requirement. We remark
that the RSB theory expects it to hold, while droplet expects it not to. 
Furthermore, this point is actually the only disagreement between
the RSB and the TNT picture. In fact, RSB expects the derivative
$\mathrm{d}\mathrm{E}(Q_\mathrm{link}|q)/\mathrm{d}q^2$ never to vanish. On the
other hand, TNT supporters expect this derivative to scale as
$L^{D_s-D}$, where $D_s$ represents the (would be) fractal dimension
of the surface of the spin-glass domains. In $D\!=\!3$, $D-D_s\approx
0.44$~\cite{palassini:00}.

\begin{table}[t]
\caption{Coefficients $c_2^{(2m)}$ in the fit to
  eq.~(\ref{eq:fit-derivative}), for various orders of the fitting
  polynomial, and $T=0.703$ and $0.625$.  This coefficient is
  interpreted as
  $\bigl[\mathrm{d}\mathrm{E}(Q_\mathrm{link}|q)/\mathrm{d}q^2\bigr]_{q^2=0}$. We
  report as well the results for fits of the form $c_2^{(4)}= A/L+c$
  (centre) and $c_2^{(4)}= B/L^{0.44}+d$ (bottom). For both fits, we
  also provide the extrapolation to $L\!=\!110$ which, according to the
  time-length dictionary, corresponds to the experimentally relevant
  length scale.}
\label{tab:c2}
\lineup
\begin{tabular*}{\columnwidth}{@{\extracolsep{\fill}}cccccccc}
\hline
&& {\bfseries $T = 0.703$} & & &&{\bfseries $T = 0.625$} &\\
\hline
{\bfseries $L$ } & $c_2^{(2)}$ & $c_2^{(4)}$ & $c_2^{(6)}$ &&
 $c_2^{(2)}$ & $c_2^{(4)}$ & $c_2^{(6)}$\\
\hline
8  & 0.403(5) & 0.405(16) & 0.43(3)& & 0.414(7) & 0.423(19) & 0.45(4)\\
12 & 0.317(5) & 0.321(14) & 0.35(3)& & 0.331(6) & 0.335(18) & 0.36(3)\\
16 & 0.271(4) & 0.262(11) & 0.26(2)& & 0.282(6) & 0.275(16) & 0.28(3)\\
24 & 0.224(5) & 0.222(15) & 0.22(3)& & 0.231(5) & 0.220(14) & 0.22(3)\\
32 & 0.199(6) & 0.201(18) & 0.20(4)& & --- & --- & --- \\
\hline
$\chi^2/\mathrm{d.o.f.}$ && 0.57/3   &&&& 0.46/2 & \\
$A$                      && 2.23(21)  &&&& 2.46(27) & \\
$c$                      && $\ \ \, 0.129(16)$ &&&& $\ \ \,0.121(21)$ & \\
$L=110$                  && $\ \ \, 0.149(14)$ &&&& $\ \ \,0.143(19)$ &\\
\hline
$\chi^2/\mathrm{d.o.f.}$ && 2.39/3   &&&& 0.18/2 & \\
$B$                      && 1.45(11)  &&&&  1.32(15) & \\
$d$                      && $-0.06(3)$ &&&& $-0.11(5)$ & \\
$L=110$                  && $\ \ \, 0.082(20)$&&&&    $\ \ \, 0.058(28)$ &\\
\hline
\end{tabular*}
\end{table}
\begin{table}
\caption{$C(r=1 | q)$ for $q=0$ and $q=q_\mathrm{EA}$ for all our system sizes 
at $T=0.703$. For each $L$, we include the correlation coefficient
between both values of $q$. Specifically, for two quantities $A$ and $B$, 
$\mathcal{R}_{AB} = \overline{( \langle A\rangle - \overline{\langle A\rangle})
( \langle B\rangle - \overline{\langle B\rangle})} /
 \sqrt{\overline{(\langle A\rangle - \overline{\langle A\rangle})^2}
 \  \overline{(\langle B\rangle - \overline{\langle B\rangle})^2}} $
 }\label{tab:C4_r_1}
\lineup
\begin{tabular*}{\columnwidth}{@{\extracolsep{\fill}}cccr}
\hline
\multicolumn{1}{c}{$L$ } & \multicolumn{1}{c}{ $C(1 |0)$}
& \multicolumn{1}{c}{ $C(1 |q_\mathrm{EA})$} & \multicolumn{1}{c}{$\mathcal R$}\\
\hline
8  & 0.46138(82) & 0.57253(33) & 0.134\\
12 & 0.51649(71) & 0.60390(28) & 0.051\\
16 & 0.54552(60) & 0.62089(22) & 0.060\\
24 & 0.57573(77) & 0.63742(17) & $-0.119$ \\
32 & 0.59131(94) & 0.64579(24) & 0.063\\
\hline
\end{tabular*}
\end{table}

To estimate the derivative
$\mathrm{d}\mathrm{E}(Q_\mathrm{link}|q)/\mathrm{d}q^2$, we observe
that $E(Q_\mathrm{link}|q)$ is an extremely smooth function of $q^2$
(see the $r\!=\!1$ curves in
\fref{fig:Estatica-vs-Dinamica}). Hence we can attempt a
polynomial fit:
\begin{equation}\label{eq:fit-derivative}
\mathrm{E}(Q_\mathrm{link}|q)-\mathrm{E}(Q_\mathrm{link}|q=0)=\sum_{k=1}^m\, c_{2k}^{(2m)} q^{2k}\,.
\end{equation}
In particular, the coefficient $c_2^{(2m)}$ provides an estimate of
$\mathrm{d}E(Q_\mathrm{link}|q)/\mathrm{d}q^2$ at $q^2=0$. Playing
with the order $2m$ of the polynomials, one can control systematic
errors. Mind that it is very important to fit the {\em difference}
$\mathrm{E}(Q_\mathrm{link}|q)-\mathrm{E}(Q_\mathrm{link}|q=0)$,
which, due to statistical correlations, has much reduced statistical
errors. On the other hand, data for different $q$ are so strongly
correlated that standard fitting techniques are inappropriate.  We
thus used the approach explained in ref.~\cite{janus:09b}.  The
results, see \tref{tab:c2}, indicate that $c_2^{(4)}$ offers a
reasonable compromise between systematic and statistical errors.

Once we have the derivatives in our hands, we may try to extrapolate
them to large $L$ by means of an RSB fit ($A/L+b$, middle part of
\tref{tab:c2}) or using a TNT fit ($B/L^{0.44}+d$, bottom part of
\tref{tab:c2}). The two functional forms produce a reasonable fit. As
expected, the $1/L$ extrapolation to $L\!=\!\infty$ yields a
non-vanishing derivative, while the $1/L^{0.44}$ extrapolation
suggests that, for large $L$, $\mathrm{E}(Q_\mathrm{link}|q)$ is
constant as $q^2$ varies.  We remark as well that the very same
conclusion was reached in the analysis of the non-equilibrium temporal
correlation functions~\cite{janus:09b}.

However, we have far more accurate data at our disposal than the
derivative $\mathrm{d}E(Q_\mathrm{link}|q)/\mathrm{d}q^2$ at $q^2=0$,
namely the correlation functions themselves. In table~\ref{tab:C4_r_1}
we give our estimates for $C(r=1 | q=0)$ and $C(r=1 | q=0.523\approx
q_\mathrm{EA})$. According to a TNT picture of the SG phase, the two
correlation functions should be equal. As the reader can check, an
infinite volume extrapolation as $L^{-0.44}$ is unbearable for both
correlation functions (even if we discard the two smallest sizes). The
same conclusions hold substituting $L$ by
\begin{equation}
\ell=\pi/\sin(\pi/L)\,,\label{eq:def-ell}
\end{equation}
which is more natural for lattice systems.  Yet, it could be argued
that our data are preasymptotic. Hence, we may try a TNT extrapolation
including scaling corrections.
\begin{equation}\label{eq:TNT-corrections}
C(r=1 |q)  = C_\infty + A_q L^{-0.44} ( 1+ B_q L^{-y})\,.
\end{equation}
We have performed a joint fit of the data on table~\ref{tab:C4_r_1} to
eq.~\eref{eq:TNT-corrections}. The fitting parameters were the four
amplitudes $A_0,B_0,A_{0.523}$ and $B_{0.523}$, the common scaling
corrections exponent $y$ and the common large-$L$ extrapolation
$C_\infty$. We take into account the (almost negligible) correlation
in data for the same $L$ by computing $\chi^2$ with the covariance
matrix, which can be reconstructed from the data on \tref{tab:C4_r_1}.
The result is (notice the highly asymmetric errors)
\begin{equation}
C_\infty =  0.677^{+0.012}_{-0.005}, \qquad y = 0.57^{+0.26}_{-0.08}, \qquad \chi^2 /\mathrm{d.o.f.} = 9.1/4.
\end{equation}
Were the functional form in eq.~\eref{eq:TNT-corrections} correct, the
probability of $\chi^2$ being even larger than we found would be only
$6\%$.

On the other hand, in an RSB setting, one would expect $C(r\!=\!1|q)$
to scale as $1/L$, with a $q$-dependent infinite volume value
$C_\infty(q)$. Indeed, if we fit the data on \tref{tab:C4_r_1} to
$C(1|q) = C_\infty(q) + A /\ell$ we obtain
\begin{eqnarray}
C_\infty(q=0) &=& 0.6349(8), \qquad \chi^2/\mathrm{d.o.f.} = 3.63/3,\\
C_\infty(q=q_\mathrm{EA}) &=& 0.6711(2),\qquad \chi^2/\mathrm{d.o.f.} = 2.86/3.
\end{eqnarray}
We note as well that
$[C_\infty(q=q_\mathrm{EA})-C_\infty(q=0)]/q_\mathrm{EA}^2\approx
0.132$, in fair agreement with the $1/L$ extrapolation for the
derivative in \tref{tab:c2}.

However, more important than the extrapolation to $L\!=\!\infty$ is
the extrapolation to $L\!=\!110$, the length scale that, for
$T\!=\!0.7$, matches the experimental time scales. For $T\!=\! 0.625$,
$L\!=\!110$ is surely larger than the relevant length scale but,
unfortunately, the time-length dictionary at such a low temperature
still needs to be tuned. As it can be seen in the middle and bottom
parts of \tref{tab:c2}, the two extrapolations yield a non-vanishing
derivative.

Thus, whichever the standpoint adopted, the conclusion is identical
for RSB and TNT theories: at the experimentally relevant length
scales, overlap equivalence can be assumed.

\subsection{Replica equivalence}\label{SECT:REPLICA-EQUIVALENCE}
\begin{figure}[b]
\centering
\begin{minipage}{.48\linewidth}
\includegraphics[height=\linewidth,angle=270]{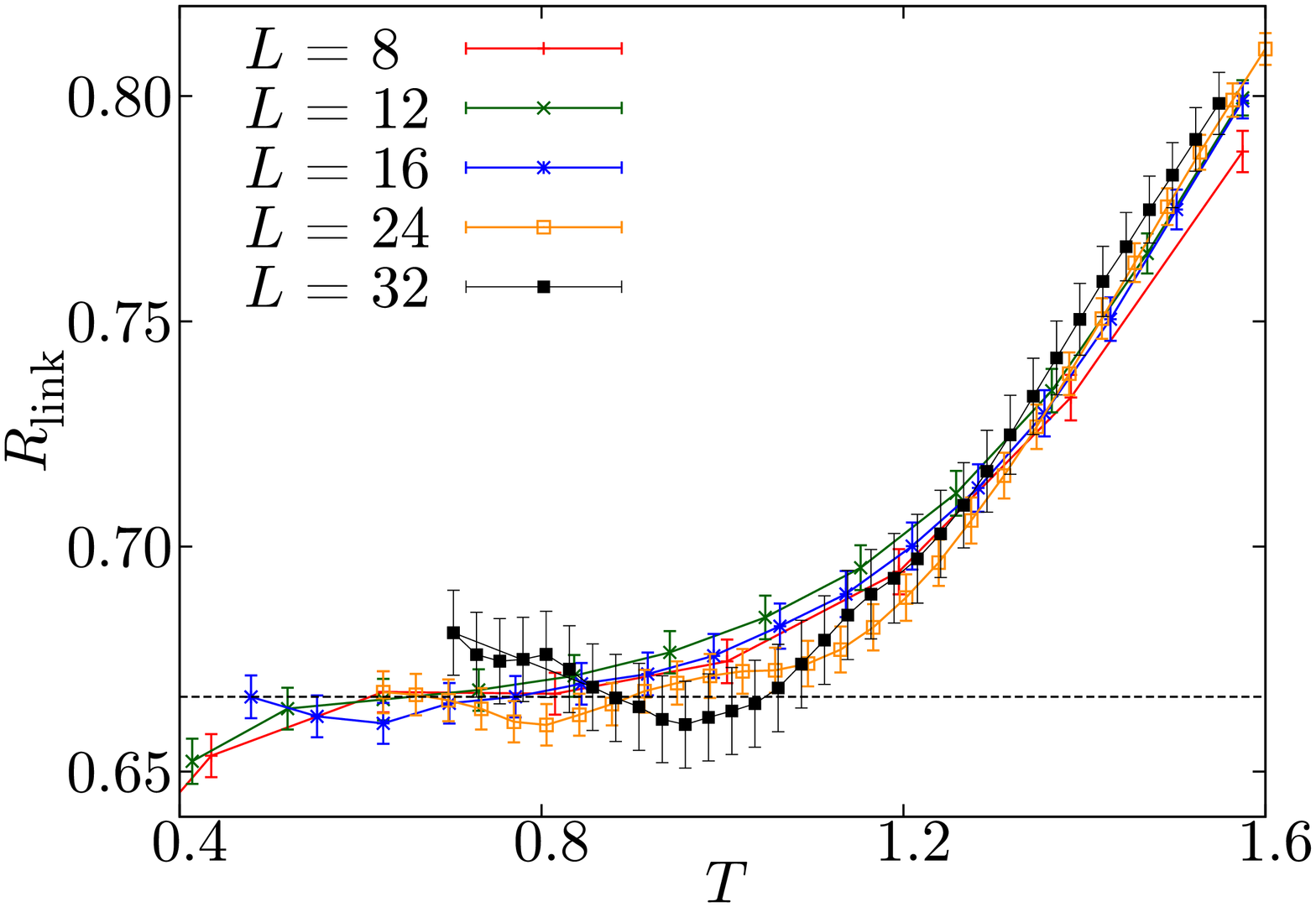}
\end{minipage}
\begin{minipage}{.48\linewidth}
\includegraphics[height=\linewidth,angle=270]{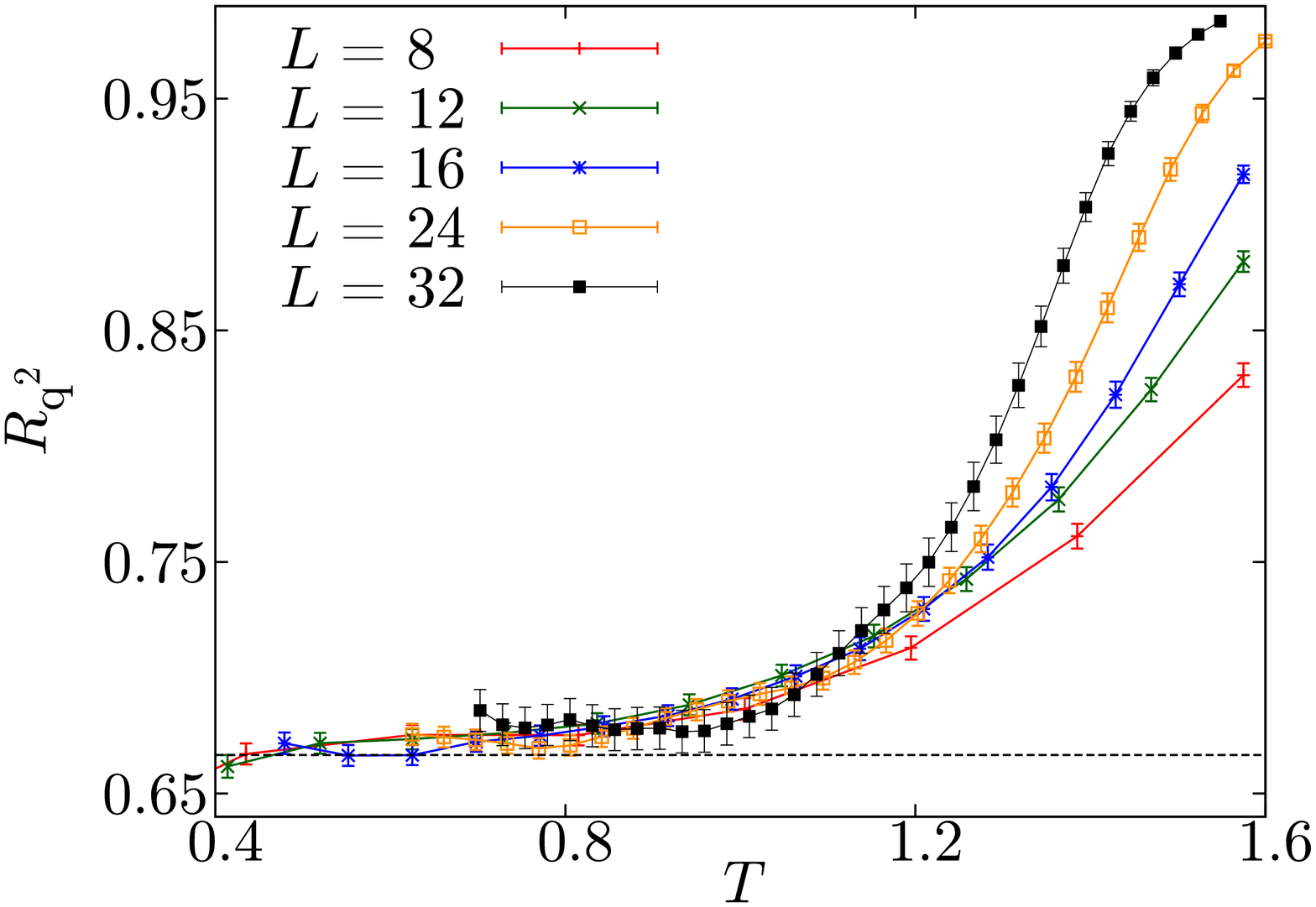}
\end{minipage}
\caption{The ratios $R_\mathrm{link}$, eq.~(\ref{eq:R-link}), (left panel)
  and $R_\mathrm{q^2}$, eq.~(\ref{eq:R-q2}), (right panel) versus $T$
for the different system sizes. The replica equivalence property implies
that, in an RSB system below $T_\mathrm{c}$, $R^\mathrm{link} = 2/3$ 
in the large-$L$ limit. Recall that $T_\mathrm{c} \approx 1.1$.}
\label{fig:estabilidad}
\end{figure}

We consider now the ratio
\begin{equation}\label{eq:R-link2}
R_{\mathrm{link}}=\frac{\overline{\langle Q_\mathrm{link}^2 \rangle\ -\ \langle Q_\mathrm{link} \rangle^2}}{\overline{\langle Q_\mathrm{link}^2 \rangle}\ -\ \overline{\langle Q_\mathrm{link} \rangle}^2}\,,
\end{equation}
defined in \sref{SECT:DEF-QLINK}. As was explained there, the RSB
theory expects it to reach a constant value $2/3$ below
$T_\mathrm{c}$, whereas the droplet and TNT theories lack a definite
prediction. Our numerical data fit very well the RSB expectation (see
\fref{fig:estabilidad}--left).

Besides, we can also study a similar ratio, in which the mean-field
substitution $ Q_\mathrm{link}\rightarrow q^2$ is performed:
\begin{equation}\label{eq:R-q2}
R_{q^2}=\frac{\overline{\langle q^4 \rangle\ -\ \langle q^2 \rangle^2}}{\overline{\langle q^4 \rangle}\ -\ \overline{\langle q^2 \rangle}^2}\,.
\end{equation}
Overlap equivalence suggests that $R_{q^2}$ approaches $2/3$ in the
large $L$ limit (again neither the droplet nor the TNT theories have a
definite prediction). Our data at low temperatures seem compatible
with the $2/3$ expectation, see \fref{fig:estabilidad}--right. On
the other hand, the convergence to the thermodynamic limit seems
fairly slower close to $T_\mathrm{c}$. We recall that a previous
computation concluded as well that violations of $R_{q^2}\!=\!2/3$ are
due to critical fluctuations~\cite{marinari:00}.

\subsection{Link susceptibility}\label{SECT:LINK-SUSCEPTIBILITY}
\begin{figure}
\centering
\begin{minipage}{.48\linewidth}
\includegraphics[height=\linewidth,angle=270]{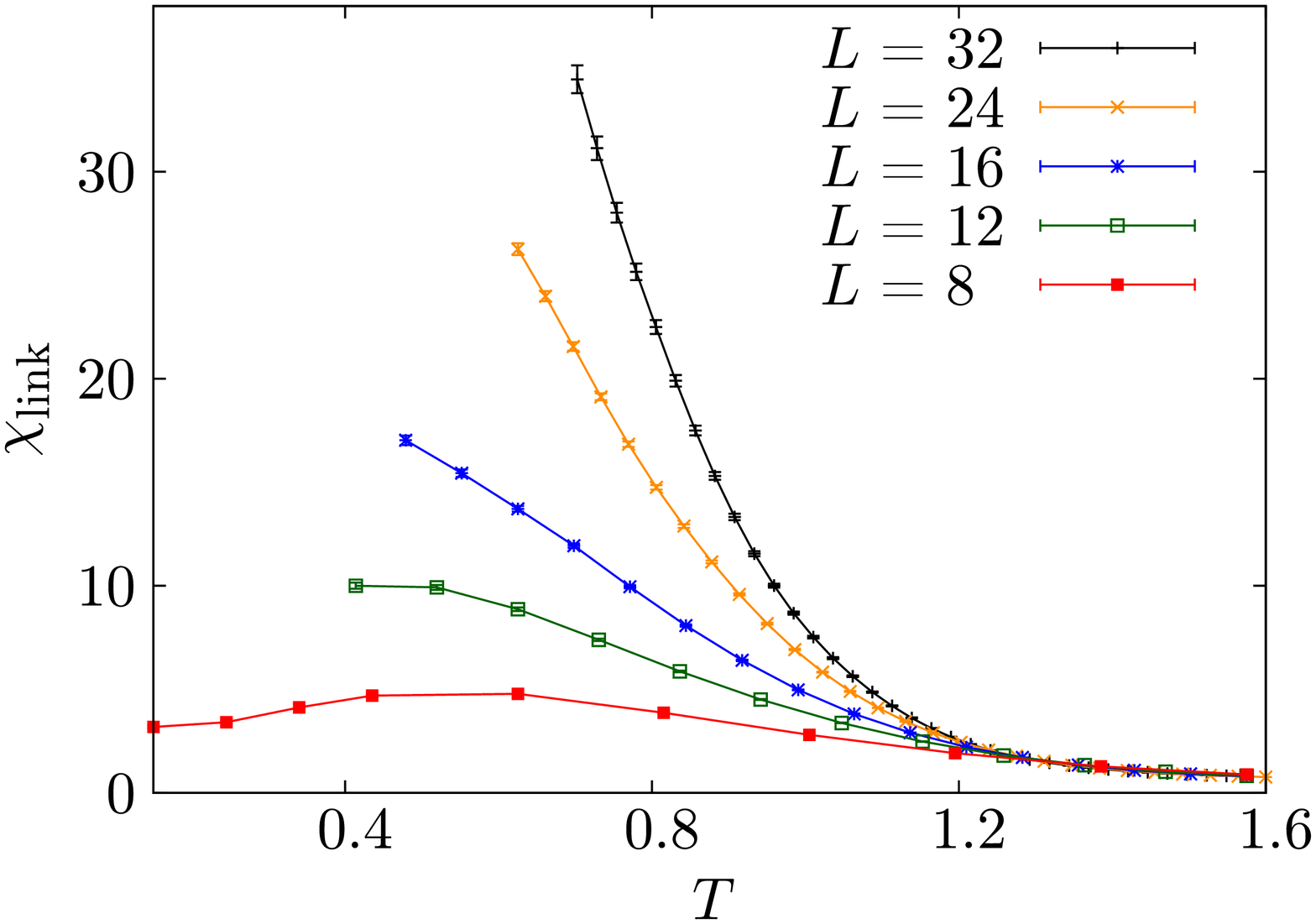}
\end{minipage}
\begin{minipage}{.48\linewidth}
\includegraphics[height=\linewidth,angle=270]{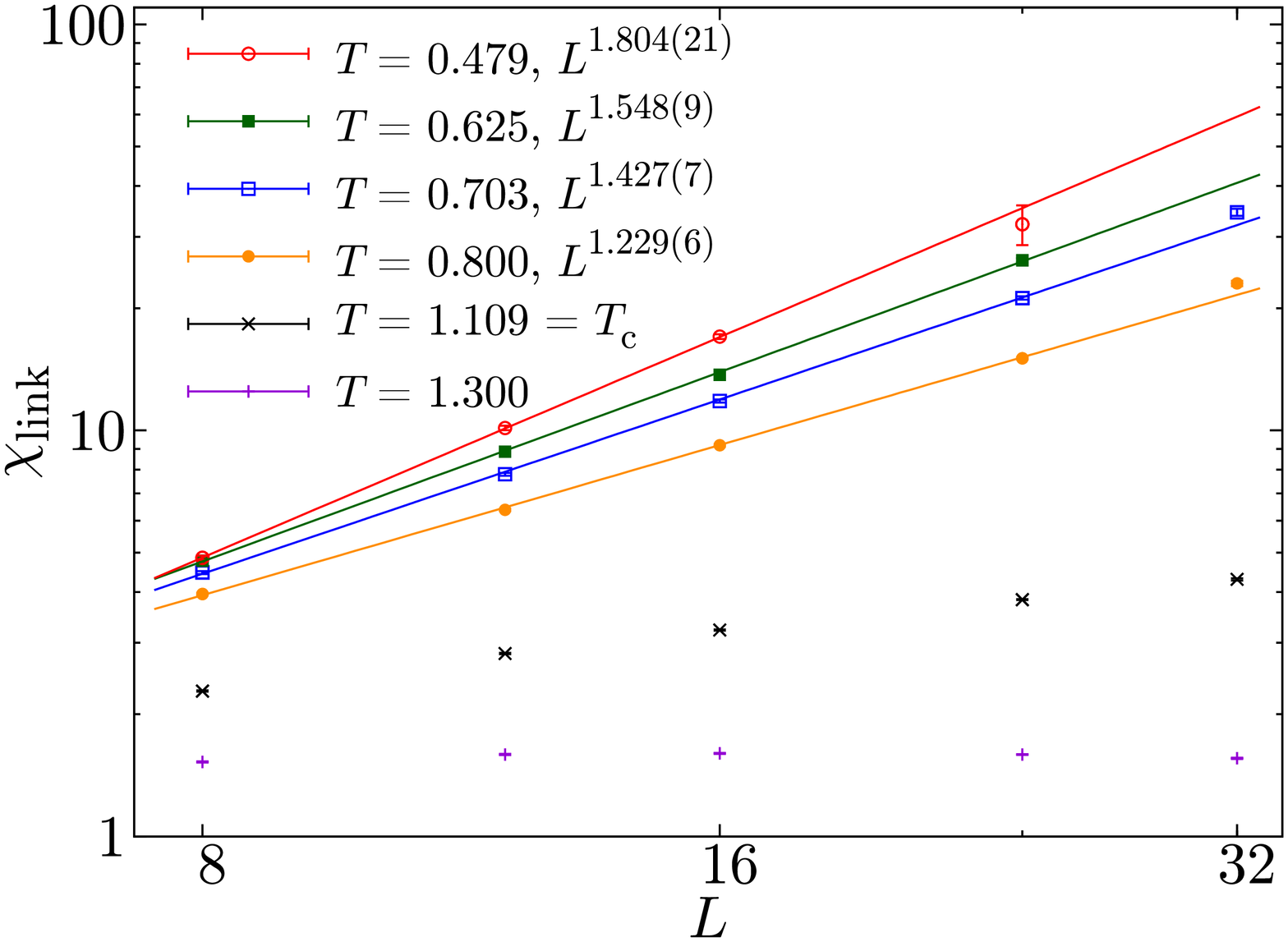}
\end{minipage}
\caption{(Left) Susceptibility $\chi_\mathrm{link}$ vs. temperature
  for the different system sizes. (Right) Behaviour of
  $\chi_\mathrm{link}$ with $L$ for different temperatures. Lines are
  power-law fits. The effective exponents found in each of these fits
  is reported in the legends.}
\label{fig:chi_link}
\end{figure}

We show in \fref{fig:chi_link}--left the link susceptibility
$\chi_\mathrm{link}$, eq.~(\ref{DEF:CHI-LINK}), as a function of
temperature for different lattice sizes. It is clear enough that this
susceptibility is divergent in the spin-glass phase and that the lower
the temperature, the more violent the divergence. Hence, it is clear
that this particular effect is not due to critical fluctuations.

We perform a more quantitative study in
\fref{fig:chi_link}-right. As discussed in
\sref{SECT:DEF-QLINK}, according to RSB theory, one would expect
$\chi_\mathrm{link}\sim L^D$ in the SG phase.

We find evidence of a critical divergence. At and above
$T_\mathrm{c}$, our data grow very softly with $L$ (at
$T=1.3\approx 1.17 T_\mathrm{c}$, data seem to reach a limiting
value). However, below $T_\mathrm{c}$, we observe an effective
exponent that grows when we lower the temperature 
We observe that the effective exponent, for our lattice
sizes and temperatures, has already grown beyond the Chayes bound of
$D/2$ but still has not reached the RSB expectation of $D$.  Note that
no existing theory of the spin-glass phase can accommodate a
temperature-dependent exponent. Therefore, the most economic scenario
is that our lattice sizes are not large enough, so we are still
in a preasymptotic regime for this quantity.

\begin{table}[t]
\caption{ Ratio $S_\mathrm{link}^{(2m)}$, eq.~\eref{eq:Slink}, for all our
lattice sizes at $T=0.625,0.703$, using the coefficients from \tref{tab:c2}.}
\label{tab:Slink}
\lineup
\begin{tabular*}{\columnwidth}{@{\extracolsep{\fill}}cllcll}
\cline{2-6}
&\multicolumn{2}{c}{$T = 0.703$} & &\multicolumn{2}{c}{$T = 0.625$} \\
\hline
{\bfseries $L$ } &\multicolumn{1}{c}{$S_\mathrm{link}^{(2)}$} &\multicolumn{1}{c}{$S_\mathrm{link}^{(4)}$} &&
\multicolumn{1}{c}{$S_\mathrm{link}^{(2)}$} & \multicolumn{1}{c}{$S_\mathrm{link}^{(4)}$} \\
\hline
8  & 0.838(21)  & 0.846(67)  & & 0.859(29) & 0.897(81)   \\
12 & 0.777(25)  & 0.797(70)  & & 0.801(29) & 0.821(88)   \\
16 & 0.755(22)  & 0.706(59)  & & 0.766(33) & 0.729(85)   \\
24 & 0.776(35)  & 0.76(10)   & & 0.745(32) & 0.675(86)   \\
32 & 0.816(49)  & 0.83(15)   & & \multicolumn{1}{c}{ ---} & \multicolumn{1}{c}{---}  \\
\hline
\end{tabular*}
\end{table}

Let us take a slightly different point of view. Rigorous theorems
discussed in \sref{SECT:DEF-QLINK} tell us that,
eq.~(\ref{REP-EQUIVALENCE-SENCILLA}), if
\begin{equation}
\lim_{L\to\infty} \chi_\mathrm{link}/ L^D >0\,,
\end{equation} also the width $\sigma_{Q_\mathrm{link}}$ of the
probability density function for $Q_\mathrm{link}$,
\begin{equation}
\sigma^2_{Q_\mathrm{link}}=\overline{\langle Q_\mathrm{link}^2\rangle} \ -\ \overline{\langle
  Q_\mathrm{link}\rangle}^2\,,
\end{equation}
 will be non-vanishing in the thermodynamic limit (see also
 \sref{SECT:REPLICA-EQUIVALENCE}). It is very important that the
converse statement also holds.

Now, using the identity (\ref{eq:var-q-anchura}), we can split up this
variance in two different contributions:
\begin{eqnarray}
\sigma^2_{Q_\mathrm{link}}&=&\int_{-\infty}^{\infty}\mathrm{d}q\ P(q)\left(\mathrm{Var}(
Q_\mathrm{link}|q)\ +\ 
 \left[\mathrm{E}\left(Q_\mathrm{link}|q\right)-\overline{\left\langle
  Q_\mathrm{link}\right\rangle}\right]^2\right)\,.
\end{eqnarray}
Since $\mathrm{Var}(Q_\mathrm{link}|q)$ scales as $L^{-D/2}$ (see
~\cite{contucci:06} and \fref{fig:var-qlink}), only the second
term may survive the large $L$ limit.

This suggests the definition of a modified link susceptibility:
\begin{equation}\label{DEF:HAT-CHI-LINK}
\hat\chi_\mathrm{link}=\frac{L^D\int_{-\infty}^{\infty}\mathrm{d}q\ P(q)\left[\mathrm{E}\left(Q_\mathrm{link}|q\right)-\overline{\left\langle
  Q_\mathrm{link}\right\rangle}\right]^2}{\overline{\left\langle q^4\right\rangle-\overline{\left\langle
  q^2\right\rangle}^2}}.
\end{equation}
According to RSB theory, $\hat\chi_\mathrm{link}$ should scale as
$L^D$ whereas it would not diverge as violently in a droplet or TNT
scenario.  The rationale for dividing out the $\overline{\langle q^4\rangle-\overline{\langle
  q^2\rangle}^2}$ can be found in
eq.~(\ref{eq:fit-derivative}). Assuming that the lowest order
polynomial is adequate, one finds that (of course, the particular value
of the index $m$ should be immaterial)
\begin{equation}\label{eq:estimacion}
\hat\chi_\mathrm{link}\approx L^D \bigl[c_2^{(2m)}\bigr]^2\,.
\end{equation}
Hence, the TNT theory would expect $\chi_\mathrm{link}/L^D$ to
tend to zero, just because it predicts that in the large-$L$ limit
$c_2^{(2m)}=0$. Note that the droplet theory would predict a vanishing
$\chi_\mathrm{link}/L^D$ for a different reason, namely because
they expect that $\overline{\langle q^4\rangle-\overline{\langle
    q^2\rangle}^2}$ should vanish.

Let us check to what extent the estimate \eref{eq:estimacion} is accurate.
We show on \tref{tab:Slink} the ratios
\begin{equation}\label{eq:Slink}
S_\mathrm{link}^{(2)}=\frac{L^D[c_2^{(2)}]^2}{\hat\chi_\mathrm{link}}\,,\qquad
S_\mathrm{link}^{(4)}=\frac{L^D[c_2^{(4)}]^2}{\hat\chi_\mathrm{link}}\,.
\end{equation}
Referring again to eq.~\eref{eq:fit-derivative}, it is clear that the
contribution linear in $q^2$ explains a large fraction of
$\hat\chi_\mathrm{link}$, and that this fraction is not likely to
vanish in the large-$L$ limit. 

Hence the question of whether $\chi_\mathrm{link}$ diverges as $L^D$
or not, turns out to be strictly equivalent to that of overlap
equivalence that we discussed at length in
Sect.~\ref{SECT:OVERLAP-EQUIVALENCE}. Our interpretation is that the
effective scaling in \fref{fig:chi_link}--right is mostly due to
strong finite size effects in $c_2^{(2m)}$. Under this light, the
effective exponents reported in \fref{fig:chi_link}--right are
preasymptotic.
In fact, the ratio $A/c$ is large ($c^{(2m)}_2(L) = c + A/L$,
see \tref{tab:c2}), which tells us that
for $\chi_\mathrm{link}$ and related quantities finite volume corrections are
particularly large and naive power law fits may give wrong results.

Let us conclude this section by checking how these quantities behave
in a 2$D$ Ising ferromagnet (i.e. with no disorder built
in). Although this model is clearly too simple, it is also true that,
up to our knowledge, the quantities investigated here have not been
looked at before. Hence, it is interesting to see what happens even in
this simple case. We use two replicas to compute
$\chi_\mathrm{link}$. Results for $\chi_\mathrm{link}$ are presented
on \tref{tab:2Dising} for two different temperatures below the
critical temperature $T_\mathrm{c}$. There we can see that
$\chi_\mathrm{link}$ approaches a limiting $\mathcal O(L^0)$ value
when $L$ grows. Furthermore, the limiting value decreases when
lowering the temperature away from $T_\mathrm{c}$. Hence, a divergent
link susceptibility below $T_\mathrm{c}$ is something that should {\em not} be taken for granted.

\begin{table}[h]
\caption{$\chi_\mathrm{link}$ in the 2$D$-Ising model
  ($T_\mathrm{c}=2/\log(1+\sqrt{2})\approx 2.26918531\ldots$).}\label{tab:2Dising}
\lineup
\begin{tabular*}{\columnwidth}{@{\extracolsep{\fill}}ccc}
\hline
\multicolumn{1}{c}{\bfseries $L$ } & \multicolumn{1}{c}{\bfseries $T = 0.992T_\mathrm{c}$}
& \multicolumn{1}{c}{\bfseries $T = 0.986T_\mathrm{c}$}\\
\hline
8  & 7.16(1)\0 & 7.091(8)\0 \\
12 & 8.776(6) & 8.614(15)\\
16 & 10.247(12) & 9.868(9)\0\\
24 & 11.47(2)\0\0 & 10.51(2)\0\0\0\\
32 & 12.058(15) & 10.639(12)\0\\
\hline
\end{tabular*}
\end{table}

\section{Conclusions}\label{SECT:CONCLUSIONS}

We have obtained equilibrium configurations of the Ising spin glass
($D\!=\!3$,  $\pm 1$ Edwards-Anderson model) on large lattices at low
temperatures ($T=0.64 T_\mathrm{c}$ for $L=32$, $T=0.56
T_\mathrm{c}$ for $L=24$, and even lower temperatures for smaller
systems, see \tref{tab:parameters}).  This unprecedented
computation has been made possible by the Janus computer. However, the
parallel tempering had never before been put to such stress, and we
have devoted a large effort to convince ourselves that thermalisation
was achieved. New thermalisation tests were devised. Furthermore, a
new simulation strategy had to be employed: the simulation time needs
to be tailored sample by sample (for one cannot afford adopting
worst-case parameters).

The main conclusion we draw is that the correspondence between
equilibrium results and non-equilibrium dynamics (much
easier to compare with experimental work), is deeper than anticipated. In fact,
one can construct a time-length dictionary, such that equilibrium
correlation functions on finite systems match non-equilibrium
correlators at finite time (but infinite system size). The evidence
for this correspondence consists of: (i) quantitative comparison of
the spatial correlation functions and (ii) the analysis of overlap
equivalence on equilibrium (this work) and non-equilibrium
settings~\cite{janus:09b}. In addition, there is a remarkable
coincidence between the replicon exponent obtained from equilibrium
methods~\cite{janus:10b}, and from non-equilibrium
dynamics~\cite{janus:08b,janus:09b}.

The unavoidable consequence of this time-length correspondence is
that the system size that is relevant for the experimental work (time
scales of one hour, say) at $T\!=\!0.64 T_\mathrm{c}$ is not infinite,
but $L\!=\!110$. Note that this correspondence was obtained assuming a
power-law growth with time of the spin-glass coherence length in
experimental samples. Should the modified droplet scaling for
$\xi(\tw)$ hold~\cite{bouchaud:01}, the relevant equilibrium system
size would be even smaller.  It is obvious that extrapolating
numerical data from $L\!=\!32$ to $L\!=\!110$ is far less demanding
than extrapolating them to infinite size. All such extrapolations in
this work (even those assuming droplet scaling) were conclusive.  The only
effective theory that is relevant at experimental time scales is
Replica Symmetry Breaking.

However, the question of whether RSB is only an effective theory in
$D\!=\!3$ or a fundamental one does not lack theoretical interest.  We
have attempted several extrapolations to infinite system size in this
work, finding that droplet theory is ruled out, unless a change of
regime arises for system sizes much larger than our reached
$L\!=\!32$. We remark that in Sect.~\ref{sect:picos} we have
numerically determined a crossover length that rules finite size
effects. As expected for a large enough system, it scales with
temperature as a {\em bulk} correlation length. However, on the basis
of numerical data alone, one can never discard that new behaviour
might appear for much larger system sizes, irrelevant for current
experimental work.

We found three contradictions with droplet theory. First, in order to
have a trivial Binder cumulant, finite-size corrections had to be of
order $\sim L^{-0.11}$. Such finite size corrections would imply a
vanishing, or even negative, spin-glass order parameter
$q_\mathrm{EA}$. Second, according to droplet theory
(see~\cite{bray:87}, page 139) finite size corrections $\sim
L^{-0.11}$ imply that the connected spatial correlation function at
$q\!=\!q_\mathrm{EA}$ decays as $1/r^{0.11}$. A direct estimate
indicates that, at $q\!=\!q_\mathrm{EA}$, correlations decay as
$1/r^{0.6}$~\cite{janus:10b}. Third, the probability density function
$P(q\!=\!0)$ does not decrease with increasing system size (a similar
conclusion was reached in~\cite{katzgraber:01,katzgraber:03}).

Our analysis of overlap equivalence is compatible with the RSB
picture, without invoking sophisticated finite-size effects. On the
other hand, the statistical likelihood for TNT theory, as formulated
in~\cite{palassini:00}, has been quantified to be $6\%$. In any case,
TNT scaling predicts that for $L\!=\!110$ the surface-to-volume ratio
of the magnetic domains is still of order one (in agreement with
RSB). In addition, we find that replica equivalence is consistent with
the RSB picture (while TNT lacks a definite prediction). Furthermore,
the link susceptibility, $\chi_\mathrm{link}$, is definitively
divergent in the spin-glass phase (since the divergence is stronger
the lower the temperature, its origin is obviously non-critical). We
are aware of no argument in TNT theory implying the divergence of the
link susceptibility. On the other hand, RSB theory does require a
divergent $\chi_\mathrm{link}$. However, RSB demands a scaling
$\chi_\mathrm{link}\sim L^D$. Such growth regime has still not been
reached for our system sizes, although we have identified the origin
of this preasymptotic behaviour.

A final lesson from the present numerical study is that careful
non-equilibrium simulations~\cite{janus:09b} are almost as rewarding
as the equilibrium work. Indeed, our previous non-equilibrium
study~\cite{janus:08b,janus:09b} reached a time scale that corresponds
to the present equilibrium $L\!=\!32$ simulation. Yet, the numerical
effort to obtain the data in Fig.~\ref{fig:Estatica-vs-Dinamica} has
been larger by, roughly, a factor of 20 in the case of the equilibrium
work. It is true that the equilibrium approach allows to investigate
directly the crucial $q=0$ region, where in the nonequilibrium case
one would need to rely on difficult extrapolations to infinite time.
However, we do not think that there is much road ahead for equilibrium
studies, due to the failure of the parallel tempering
algorithm. Indeed, see \tref{tab:parameters}, it takes about 3.5 times
more numerical work to equilibrate 1000 samples of $L=32$ at $T=0.64
T_\mathrm{c}$ than 4000 samples of $L=24$ down to $T=0.56
T_\mathrm{c}$. Clearly enough, the temperature window accesible with
the parallel tempering algorithm decreases very fast as the system
size grows.  We believe this failure to be due to a genuine
temperature-chaos effect. However, in order to analize quantitatively
the effect one needs to correlate the (sample dependent) temperature
bottlenecks, see Fig.~\ref{fig:historiabetas}--left, with the spin
overlap at different temperatures. This analysis is left for future
work~\cite{janus:xx}.

\section*{Acknowledgments}
We acknowledge support from MICINN, Spain, through research contracts
No. TEC2007-64188, FIS2006-08533-C03, FIS2007-60977, FIS2009-12648-C03
and from UCM-Banco de Santander. B.S. and D.Y. are FPU
fellows (Spain) and R.A.B. and J.M.-G. are DGA fellows. S.P.-G. was
supported by FECYT (Spain). The authors would
like to thank the Ar\'enaire team, especially J.~Detrey and 
F.~de~Dinechin for the VHDL code of the logarithm function~\cite{detrey:07}.
M. Moore posed interesting questions that helped us sharpen the 
discussion in \sref{sect:picos}.

\appendix

\section{Our thermalisation protocol}\label{AP:THERMALISATION}
\label{sec:protocol}
We have followed a three-step procedure to thermalise each sample:
\begin{enumerate}
\item We simulate for a fixed minimum length of $N_\mathrm{HB}^\mathrm{min}$ 
MCS, chosen to be enough to thermalise most of the samples. Notice that
most published parallel-tempering simulations stop here, assessing the thermalisation
only through the time evolution of disorder-averaged observables.
\item We discard the first sixth of the measurements and 
compute the integrated autocorrelation time, choosing the self-consistent 
window $W$ of~(\ref{eq:tau-int}) so that $W>6\tau_\mathrm{int}$. 
Using this first estimate of the integrated time, we enlarge the simulation
until $N_\mathrm{HB} > 22 \tau_\mathrm{int}$, always discarding its first sixth. 
A criterion based on $\tau_\mathrm{int}$ was first used in~\cite{fernandez:09b}.
\item Now that we have a reasonably dimensioned simulation, we can 
compute the exponential autocorrelation time  (which is typically bigger than, 
but of the same order of magnitude of, the integrated time).
We demand that $N_\mathrm{HB}$ be larger than $12\tau_\mathrm{exp}$.
\end{enumerate}
This last step is the main innovation of these simulations. Notice that we
have to perform non-linear fits in some $10^4$ autocorrelation functions, with
a sample-dependent fitting range. This is a somewhat delicate procedure, so we
have taken great care to ensure it is failsafe.

We start by assuming that the correlation function~(\ref{eq:corr}) can be 
approximated by the sum of two exponentials:
\begin{equation}\label{eq:fit}
\hat C(t) \simeq A_1 \rme^{-t/\tau_1} + A_2 \rme^{-t/\tau_2},\qquad \tau_1 = \tau_\mathrm{exp} > \tau_2.
\end{equation}
Usually, one chooses the range for such a fit manually but this is not
practical here, due to the sheer number of correlation functions 
we have to study. Hence, assuming that the exponential time is not much larger 
than the integrated one, we have used the latter in order to define 
our fitting range (notice that if $A_2=0$, $A_1=1$ and $\tau_1 = \tau_\mathrm{int}$).
Our fitting procedure has three steps
\begin{enumerate}
\item[(a)] We perform a first fit to a single exponential in the range $[2\tau_\mathrm{int},3\tau_\mathrm{int}]$, from
which we obtain an amplitude $A$ and a time $\tau$.
\item[(b)] Using $\tau_1=\tau$, $\tau_2=\tau/10$, $A_1=A$ and $A_2=1-A$ as a starting point we
perform the non-linear fit to~(\ref{eq:fit}) with a  Levenberg-Marquardt scheme~\cite{press:92}.
The fitting range is chosen as $[\tau_\mathrm{int}/10, 10\tau_\mathrm{int}]$.
\item[(c)] Sometimes $\tau_2$ is very small and $\hat C(t)$ is indistinguishable
from a single exponential in $[\tau_\mathrm{int}/10, 10\tau_\mathrm{int}]$.
In these occasions the fit in step 2 fails, which can be detected in a  number of ways
(very large or even negative values for one of the $A_i$, absurdly large values 
for $\tau_1$ or even a complete breakdown of the iterative method). 
For these samples, a third fit to a single exponential is performed in the 
range $[5 \tau_\mathrm{int}, 10\tau_\mathrm{int}]$. There is one 
exception: when one of the $A_i$ is negative, there appears a very pronounced 
downwards fluctuation in $\hat C(t)$ for large times, which can lead
to an underestimation of $\tau_\mathrm{exp}$. In these occasions, the third
fit is performed in $[2.5 \tau_\mathrm{int}, 5\tau_\mathrm{int}]$.
\end{enumerate}
This automatic and fully quantitative procedure works for most samples, but there
are some potential pitfalls which may lead to our underestimating $\tau_\mathrm{exp}$.
Sometimes, the exponential time is much larger than $\tau_\mathrm{int}$. 
This can result in a failure of the automatic method for two reasons:
(1) as $\tau_\mathrm{exp} \gg \tau_\mathrm{int}$, the fitting ranges 
are no longer well adjusted (2) a very large $\tau_\mathrm{exp} /\tau_\mathrm{int}$ 
implies a very low value for $A_1$. We address this problem
by enlarging the measurement bins by a factor of 10 in 
case $\tau_\mathrm{exp} > 10 \tau_\mathrm{int}$. This way, both $\tau_\mathrm{int}$ 
and $A_1$ grow, and the fit works much better.

The possibility also exists that $\hat C(t)$ may be misleading,
because the simulation is so much shorter than the exponential time
that some of the configurations have not yet explored the relevant
minima of the free energy (i.e., the $\hat C(t)$ we are measuring is
not yet the \emph{equilibrium} one). This happens when some of the
$4N_T$ configurations have not crossed the critical temperature in the
parallel-tempering dynamics.  The assumption here, key to the parallel
tempering method, is that once a configuration spends a few MCS at
high temperatures it becomes completely decorrelated (remember that,
due to Janus' special characteristics, the interval between
measurements is very large).  To prevent this from happening, we
measure the time $t_\mathrm{hot}$ that each configuration spends at
temperatures greater than $T_\mathrm{c}$.  In case any of the $4N_T$
configurations has a value of $t_\mathrm{hot}$ smaller than one third
of the median, we consider that the simulation is far too short for us
to measure $\tau_\mathrm{exp}$ and we simply double
$N_\mathrm{HB}$. Notice that this last criterion is unlike the others
in that it is not completely quantitative. It simply detects that our
starting point is very badly dimensioned.

As a final test, we have increased $N_\mathrm{min}$ by a factor of
$10$ for the first $1\%$ of the samples in all lattices. None of the
$\tau_\mathrm{exp}$ estimates changed within errors.

\section{Unbiased estimators of non-linear functions}\label{AP:NON-BIAS}

Non linear functions of thermal mean values, that are computed sample
by sample and afterwards averaged over disorder, are prone to suffer
systematic errors larger than the statistical ones. General cures for
this problem are known~\cite{ballesteros:97,hasenbusch:08b}. In our
case, the only such quantity is $\chi_\mathrm{link}$, defined in
eq.~(\ref{DEF:CHI-LINK}). Since we have 4 replicas,
the bias problem could be avoided for $\chi_\mathrm{link}$, see
eq.~(\ref{BIAS-CORRECTION2}) below.  However, if one decides instead to
face it, a nice test for the statistical quality of the data
is obtained.

Indeed, consider eq.~(\ref{DEF:CHI-LINK}), and let $[Q_\mathrm{link}]$
be our Monte Carlo estimate of $\langle Q_\mathrm{link}\rangle$ as
computed from $N$ measurements for a given sample. The expectation
value $\langle [Q_\mathrm{link}]^2\rangle$ is {\em not} $\langle
Q_\mathrm{link}\rangle^2$. To quantify the effect, we need some
notation~\cite{amit:05,sokal:97}.  The normalised equilibrium
autocorrelation function for $Q_\mathrm{link}$, at a given temperature
and for a given sample, is
\begin{equation}
\hat C_{Q_\mathrm{link}}(t)=\frac{\left\langle(Q_\mathrm{link}^{(s+t)}-\langle Q_\mathrm{link}\rangle) (Q_\mathrm{link}^{(s)}-\langle Q_\mathrm{link}\rangle) \right\rangle}{\langle Q_\mathrm{link}^2\rangle-\langle Q_\mathrm{link}\rangle^2},
\end{equation}
where $Q_\mathrm{link}^{(s)}$ stands for the value taken by
$Q_\mathrm{link}$ at time $s$. Two characteristic time scales are relevant
to us:
\begin{equation}
\tau_\mathrm{int,Q_\mathrm{link}}=\frac12\sum_{t=-\infty}^{t=+\infty} \hat C_{Q_\mathrm{link}}(t),\quad \tau_\mathrm{avg,Q_\mathrm{link}}=\frac{\sum_{t=-\infty}^{t=+\infty} |t|\hat C_{Q_\mathrm{link}}(t)}{\sum_{t=-\infty}^{t=+\infty} \hat C_{Q_\mathrm{link}}(t)}.
\end{equation}
Then a straightforward computation shows that 
\begin{equation}
\langle [Q_\mathrm{link}]^2\rangle= \langle Q_\mathrm{link}\rangle^2+
\frac{2 \tau_\mathrm{int,Q_\mathrm{link}} [\langle
    Q^2_\mathrm{link}\rangle-\langle
    Q_\mathrm{link}\rangle^2]}{N}\left(1-\frac{\tau_\mathrm{avg,Q_\mathrm{link}}}{N}\right),\label{BIAS}
\end{equation}
up to corrections of order ${\cal
  O}(\mathrm{e}^{-N/\tau_\mathrm{exp}})$ (the exponential
autocorrelation time $\tau_\mathrm{exp}$ was discussed in
\sref{SECT:THERMALIZATION-CRITERIA}). This computation is
performed in textbooks~\cite{amit:05,sokal:97} only to order $1/N$, and
with a rather different aim: it provides an estimate of the (squared)
statistical error in the Monte Carlo estimation of $\langle
Q_\mathrm{link}\rangle$. Our interest in eq.(\ref{BIAS}) is different.
It tells us that, when taking $[Q_\mathrm{link}]^2$ as $\langle
Q_\mathrm{link}\rangle^2$  we are incurring
in bias not only of order $1/N$, but also of order $1/N^2$.

It is easy to obtain bias-corrected estimators~\cite{ballesteros:97}:
one divides the Monte Carlo history in two halves, four quarters, and
eight eighths. Recall that we will be dropping in the analysis the
full first half of the Monte Carlo history.  Then, one computes
$[Q_\mathrm{link}]_{2/2}^2$ from the last half of the data as well as
$[Q_\mathrm{link}]_{3/4}^2$ and $[Q_\mathrm{link}]_{4/4}^2$ from the
third and fourth quarters respectively. Similarly, we compute
$[Q_\mathrm{link}]_{5/8}^2$, $[Q_\mathrm{link}]_{6/8}^2$,
$[Q_\mathrm{link}]_{7/8}^2$ and $[Q_\mathrm{link}]_{8/8}^2$.  Then,
eq.(\ref{BIAS}), the thermal expectation value of
\begin{equation}
Q^{(2)}_\mathrm{link,linear}=
2[Q_\mathrm{link}]_{2/2}^2-\frac{[Q_\mathrm{link}]_{3/4}^2+[Q_\mathrm{link}]_{4/4}^2}{2},\label{BIAS-CORRECTION-LINEAR}
\end{equation}
is $\langle Q_\mathrm{link}\rangle^2$, up to a bias of order $\tau_{\mathrm{int},Q_\mathrm{link}} \tau_{\mathrm{avg},Q_\mathrm{link}}/N^2$. We can do it even better:
\begin{eqnarray}
Q^{(2)}_\mathrm{link,quadratic}&=&
\frac{8}{3}[Q_\mathrm{link}]_{2/2}^2-2\frac{[Q_\mathrm{link}]_{3/4}^2+[Q_\mathrm{link}]_{4/4}^2}{2}\nonumber\\ &+& \frac{1}{3}\frac{[Q_\mathrm{link}]_{5/8}^2+[Q_\mathrm{link}]_{6/8}^2+[Q_\mathrm{link}]_{7/8}^2+[Q_\mathrm{link}]_{8/8}^2}{4},\label{BIAS-CORRECTION-QUADRATIC}
\end{eqnarray}
has thermal expectation value $\langle Q_\mathrm{link}\rangle^2$, up
to corrections of order ${\cal O}(\mathrm{e}^{-N/\tau_\mathrm{exp}})$.

The fact that we have 4 real replicas offers us an alternative way of
overcoming this problem. In fact, denoting by $Q_\mathrm{link}^{(ij)}$ the
link overlap computed from replicas $i$ and $j$, we have
\begin{equation}
Q^{(2)}_{\mathrm{link,4R}}=\frac{[Q_\mathrm{link}^{(12)} Q_\mathrm{link}^{(34)}+
Q_\mathrm{link}^{(13)} Q_\mathrm{link}^{(24)}+
Q_\mathrm{link}^{(14)} Q_\mathrm{link}^{(23)}]
}{3}\,,\label{BIAS-CORRECTION2}
\end{equation}
with $\langle Q^{(2)}_{\mathrm{link,4R}}\rangle= \langle
Q_\mathrm{link}\rangle^2$ (we average over the three equivalent
replica pairings to reduce statistical errors). The comparison of the
two procedures offers an interesting test on the statistical quality
of our data, because eq.(\ref{BIAS}) holds only for
$N\gg\tau_\mathrm{int,Q_\mathrm{link}},\tau_\mathrm{exp}\,.$

From the different estimators for the $\langle Q_\mathrm{link} \rangle^2$
we finally obtain four estimators of $\chi_\mathrm{link}$:
\begin{eqnarray}
\chi_\mathrm{link,2R}^\mathrm{biased}&=& V  \overline{\bigl( [Q_\mathrm{link}^2]-[Q_\mathrm{link}]^2\bigr)}\,,\label{CHI-2R-NO}\\
\chi_\mathrm{link,2R}^\mathrm{linear}&=& V  \overline{ \bigl([Q_\mathrm{link}^2]-Q^{(2)}_\mathrm{link,linear}\bigr)}\,,\label{CHI-2R-LINEAR}\\
\chi_\mathrm{link,2R}^\mathrm{quadratic}&=& V  \overline{\bigl( [Q_\mathrm{link}^2]-Q^{(2)}_\mathrm{link,quadratic}\bigr)}\,,\label{CHI-2R-QUADRATIC}\\
\chi_\mathrm{link,4R} &=& V  \overline{ \bigl([Q_\mathrm{link}^2]-Q^{(2)}_{\mathrm{link,4R}}\bigr)}\,.\label{CHI-4R}
\end{eqnarray}

\clearpage

\section*{References}

\providecommand{\newblock}{}

\end{document}